\documentclass[12pt]{article}
\usepackage{latexsym,cite,subeqn}
\input amssym.def
\input amssym.tex

\makeatletter  
\renewcommand{\theequation}{\thesection.\arabic{equation}}
\@addtoreset{equation}{section}
\makeatother

\font\msym=msbm10

\def\Com{{\mathop{\hbox{\msym \char  '103}}}} 
\def\Real{{\mathop{\hbox{\msym \char  '122}}}} 
 
\def\Z{{\mathop{\hbox{\msym\char '132}}}}

\def\tX{\tilde{{\rm X}}}

\def\tx{\tilde{{\rm x}}}
\def\x{{\rm x}} 
 
\def\ta{\tilde{{\rm a}}}
 
\def\da{\dot{\alpha}} 
\def\db{\dot{\beta}} 
\def\dg{\dot{\gamma}} 
\def\bt{\bar{\theta}}
\def\X{{\rm X}}

\begin{document}
\begin{titlepage}
\title{\vskip -60pt
{\small
\begin{flushright} 
KIAS-99019\\
hep-th/9903230
\end{flushright}}
\vskip 20pt
Superconformal Symmetry and Correlation Functions \\ ~~~\\~~~ }
\author{Jeong-Hyuck Park\thanks{E-mail address:\,\,jhp@kias.re.kr}}
\date{}
\maketitle
\vspace{-1.0cm}
\begin{center}
\textit{School of Physics, Korea Institute for Advanced Study}\\
\textit{207-43 Cheongryangri-dong, Dongdaemun-gu}\\
\textit{Seoul 130-012, Korea}\\
~~~\\
~~~\\
~~~\\
\end{center}
\begin{abstract}
\noindent Four-dimensional ${\cal N}$-extended superconformal  
symmetry  and correlation functions of quasi-primary superfields 
are  studied  within the superspace formalism. A superconformal Killing 
equation is derived and  its  solutions are  
classified in terms of   supertranslations, dilations, Lorentz
transformations, $R$-symmetry transformations and special superconformal 
transformations.  
In general, due to  the invariance under  supertranslations and special 
superconformal transformations,  superconformally  invariant   $n$-point  
functions  reduce to one  unspecified $(n-2)$-point function 
which must transform  homogeneously  under the 
remaining rigid transformations, i.e. dilations, Lorentz
transformations and  $R$-symmetry transformations. Based on this result, 
we are able to identify all the superconformal invariants and 
obtain the general form of $n$-point functions for  scalar superfields.  
In particular,  as a byproduct, 
a selection rule for correlation functions is  derived, the existence of which 
in ${\cal N}=4$ super Yang-Mills theory 
was previously  predicted  in  the context of $AdS$/CFT 
correspondence~\cite{9811047}. 
Superconformally covariant differential operators are also discussed. 
\end{abstract}
~\\
{\small
\begin{flushleft} 
\end{flushleft}}
\thispagestyle{empty}
\end{titlepage}
\newpage

\section{Introduction \& Summary\label{intro}}
Superconformal field theories have been of renewed attention after 
the Maldacena conjecture that the string/M theory on $AdS_{d+1}$ 
backgrounds is dual to a conformal field theory 
in a spacetime of dimension, $d$, which is interpreted as the boundary of  
$AdS_{d+1}$ \cite{9711200,9802109,9802150}.  
As all the known nontrivial conformal field theories in higher than 
two dimensions  are supersymmetric theories~\cite{sohniuswest,adlercollins,brink1,mandelstam,howestelle1,howestelle2,parkeswest,argyresplesser,eguchihori,howestelle3,9609161,9702038}, 
it is natural to consider a group which combines 
supersymmetry and conformal symmetry together, i.e. the 
superconformal group. In fact, the pioneering work on supersymmetry in 
four-dimensions~\cite{NPB7039} introduced the ${\cal N}=1$ superconformal
symmetry, though it is broken at quantum level. \newline 
\indent Contrary to the ordinary conformal symmetry,  
not  all spacetime dimensions allow superconformal symmetry. 
The standard supersymmetry algebra 
admits an extension to a superconformal algebra only 
if $d\leq 6$ \cite{nahm}\,(for a review see \cite{9712074}). 
In particular in four-dimensions, which is of our interest in this paper, 
the bosonic part of the superconformal algebra is
\begin{equation}
\mbox{o}(2,4)\oplus\mbox{u}({\cal N})\,.
\label{4bosonic4}
\end{equation}
Hence the four-dimensional superconformal group is identified with 
a supermatrix group, $\mbox{SU}(2,2|{\cal N})$ \cite{FP35537,paper2} or  
its complexification, 
$SL(4|{\cal N};\Com)$ \cite{howehartwell,9607161}. 
Normally ${\cal N}\geq 5$ cases are excluded from the renormalization 
point of view, as theories with more than four supercharges must have 
spins higher than one such as graviton/gravitino  and  it is unlikely 
that supergravity theories are renormalizable. 
However, such a restriction on the value of ${\cal N}$  
is not relevant to our work.\newline
\indent According to the conjecture \cite{9711200,9802109,9802150},  
four-dimensional ${\cal N}=4$ super Yang-Mills theory with gauge 
group $\mbox{SU}(N)$ is dual to type IIB string theory 
on $AdS_{5}\times S^{5}$ in the limit of small $g_{YM}$ and 
large but fixed 't Hooft coupling,  
$g_{YM}^{2}N$. In this limit, the string theory 
can be effectively described by tree level type IIB 
supergravity, while  the field theory dual is strongly coupled. 
As the  perturbative approach breaks down in the strongly coupled CFT  side,  
to check the conjectured duality  it is desirable to have non-perturbative 
understanding on super Yang-Mills theory. This motivates us to explore 
four-dimensional ${\cal N}$-extended superconformal  
symmetry and correlation functions subject to the symmetry 
as done in the present paper. \newline
\indent In our previous work~\cite{paper2}, six-dimensional $({\cal N},0)$ 
superconformal symmetry 
was analyzed in terms of  coordinate transformations on  superspace   
and through dimensional reduction basic features of four-dimensional 
${\cal N}$-extended 
superconformal symmetry were  obtained. In the present paper, 
in a similar fashion to \cite{paper2,paper1} but in a self-contained manner, 
we analyze 
four-dimensional  ${\cal N}$-extended superconformal symmetry 
on superspace.   Our main results concern the general forms of 
superconformally invariant $n$-point  functions 
for quasi-primary superfields. In particular, as a byproduct, we obtain 
a selection rule for correlation 
functions of the component fields, $\psi^{I}(x)$,   appearing 
in the power series expansions of quasi-primary superfields 
in Grassmann coordinates, $\theta$ and $\bar{\theta}$. 
The selection rule  states that  
if the the sum of the $R$-symmetry charge, $\kappa_{i}$,  is  not zero then 
the correlation function  of the component fields vanishes 
\begin{equation}
\begin{array}{cc}
\langle \psi^{I_{1}}_{1}(x_{1})\cdots\psi^{I_{n}}_{n}(x_{n})\rangle=0
~~~~~~&~~~~~~\mbox{if~~~}
\displaystyle{\sum_{i=1}^{n}\kappa_{i}\neq 0}\,.
\end{array}
\label{selection}
\end{equation}
The existence of this kind of selection rule in ${\cal N}=4$ super Yang-Mills 
theory was previously predicted  by  Intriligator 
within  the context of $AdS$/CFT correspondence, as the dual IIB supergravity contains a corresponding $\mbox{U}(1)$ symmetry~\cite{9811047}. 
Therefore our results provide a supporting evidence for the Maldacena 
conjecture, as the selection rule here is derived 
by purely considering the symmetry on CFT side 
without referring to the string side. \newline
\newline
\indent  The contents of the present paper are as follows.  
In section~\ref{4scfs}, 
we  first define the four-dimensional ${\cal N}$-extended  
superconformal group in terms of 
coordinate transformations on superspace as  a generalization
of the definition of ordinary conformal transformations. We then
derive a superconformal Killing equation,  
which is a necessary and sufficient  condition for a supercoordinate 
transformation to be superconformal.  
The general solutions are  identified in terms of  
supertranslations, dilations, Lorentz 
transformations, $R$-symmetry transformations and special superconformal 
transformations, where $R$-symmetry is given by  $\mbox{U}({\cal N})$ 
as in eq.(\ref{4bosonic4}) and supertranslations and special superconformal transformations are dual to each other through superinversion map.  
The four-dimensional ${\cal N}$-extended  superconformal 
group is then  identified with  a  
supermatrix group, $\mbox{SU}(2,2|{\cal N})$, having  dimensions 
$(15+{\cal N}^{2}|8{\cal N})$ as known.  However, we point out that 
for ${\cal N}=4$ case       
an equivalence relation must be imposed on the supermatrix group and so 
the four-dimensional ${\cal N}=4$ superconformal 
group is isomorphic to a  quotient group of the  supermatrix group.  \newline
\indent In section~\ref{4coset}, we obtain an explicit formula for
the finite nonlinear 
superconformal transformations of the supercoordinates, $z$,
parameterizing superspace and discuss several 
representations of the superconformal group.  
We also construct matrix or vector valued 
functions depending on  two or   three points 
in superspace which transform 
covariantly under superconformal transformations.  
For two points, $z_{1}$ and $z_{2}$, we find a matrix, 
$I(z_{1},z_{2})$, which transforms covariantly  like a
product of two tensors at $z_{1}$ and $z_{2}$.   
For three points, $z_{1},z_{2},z_{3}$, we find `tangent' vectors, 
$Z_{i}$, which transform homogeneously at $z_{i},\,i=1,2,3$. 
These  are crucial variables for  
obtaining two-point, three-point and general $n$-point 
correlation functions later. \newline
\indent In section~\ref{4correlation}, we discuss the    
superconformal invariance of correlation 
functions for  quasi-primary superfields and exhibit general forms of   
two-point, three-point and $n$-point functions. 
Explicit formulae for two-point  functions of   
superfields in various cases are given.  
In general, due to  the  invariance under    
supertranslations and special superconformal transformations,  
$n$-point functions  reduce to one  unspecified $(n-2)$-point function 
which must transform  homogeneously  under the 
rigid transformations - dilations, Lorentz
transformations and  $R$-symmetry transformations.  
We   then identify all the superconformal invariants and 
obtain the general form of $n$-point functions of scalar 
superfields. As a byproduct, we derive  
the  selection rule for correlation functions~(\ref{selection}).\newline
\indent In section \ref{4operator},  superconformally  
covariant differential operators are  discussed.  The 
conditions for 
superfields, which are  formed by the action of  
spinor derivatives on quasi-primary superfields, to remain 
quasi-primary are obtained.  In general, the action of 
differential operator on quasi-primary fields generates an anomalous term 
under superconformal transformations. However, with a suitable choice of 
scale dimension and $R$-symmetry charge, we show that 
the anomalous term may be cancelled. 
We regard this analysis as a necessary step to  write superconformally 
invariant actions on superspace,  as the kinetic terms in such theories may 
consist of superfields formed by the action of  
spinor derivatives on quasi-primary superfields.  \newline
\indent   In the appendix,  the explicit form of superconformal algebra and a 
method of solving the superconformal Killing equation are exhibited. \newline 
\newline
\indent Recent review on the implications of ${\cal N}=1$ superconformal symmetry for four-dimensional quantum field theories is contained in \cite{9808041} 
and some related works on superconformally invariant  correlation 
functions can be found in \cite{molotkov,aneva3,howewest1,howewest2,howewest3,howewest4,anselmifgj,9901128}. ${\cal N}=1$ superconformal symmetry on curved 
superspace is studied in \cite{9804053,9809090,9811209} and  
conformally covariant differential 
operators in  non-supersymmetric theories are 
discussed in \cite{9704108,9708040}.

\section{Superconformal Symmetry in Four-dimensions\label{4scfs}}
In this section  we  first define the four-dimensional ${\cal N}$-extended  
superconformal group on superspace and then discuss  its    
superconformal Killing equation  along with the solutions.
\subsection{Four-dimensional Superspace}
The four-dimensional supersymmetry algebra has the standard\footnote{See Appendix \ref{AppendixA} for our notations and some useful equations.}  form 
with $P_{\mu}=(H,-\textbf{P})$ 
\begin{equation}
\begin{array}{c}
\{Q_{a\alpha}, \bar{Q}^{b}_{\dot{\alpha}}\}
=2\delta_{a}^{~b}\sigma^{\mu}_{\alpha\dot{\alpha}}P_{\mu}\,,\\
{}\\
{}[{P}_{\mu},{P}_{\nu}]=[{P}_{\mu},Q_{a\alpha}]=
[P_{\mu},\bar{Q}^{a}_{\dot{\alpha}}]=\{Q_{a\alpha},Q_{b\beta}\}=
\{\bar{Q}^{a}_{\dot{\alpha}},\bar{Q}^{b}_{\dot{\beta}}\}=0\,,
\end{array}
\label{4Dsusy}
\end{equation}
where $1\leq \alpha,\da\leq 2$,\,   $1\leq a\leq {\cal N}$ and  
$Q_{a\alpha}$ satisfies
\begin{equation}
Q_{a\alpha}{}^{\dagger}=\bar{Q}^{a}_{\da}\,.
\label{4QbarQ}
\end{equation}
$P_{\mu},Q_{a\alpha}$ and $\bar{Q}^{a}_{\dot{\alpha}}$ generate a
supergroup, $\mbox{G}_{T}$, 
with parameters,  
$z^{M}= (x^{\mu},\theta^{a\alpha},\bar{\theta}^{\dot{\alpha}}_{a})$,  
which
are coordinates on superspace.  The  general element of $\mbox{G}_{T}$
is written in terms of these coordinates as
\begin{equation}
g(z)=\displaystyle{e^{i(x{\cdot P}+\theta^{a}Q_{a}
+\bar{Q}^{a}\bar{\theta}_{a})}}\,.
\label{4generalelement}
\end{equation}
Corresponding to eq.(\ref{4QbarQ}) we may assume $\theta^{a\alpha}$ to satisfy 
\begin{equation}
\theta^{a\alpha}{}^{\ast}=\bt_{a}^{\da}\,,
\end{equation}
so that 
\begin{equation}
g(z)^{\dagger}=g(z)^{-1}=g(-z)\,.
\end{equation}
\indent The Baker-Campbell-Haussdorff formula  with the
supersymmetry algebra~(\ref{4Dsusy})  gives
\begin{equation}
g(z_{1})g(z_{2})=g(z_{3})\,,
\label{4groupproperty1}
\end{equation}
where
\begin{equation}
\begin{array}{ccc}
x^{\mu}_{3}=x^{\mu}_{1}+x^{\mu}_{2}+i\theta_{1}^{a}\sigma^{\mu}\bt_{2a}
-i\theta_{2}^{a}\sigma^{\mu}\bar{\theta}_{1a}\,,~~~&~~~
\theta^{a}_{3}=\theta_{1}^{a}+\theta_{2}^{a}\,,~~~&~~~
\bt_{3a}=\bt_{1a}+\bt_{2a}\,.
\end{array}
\label{4groupproperty2}
\end{equation}
Letting $z_{1}\rightarrow -z_{2}$ we may get the 
supertranslation  invariant  one forms,  $e^{M}=(e^{\mu},
{\rm d}\theta^{a\alpha},{\rm
d}\bar{\theta}{}_{a}^{\dot{\alpha}})$, where 
\begin{equation}
e^{\mu}(z)={\rm d}x^{\mu}+i{\rm
d}\theta^{a}\sigma^{\mu}\bar{\theta}_{a}-i\theta^{a}\sigma^{\mu}
{\rm d}\bar{\theta}_{a}\,.
\label{4infint}
\end{equation}
\indent The exterior derivative,~${\rm d}$, on superspace  is 
defined as
\begin{equation}
{\rm d}
\equiv {\rm d}z^{M}\frac{\partial~}{\partial z^{M}}
=e^{M}D_{M}=e^{\mu}\partial_{\mu}+{\rm 
d}\theta^{a\alpha}D_{a\alpha}
-{\rm d}\bar{\theta}^{\dot{\alpha}}_{a}\bar{D}{}^{a}_{\dot{\alpha}}\,,
\end{equation}
where $D_{M}
=(\partial_{\mu},D_{a\alpha},-\bar{D}{}_{\dot{\alpha}}^{a})$ are 
covariant  derivatives  
\begin{equation}
\begin{array}{ccc}
\displaystyle{\partial_{\mu}=\frac{\partial~}{\partial x^{\mu}}}\,,~~~&~~~
\displaystyle{
D_{a\alpha}=\frac{\partial~}{\partial\theta^{a\alpha}}
-i(\sigma^{\mu}\bar{\theta}_{a})_{\alpha}\frac{\partial~}{\partial x^{\mu}}}\,,~~~&~~~\displaystyle{
\bar{D}{}^{a}_{\dot{\alpha}}=
-\frac{\partial~}{\partial\bar{\theta}^{\dot{\alpha}}_{a}}+
i(\theta^{a}\sigma^{\mu})_{\dot{\alpha}}\frac{\partial~}{\partial x^{\mu}}}\,,
\end{array}
\end{equation}
satisfying the anti-commutator relations
\begin{equation}
\displaystyle{
\{D_{a\alpha},\bar{D}{}^{b}_{\dot{\alpha}}\}=
2i\delta_{a}^{~b}\sigma_{\alpha\dot{\alpha}}^{\mu}\partial_{\mu}}\,.
\label{4anticomD}
\end{equation}
Under an arbitrary superspace coordinate transformation,   
$z \longrightarrow z^{\prime}$,   
$e^{M}$ and $D_{M}$  transform as
\begin{equation}
\begin{array}{cc}
e^{M}(z^{\prime})
=e^{N}(z){\cal R}_{N}{}^{M}(z)\,,~~~~&~~~~
D^{\prime}_{M}={\cal R}^{-1}{}_{M}{}^{N}(z)D_{N}\,,
\end{array}
\label{4ReD}
\end{equation}
so that the exterior derivative is left invariant
\begin{equation}
e^{M}(z)D_{M}=e^{M}(z^{\prime})D^{\prime}_{M}\,,
\end{equation}
where ${\cal R}_{M}{}^{N}(z)$ is a $(4+4{\cal N})\times (4+4{\cal N})$ 
supermatrix of the form 
\begin{equation}
{\cal R}_{M}{}^{N}(z)=
\left(\begin{array}{rrr}
R^{~\nu}_{\mu}(z)&\partial_{\mu}\theta^{\prime b\beta}&
\partial_{\mu}\bar{\theta}_{b}^{\prime\db}\\
B^{\mu}_{a\alpha}(z)&D_{a\alpha}\theta^{\prime b\beta}&
D_{a\alpha}\bar{\theta}_{b}^{\prime\db}\\
-\bar{B}^{a\mu}_{\da}(z)&-\bar{D}{}^{a}_{\da}\theta^{\prime b\beta}&
-\bar{D}{}^{a}_{\da}\bar{\theta}_{b}^{\prime\db}
\end{array}\right)\,,
\label{4calRorig}
\end{equation}
with
\begin{eqnarray}
&\displaystyle{R^{~\nu}_{\mu}(z)=\frac{\partial 
x^{\prime \nu}}{\partial x^{\mu}}
+i\frac{\partial\theta^{\prime a}}{\partial 
x^{\mu}}\sigma^{\nu}\bar{\theta}{}^{\prime}_{a}   
-i\theta^{\prime a}\sigma^{\nu}
\frac{\partial\bar{\theta}{}^{\prime}_{a}}{\partial 
x^{\mu}}}\,,~~~~~~~~~~~~~~~~~~~~~~~~~\label{4Rh}\\
&{}\nonumber\\
&B^{\mu}_{a\alpha}(z)=D_{a\alpha}x^{\prime\mu}
+iD_{a\alpha}\theta^{\prime b}\sigma^{\mu}\bar{\theta}^{\prime}_{b}+
i\theta^{\prime b}\sigma^{\mu}D_{a\alpha}\bar{\theta}^{\prime}_{b}\,,
~~~~~~~~~~~~~~~\label{4B}\\
&{}\nonumber\\
&\bar{B}^{a\mu}_{\da}(z)=\bar{D}{}^{a}_{\dot{\alpha}}x^{\prime\mu}
+i\bar{D}{}^{a}_{\dot{\alpha}}
\theta^{\prime b}\sigma^{\mu}\bar{\theta}^{\prime}_{b}+
i\theta^{\prime b}\sigma^{\mu}\bar{D}{}^{a}_{\dot{\alpha}}
\bar{\theta}^{\prime}_{b}=(B^{\mu}_{a\alpha}(z))^{\dagger}\,.\label{4Bbar}
\end{eqnarray}

\subsection{Superconformal Group \&  Killing Equation}
The superconformal group is  defined here as a group of 
superspace coordinate
transformations, $z \stackrel{g}{\longrightarrow} z^{\prime}$,  that 
preserve the 
infinitesimal supersymmetric interval length, $e^{2}=\eta_{\mu\nu}
e^{\mu}e^{\nu}$ up to a local scale factor, so that
\begin{equation}
{e^{2}}(z)~\rightarrow~e^{2}(z^{\prime})=\Omega^{2}(z;g)e^{2}(z)\,, 
\label{4scdef}
\end{equation}
where $\Omega(z;g)$ is a local scale factor.\newline
This requires $B^{\mu}_{a\alpha}(z)=\bar{B}^{a\mu}_{\da}(z)=0$ 
\begin{equation}
\begin{array}{l}
D_{a\alpha}x^{\prime\mu}
+iD_{a\alpha}\theta^{\prime b}\sigma^{\mu}\bar{\theta}^{\prime}_{b}+
i\theta^{\prime b}\sigma^{\mu}D_{a\alpha}\bar{\theta}^{\prime}_{b}=0\,,\\
{}\\
\bar{D}{}^{a}_{\dot{\alpha}}x^{\prime\mu}
+i\bar{D}{}^{a}_{\dot{\alpha}}
\theta^{\prime b}\sigma^{\mu}\bar{\theta}^{\prime}_{b}+
i\theta^{\prime b}\sigma^{\mu}\bar{D}{}^{a}_{\dot{\alpha}}
\bar{\theta}^{\prime}_{b}=0\,,
\end{array}
\label{4BB=0}
\end{equation}
and
\begin{eqnarray}
&e^{\mu}(z^{\prime})=e^{\nu}(z)R^{~\mu}_{\nu}(z;g)\,,&\label{4ehomogeneous}\\
&{}&\nonumber\\
&R^{~\lambda}_{\mu}(z;g)R^{~\rho}_{\nu}(z;g)\eta_{\lambda\rho}
=\Omega^{2}(z;g)\eta_{\mu\nu}\,,~~~~~~~~~
\det R(z;g)=\Omega^{4}(z;g)\,.\label{4R2}&
\end{eqnarray}
Hence ${\cal R}_{M}{}^{N}$ in eq.(\ref{4calRorig}) is of the 
form\footnote{More explicit form of ${\cal R}_{M}{}^{N}$ is obtained later 
in eq.(\ref{4calRR}).}
\begin{equation}
{\cal R}_{M}{}^{N}(z;g)=
\left(\begin{array}{crr}
R^{~\nu}_{\mu}(z;g)&\partial_{\mu}\theta^{\prime b\beta}&
\partial_{\mu}\bar{\theta}_{b}^{\prime\db}\\
0&D_{a\alpha}\theta^{\prime b\beta}&
D_{a\alpha}\bar{\theta}_{b}^{\prime\db}\\
0&-\bar{D}{}^{a}_{\da}\theta^{\prime b\beta}&
-\bar{D}{}^{a}_{\da}\bar{\theta}_{b}^{\prime\db}
\end{array}\right)\,.
\label{4calR}
\end{equation}
$R_{\mu}^{~\nu}(z;g)$ is a representation of the superconformal group. Under  
the successive superconformal transformations, 
$z\stackrel{g}{\longrightarrow}z^{\prime}
\stackrel{g^{\prime}}{\longrightarrow}z^{ \prime\prime}$ giving 
$z\stackrel{g^{\prime\prime}}{\longrightarrow}z^{\prime\prime}$, we have
\begin{equation}
R(z;g)R(z^{\prime};g^{\prime})=R(z;g^{\prime\prime})\,.
\label{4Rrep}
\end{equation}
\indent Infinitesimally   
$z^{\prime}\simeq z+\delta z$, eq.(\ref{4BB=0})  gives
\begin{equation}
\begin{array}{cc}
D_{a\alpha}h^{\mu}=-2i(\sigma^{\mu}\bar{\lambda}_{a})_{\alpha}\,,~~~~&~~~~
\bar{D}{}^{a}_{\dot{\alpha}}h^{\mu}
=2i(\lambda^{a}\sigma^{\mu})_{\dot{\alpha}}\,,
\end{array}
\label{4enough}
\end{equation}
where we define
\begin{equation}
\begin{array}{cc}
\lambda^{a}=\delta\theta^{a}\,,~~~&~~~
\bar{\lambda}_{a}=\delta\bar{\theta}_{a}\,,\\
{}&{}\\
\multicolumn{2}{c}{
h^{\mu}=\delta
x^{\mu}+i\delta\theta^{a}\sigma^{\mu}\bar{\theta}_{a}-
i\theta^{a}\sigma^{\mu}\delta\bar{\theta}_{a}\,.}
\end{array}
\end{equation}
Infinitesimally  $R_{\mu}^{~\nu}$ from eq.(\ref{4Rh}) is of the form
\begin{equation} 
R_{\mu}^{~\nu}\simeq\delta_{\mu}^{~\nu}+\partial_{\mu}h^{\nu}\,,
\label{4infR}
\end{equation}
so that the condition~(\ref{4R2}) reduces to the ordinary conformal
Killing equation
\begin{equation}
\partial_{\mu}h_{\nu}+\partial_{\nu}h_{\mu}\propto \eta_{\mu\nu}\,.
\label{4ordiKi}
\end{equation}
Eq.(\ref{4ordiKi}) follows from eq.(\ref{4enough}).  Using   
the anti-commutator relation for  
$D_{a\alpha}$ and $\bar{D}{}^{b}_{\da}$~(\ref{4anticomD})
we get from eq.(\ref{4enough})
\begin{equation}
\delta_{a}^{~b}\partial_{\nu}h^{\mu}
=\textstyle{\frac{1}{2}}\left(
D_{a\alpha}(\lambda^{b}\sigma^{\mu}\tilde{\sigma}_{\nu})^{\alpha}
-(\tilde{\sigma}_{\nu}\sigma^{\mu}\bar{D}{}^{b}_{\da}\bar{\lambda}_{a})^{\da}
\right)\,,
\end{equation}
and hence
\begin{equation}
\delta_{a}^{~b}(\partial_{\mu}h_{\nu}+\partial_{\nu}h_{\mu})
=(D_{a\alpha}\lambda^{b\alpha}
-\bar{D}{}^{b}_{\da}\bar{\lambda}^{\da}_{a})\eta_{\mu\nu}\,,
\label{4ordinaryKilling}
\end{equation}
which implies eq.(\ref{4ordiKi}).   
Thus eq.(\ref{4enough}) is  a necessary and  sufficient  condition for a 
supercoordinate transformation to be superconformal.\newline
\indent   With the notation as in eq.(\ref{4vvt}) we write 
\begin{equation}
\begin{array}{cc}
{\rm h}_{\alpha\da}=h^{\mu}\sigma_{\mu\alpha\da}\,,~~~~~~&~~~~~
\tilde{{\rm h}}^{\da\alpha}=h^{\mu}\tilde{\sigma}_{\mu}^{\da\alpha}\,,
\end{array}
\end{equation}
and using eq.(\ref{4contract}), eq.(\ref{4enough})
is equivalent to
\begin{equation}
\begin{array}{cc}
D_{a(\alpha}{\rm h}_{\beta)\db}=0\,,~~~~&~~~~
\bar{D}^{a}_{(\da}{\rm h}_{|\beta|\db)}=0\,,
\end{array}
\label{4masternew}
\end{equation}
or 
\begin{equation}
\begin{array}{cc}
D_{a\alpha}\tilde{{\rm h}}^{\db\beta}=
\textstyle{\frac{1}{2}}\delta_{\alpha}^{~\beta}D_{a\gamma}
\tilde{{\rm h}}^{\dot{\beta}\gamma}\,,~~~~&~~~~
\bar{D}^{a}_{\da}\tilde{{\rm h}}^{\db\beta}=\textstyle{\frac{1}{2}}
\delta_{\da}^{~\db}\bar{D}^{a}_{\dot{\gamma}}
\tilde{{\rm h}}^{\dot{\gamma}\beta}\,,
\end{array}
\label{4master}
\end{equation}
while $\lambda^{a\alpha},\bar{\lambda}_{a}^{\da}$ are given by
\begin{equation}
\begin{array}{cc}
\lambda^{a\alpha}=-i\textstyle{\frac{1}{8}}\bar{D}^{a}_{\da}\tilde{{\rm
h}}^{\da\alpha}\,,~~~~&~~~~
\bar{\lambda}_{a}^{\da}=i\textstyle{\frac{1}{8}}D_{a\alpha}\tilde{{\rm
h}}^{\da\alpha}\,.
\end{array}
\end{equation}
Eq.(\ref{4masternew}) may therefore be regarded as the fundamental
superconformal Killing equation and its solutions  give 
the generators of  extended superconformal transformations in 
four-dimensions. The general solution is\footnote{A method of obtaining 
the solution~(\ref{4solutionforh}) is demonstrated  
in Appendix \ref{4AppendixB}.} 
\begin{equation}
\begin{array}{ll}
\tilde{{\rm h}}(z)=&\tx_{-}{\rm b}\tx_{+}
-\tx_{-}(\tilde{w}-\textstyle{\frac{1}{2}}\lambda+4\rho_{a}\theta^{a})
+(w+\textstyle{\frac{1}{2}}\lambda-4\bt_{a}\bar{\rho}^{a})\tx_{+}
\\
{}&{}\\
{}&\,-4i\bt_{a}t^{a}_{~b}\theta^{b}
+2\Omega\bt_{a}\theta^{a}+4i(\bar{\varepsilon}_{a}\theta^{a}
-\bt_{a}\varepsilon^{a})+\tilde{{\rm a}}\,,
\end{array}
\label{4solutionforh}
\end{equation}
where $a^{\mu},b^{\mu},\lambda\in\Real,\,
\Omega\in S^{1},\,t\in\mbox{su}({\cal N})$ and for 
$w_{\mu\nu}=-w_{\nu\mu}$ we define
\begin{equation}
\begin{array}{cc}
w=\textstyle{\frac{1}{4}}w_{\mu\nu}\tilde{\sigma}^{\mu}\sigma^{\nu}\,,
~~~~&~~~~
\tilde{w}=\textstyle{\frac{1}{4}}w_{\mu\nu}\sigma^{\mu}\tilde{\sigma}^{\nu}\,.
\end{array}
\label{4wdef}
\end{equation}
For later use it is worth to note
\begin{equation}
\begin{array}{ccc}
\epsilon\tilde{w}{}^{t}\epsilon^{-1}=-\tilde{w}\,,~~~~~&~~~~
\bar{\epsilon}^{-1}w^{t}\bar{\epsilon}=-w\,,~~~~&~~~\tilde{w}{}^{\dagger}=-w\,.
\end{array}
\label{4wt}
\end{equation}
Eq.(\ref{4solutionforh}) also gives
\begin{equation}
\lambda^{a}=\varepsilon^{a}+\textstyle{\frac{1}{2}}(\lambda+i\Omega)
\theta^{a}-\theta^{a}\tilde{w}
+t^{a}_{~b}\theta^{b}+\theta^{a}
{\rm b}
\tx_{+}-i\bar{\rho}^{a}\tx_{+}-4(\theta^{a}\rho_{b})\,\theta^{b}\,,
\label{4sol1}
\end{equation}
and 
\begin{equation}
\delta \tx_{+}=
\tx_{+}{\rm b}\tx_{+}-4
\tx_{+}\rho_{a}\theta^{a}+\lambda
\tx_{+}+w\tx_{+}
-\tx_{+}\tilde{w}
+4i\bar{\varepsilon}_{a}\theta^{a}+\tilde{{\rm a}}\,.~~~~~~~~~
\label{4sol2}
\end{equation} 
Note that $\delta\tx_{+},\lambda^{a}$ are functions of
$\tx_{+},\theta^{a}$ only, which can be also directly shown from
eq.(\ref{4enough}). In fact, the superconformal group can be obtained
alternatively by imposing the super-diffeomorphisms to leave the
chiral subspaces of superspace invariant. The chiral structures are
given by
\begin{equation}
\begin{array}{cc}
z_{+}^{M_{+}}=(x^{\mu}_{+},\theta^{a\alpha})\,,~~~~&~~~~
z_{-}^{M_{-}}=(x^{\mu}_{-},\bt^{a}_{\da})\,.
\end{array}
\end{equation}
In this approach, one needs to  solve a reality 
condition~\cite{buchbinder,westscf} 
\begin{equation}
\delta\tx_{+}(\tx_{+},\theta)-\delta\tx_{-}(\tx_{-},\bt)=
4i\bar{\lambda}_{a}(\tx_{-},\bt)\theta^{a}
+4i\bt_{a}\lambda^{a}(\tx_{+},\theta)\,.
\end{equation}

\subsection{Extended Superconformal Transformations} 
In summary, the generators of  extended superconformal transformations in
four-dimensions acting  on the  four-dimensional  superspace, 
$\Real^{4|4{\cal N}}$, with coordinates, $z^{M}
=(x^{\mu},\theta^{a},\bt_{a})$, 
can be classified as  
\begin{enumerate}
\item  Supertranslations, $a,\varepsilon,\bar{\varepsilon}$
\begin{equation}
\begin{array}{ccc}
\delta x^{\mu}=a^{\mu}+i\varepsilon^{a}\sigma^{\mu}\bt_{a}
-i\theta^{a}\sigma^{\mu}\bar{\varepsilon}_{a}\,,~~~&~~~
\delta\theta^{a}=\varepsilon^{a}\,,~~~&~~~
\delta\bt_{a}=\bar{\varepsilon}_{a}\,.
\end{array}
\label{4trans}
\end{equation}
This is consistent with  eq.(\ref{4groupproperty2}).
\item Dilations, $\lambda$
\begin{equation}
\begin{array}{ccc}
\delta x^{\mu}=\lambda x^{\mu}\,,~~~&~~~
\delta\theta^{a}=\textstyle{\frac{1}{2}}\lambda\theta^{a}\,,~~~&~~~
\delta\bt_{a}=\textstyle{\frac{1}{2}}\lambda\bt_{a}\,.
\end{array}
\end{equation}
\item Lorentz transformations, with $w,\tilde{w}$ 
defined in eq.(\ref{4wdef})
\begin{equation}
\begin{array}{ccc}
\delta x^{\mu}=w^{\mu}_{~\nu}x^{\nu}\,,~~~&~~~ 
\delta\theta^{a}=-\theta^{a}\tilde{w}\,,~~~&~~~
\delta\bt_{a}=w\bt_{a}\,.
\end{array}
\end{equation}
\item $R$-symmetry transformations, $\mbox{U}({\cal N})$, of dimension
${\cal N}^{2}$, $t,\Omega$
\begin{equation}
\begin{array}{ccc}
\delta x^{\mu}=0\,,~~~&~~~
\delta\theta^{a}=t^{a}_{~b}\theta^{b}+
i\textstyle{\frac{1}{2}}\Omega\theta^{a}\,,~~~&~~~
\delta\bt_{a}=-\bt_{b}t^{b}_{~a}-i\textstyle{\frac{1}{2}}\Omega\bt_{a}\,,
\end{array}
\end{equation}
where the ${\cal N}\times {\cal N}$ matrix,~$t$, is a $\mbox{SU}({\cal N})$ generator, i.e. 
$t^{\dagger}=-t,~t^{a}_{~a}=0$ and $\Omega\in S^{1}$. 
\item Special superconformal transformations, $b,\rho,\bar{\rho}$ 
\begin{equation}
\begin{array}{l}
\delta x^{\mu}=2x{\cdot b}\,x^{\mu}-x^{2}b^{\mu}
+\theta^{a}\sigma^{\mu}\tx_{+}\rho_{a}+\bar{\rho}^{a}\tx_{-}\sigma^{\mu}\bt_{a}
+2\theta^{a}{\rm b}\bt_{b}\theta^{b}\sigma^{\mu}\bt_{a}\,,\\
{}\\
\delta\theta^{a}=\theta^{a}{\rm b}\tx_{+}-i\bar{\rho}^{a}\tx_{+}
-4(\theta^{a}\rho_{b})\theta^{b}\,,\\
{}\\
\delta\bt_{a}=\tx_{-}{\rm b}\bt_{a}+i\tx_{-}\rho_{a}
-4\bt_{b}(\bar{\rho}^{b}\bt_{a})\,.
\end{array}
\label{4special}
\end{equation}
\end{enumerate}

\subsection{Superinversion}
In four-dimensions we define superinversion, 
$z^{M}\stackrel{i_{s}}{\longrightarrow}z^{\prime M}
=(x^{\prime\mu},\theta^{\prime a\alpha},\bt^{\prime\da}_{a})
\in\Real^{4|4{\cal N}}$, by
\begin{equation}
\begin{array}{ccc}
x^{\prime\mu}_{\pm}=-\displaystyle{\frac{x_{\mp}^{\mu}}{x_{\mp}^{2}}}
\,,~~~~&~~~~
\theta^{\prime}{}^{at}=\displaystyle{
-i\frac{1}{x_{-}^{2}}\,\epsilon^{-1}{\rm x}_{-}\bar{\theta}_{b}\,
\zeta^{ba}}\,,
~~~~&~~~~
\bar{\theta}^{\prime}_{a}=i\displaystyle{\frac{1}{x_{+}^{2}}\,
\bar{\epsilon}^{-1}{\rm x}_{+}^{t}\theta^{bt}\,\bar{\zeta}_{ba}}\,,
\end{array}
\label{4superinversionbeta}
\end{equation} 
where ${\cal N}\times {\cal N}$ matrices, $\zeta^{ab}, \bar{\zeta}_{ab}$ satisfy
\begin{equation}
\begin{array}{ccc}
\zeta^{ab}\bar{\zeta}_{bc}=\delta^{a}_{~c}\,,~~~~&~~~~\zeta^{ab}=\zeta^{ba}\,,
~~~~&~~~~
\bar{\zeta}_{ab}=(\zeta^{ab})^{\ast}\,.
\end{array}
\end{equation}
Eq.(\ref{4superinversionbeta}) may be  rewritten as 
\begin{equation}
\begin{array}{ll}
\theta^{\prime}{}^{a}=i\displaystyle{\frac{1}{x_{-}^{2}}
\,\tilde{\bar{\theta}}{}^{a}\tx_{-}}\,,~~~~&~~~~~
\tilde{\bar{\theta}}{}^{a}
=-\zeta^{ba}\bar{\theta}_{b}^{t}\bar{\epsilon}\,,\\
{}&{}\\
\bar{\theta}^{\prime}_{a}=-i\displaystyle{\frac{1}{x_{+}^{2}}
\,\tx_{+}\tilde{\theta}_{a}}\,,~~~~&~~~~~
\tilde{\theta}_{a}=\epsilon\theta^{b}{}^{t}\bar{\zeta}_{ba}\,.
\end{array}
\label{4superinversion}
\end{equation}
It is easy to verify that superinversion is idempotent
\begin{equation}
i_{s}^{2}=1\,.
\end{equation}
Using
\begin{equation}
\begin{array}{cc}
\tilde{{\rm e}}=e^{\mu}\tilde{\sigma}_{\mu}={\rm
d}\tx_{+}-4i\bt_{a}{\rm d}\theta^{a}\,,~~~~&~~~~
{\rm e}=e^{\mu}\sigma_{\mu}={\rm d}{\rm x}_{+}+4i{\rm d}\tilde{\theta}_{a}
\tilde{\bt}{}^{a}\,,
\end{array}
\end{equation}
we get under superinversion
\begin{equation}
\begin{array}{cc}
\tilde{{\rm e}}(z^{\prime})={\rm x}_{+}^{-1}{\rm e}(z){\rm x}_{-}^{-1}
\,,~~~~&~~~~
{\rm e}(z^{\prime})=\tx_{-}^{-1}\tilde{{\rm e}}(z)\tx_{+}^{-1}\,,
\end{array}
\label{4ee}
\end{equation}
and hence
\begin{equation}
\begin{array}{cc}
e^{2}(z^{\prime})=\Omega^{2}(z;i_{s})e^{2}(z)\,,~~~~&~~~~
\Omega(z;i_{s})=\displaystyle{\frac{1}{\sqrt{x^{2}_{+}x^{2}_{-}}}}\,.
\end{array}
\label{4Oi}
\end{equation}
Eq.(\ref{4ee}) can be rewritten as 
\begin{equation}
\begin{array}{c}
e^{\mu}(z^{\prime})=e^{\nu}(z)R_{\nu}{}^{\mu}(z;i_{s})\,,\\
{}\\
R_{\nu}{}^{\mu}(z;i_{s})=\textstyle{\frac{1}{2}}
\mbox{tr}(\tx_{-}^{-1}\tilde{\sigma}_{\nu}
\tx_{+}^{-1}\tilde{\sigma}^{\mu})=\textstyle{\frac{1}{2}}
\mbox{tr}(\rm{x}_{+}^{-1}\sigma_{\nu}\rm{x}_{-}^{-1}\sigma^{\mu})\,,
\end{array}
\label{4Rzi}
\end{equation}
or using eq.(\ref{4Tr4})
\begin{equation}
R^{\mu\nu}(z;i_{s})=\displaystyle{\frac{1}{x_{+}^{2}x_{-}^{2}}}
(x_{+}^{\mu}x_{-}^{\nu}+x^{\mu}_{-}x_{+}^{\nu}-\eta^{\mu\nu}x_{+}{\cdot
x_{-}}
+i\epsilon^{\mu\nu}{}_{\lambda\rho}x_{-}^{\lambda}x_{+}^{\rho})\,.   
\end{equation}
\indent If we consider a transformation,$~z\,
\stackrel{i_{s}\circ g\circ i_{s}}{-\!\!\!-\!\!\!-\!\!\!\longrightarrow}\,z^{\prime}$, where 
$g$ is a four-dimensional superconformal transformation, then  
we get from eqs.(\ref{4sol1},\,\ref{4sol2})
\begin{equation}
\begin{array}{l}
\delta \tx^{\prime}_{+}=
\tx^{\prime}_{+}{\rm a}\tx^{\prime}_{+}-4
\tx^{\prime}_{+}\tilde{\varepsilon}_{a}\theta^{\prime a}-\lambda
\tx^{\prime}_{+}+w\tx^{\prime}_{+}
-\tx^{\prime}_{+}\tilde{w}
+4i\tilde{\bar{\rho}}_{a}\theta^{\prime a}+\tilde{{\rm b}}\,,\\
{}\\
\delta\theta^{\prime a}
=\tilde{\rho}^{a}-\textstyle{\frac{1}{2}}(\lambda+i\Omega)
\theta^{\prime a}-\theta^{\prime a}\tilde{w}
-(\zeta t^{t}\bar{\zeta})^{a}_{~b}\theta^{\prime b}+\theta^{\prime a}
{\rm a}
\tx^{\prime}_{+}-i\tilde{\bar{\varepsilon}}{}^{a}\tx^{\prime}_{+}
-4(\theta^{\prime a}\tilde{\varepsilon}_{b})\,\theta^{\prime b}\,,
\end{array}
\end{equation}
where 
\begin{equation}
\begin{array}{ll}
\tilde{\rho}^{a}=-\zeta^{ba}\rho^{t}_{b}\epsilon^{-1}\,,~~~~&~~~~
\tilde{\bar{\rho}}_{a}=\bar{\epsilon}^{-1}\bar{\rho}^{bt}\bar{\zeta}_{ba}\,,\\
{}&{}\\
\tilde{\varepsilon}_{a}=\epsilon\varepsilon^{bt}\bar{\zeta}_{ba}\,,~~~~&~~~~
\tilde{\bar{\varepsilon}}{}^{a}=-\zeta^{ba}
\bar{\varepsilon}^{t}_{b}\bar{\epsilon}\,.
\end{array}
\end{equation}
Hence, under  superinversion, the superconformal transformations
are related by 
\begin{equation}
\begin{array}{ccc}
{\cal K}\equiv\left(\begin{array}{c}
a^{\mu}\\
b^{\mu}\\
\varepsilon^{a}\\
\rho_{a}\\
\lambda\\
\Omega\\
w^{\mu}{}_{\nu}\\
t^{a}{}_{b}
\end{array}\right)~~~~& \longrightarrow &~~~~
\left(\begin{array}{c}
b^{\mu}\\
a^{\mu}\\
\tilde{\rho}^{a}\\
\tilde{\varepsilon}_{a}\\
-\lambda\\
-\Omega\\
w^{\mu}{}_{\nu}\\
-(\zeta t^{t}\bar{\zeta})^{a}{}_{b}
\end{array}\right)\,.
\end{array}
\end{equation}
In particular, special superconformal transformations~(\ref{4special})
can be obtained by 
\begin{equation}
z\,\stackrel{i_{s}\circ (b,\tilde{\rho})\circ
i_{s}}{-\!\!\!-\!\!\!-\!\!\!-\!\!\!-\!\!\!-\!\!\!\longrightarrow}\,z^{\prime}
\equiv z_{s}(z;u)\,, 
\label{specialz}
\end{equation}
where $(b,\tilde{\rho})$ is a supertranslation and $u^{M}=
(b^{\mu},\tilde{\rho}^{a\alpha},\tilde{\bar{\rho}}{}_{a}^{\da})$.

\subsection{Superconformal Algebra\label{SCA}}
The generator of  infinitesimal  
superconformal transformations,~${\cal L}$,
is given by
\begin{equation}
{\cal L}
=h^{\mu}\partial_{\mu}+\lambda^{a\alpha}D_{a\alpha}-\bar{\lambda}^{\da}_{a}
\bar{D}^{a}_{\da}\,.
\label{4calL}
\end{equation}
If we write the commutator of two generators,~${\cal L}_{1},{\cal
L}_{2}$,  as
\begin{equation}
[{\cal L}_{2},{\cal L}_{1}]={\cal L}_{3}
=h^{\mu}_{3}\partial_{\mu}+\lambda^{a\alpha}_{3}D_{a\alpha}
-\bar{\lambda}^{\da}_{3a}\bar{D}^{a}_{\da}\,,
\label{4hcommutator}
\end{equation}
then $h_{3}^{\mu},\,\lambda^{a\alpha}_{3},\,\bar{\lambda}^{\da}_{3a}$ 
are given by 
\begin{equation}
\begin{array}{l}
h^{\mu}_{3}=h^{\nu}_{2}\partial_{\nu}h^{\mu}_{1}
+2i\lambda^{a}_{1}\sigma^{\mu}\bar{\lambda}_{2a}-
(1\leftrightarrow 2)\,,\\
{}\\
\lambda^{a\alpha}_{3}=h^{\mu}_{2}\partial_{\mu}\lambda^{a\alpha}_{1}
+\lambda^{b\beta}_{2}D_{b\beta}\lambda^{a\alpha}_{1}-
(1\leftrightarrow 2)\,,\\
{}\\
\bar{\lambda}_{3a}^{\da}=h^{\mu}_{2}\partial_{\mu}\bar{\lambda}_{1a}^{\da}
-\bar{\lambda}_{2b}^{\db}\bar{D}^{b}_{\db}\bar{\lambda}_{1a}^{\da}
-(1\leftrightarrow 2)\,,
\end{array}
\end{equation}
and $h^{\mu}_{3},\,\lambda^{a\alpha}_{3},\,\bar{\lambda}_{3a}^{\da}$ 
satisfy eq.(\ref{4enough})
verifying the closure of the Lie algebra.\newline
\indent  Explicitly with eqs.(\ref{4solutionforh}\,,\ref{4sol1}) 
we get
\begin{equation}
\begin{array}{l}
\displaystyle{a^{\mu}_{3}=w^{\mu}_{1}{}_{\nu}a^{\nu}_{2}
+\lambda_{1}a^{\mu}_{2}+
2i\varepsilon^{a}_{1}\sigma^{\mu}\bar{\varepsilon}_{2a} -
(1\leftrightarrow 2)}\,,\\
{}\\
\varepsilon^{a}_{3}=
-\varepsilon_{2}^{a}\tilde{w}_{1}+
\textstyle{\frac{1}{2}}\lambda_{1}\varepsilon^{a}_{2}-
i\bar{\rho}_{1}^{a}\tilde{\rm
a}_{2}+t_{1}^{a}{}_{b}\varepsilon^{b}_{2}
+i\textstyle{\frac{1}{2}}\Omega_{1}\varepsilon_{2}^{a}
-(1\leftrightarrow 2)\,,\\
{}\\
\bar{\varepsilon}_{3a}=
w_{1}\bar{\varepsilon}_{2a}+
\textstyle{\frac{1}{2}}\lambda_{1}\bar{\varepsilon}_{2a}+
i\tilde{\rm a}_{2}\rho_{1a}-\bar{\varepsilon}_{2b}t_{1}^{b}{}_{a}
-i\textstyle{\frac{1}{2}}\Omega_{1}\bar{\varepsilon}_{2a}
-(1\leftrightarrow 2)\,,\\
{}\\
\displaystyle{\lambda_{3}=2a_{2}{\cdot b_{1}}
+2(\bar{\rho}^{a}_{1}\bar{\varepsilon}_{2a}+\varepsilon_{2}^{a}\rho_{1a})-
(1\leftrightarrow 2)}\,,\\
{}\\
\displaystyle{w^{\mu\nu}_{3}=w^{\mu}_{1}{}_{\lambda}w^{\lambda\nu}_{2}
+2(a_{2}^{\mu}b_{1}^{\nu}-a_{2}^{\nu}b_{1}^{\mu})+
2(\varepsilon^{a}_{2}\sigma^{[\mu}\tilde{\sigma}^{\nu]}
\rho_{1a}-\bar{\rho}^{a}_{1}\tilde{\sigma}^{[\mu}\sigma^{\nu]}
\bar{\varepsilon}_{2a})
-(1\leftrightarrow 2)}\,,\\
{}\\
b^{\mu}_{3}=w^{\mu}_{1}{}_{\nu}b^{\nu}_{2}-\lambda_{1}b^{\mu}_{2}+
2i\bar{\rho}_{1}^{a}\tilde{\sigma}^{\mu}\rho_{2a} -
(1\leftrightarrow 2)\,,\\
{}\\
\rho_{3a}=\tilde{w}_{1}\rho_{2a}-
\textstyle{\frac{1}{2}}
\lambda_{1}\rho_{2a}+i{\rm b}_{2}\bar{\varepsilon}_{1a}-
\rho_{2b}t^{b}_{1}{}_{a}-i\textstyle{\frac{1}{2}}\Omega_{1}\rho_{2a}
-(1\leftrightarrow 2)\,,\\
{}\\
\bar{\rho}_{3}^{a}=-\bar{\rho}^{a}_{2}w_{1}-
\textstyle{\frac{1}{2}}
\lambda_{1}\bar{\rho}_{2}^{a}-i\varepsilon^{a}_{1}{\rm b}_{2}+
t^{a}_{1}{}_{b}\bar{\rho}^{b}_{2}+i\textstyle{\frac{1}{2}}\Omega_{1}
\bar{\rho}_{2}^{a}
-(1\leftrightarrow 2)\,,\\
{}\\
\displaystyle{t^{a}_{3}{}_{b}=(t_{1}t_{2})^{a}{}_{b}
+4(\varepsilon^{a}_{1}\rho_{2b}-\bar{\rho}^{a}_{2}\bar{\varepsilon}_{1b})
-\textstyle{\frac{4}{{\cal N}}}
(\varepsilon^{c}_{1}\rho_{2c}-\bar{\rho}^{c}_{2}\bar{\varepsilon}_{1c})
\delta^{a}{}_{b}-(1\leftrightarrow 2)}\,,\\
{}\\
\displaystyle{\Omega_{3}=2i(\textstyle{\frac{4}{{\cal N}}}-1)
(\varepsilon^{a}_{2}\rho_{1a}-\bar{\rho}^{a}_{1}\bar{\varepsilon}_{2a})
-(1\leftrightarrow 2)}\,.
\end{array}
\label{4MMcom}
\end{equation}
From eq.(\ref{4MMcom}) we can read off the explicit forms of 
four-dimensional  superconformal algebra as exhibited 
in Appendix \ref{4AppendixC}.\newline
\newline
\indent Now, we consider ${\cal N}\neq 4$ case and ${\cal N}=4$ case 
separately.  For ${\cal N}\neq 4$ case,  
if we define a $(4+{\cal N})\times (4+{\cal N})$ supermatrix,~$M$, as 
\begin{equation}
\displaystyle{
M=\left(\begin{array}{ccc}
       w+\textstyle{\frac{1}{2}}\lambda+i\textstyle{\frac{1}{2}}\psi 
&         -i\tilde{{\rm a}}&2\bar{\varepsilon}_{b}\\
       -i{\rm b}&\tilde{w}
-{\textstyle\frac{1}{2}}\lambda +i\textstyle{\frac{1}{2}}\psi &2\rho_{b}\\
       2\bar{\rho}^{a}&2\varepsilon^{a}&t^{a}_{~b}+i\frac{2}{{\cal N}}
\psi\delta^{a}_{~b}
        \end{array}\right)}\,,
\label{4Mform}
\end{equation}
where  
\begin{equation}
\displaystyle{\psi=\frac{\Omega}{\frac{4}{{\cal N}}-1}}\,, 
\label{psidef}
\end{equation}
then the relation above~(\ref{4MMcom}) agrees with the matrix commutator  
\begin{equation}
[M_{1},M_{2}]=M_{3}\,.
\label{4Mcommutator}
\end{equation}
This can be verified using
 eqs.(\ref{4contract},\,\ref{4sig4},\,\ref{4dde}).\newline
In general, for  ${\cal N}\neq 4$, 
$M$ can be defined as a $(4,{\cal N})$ supermatrix subject to 
\begin{equation}
\mbox{str}\:M=0\,,
\label{Mstr}
\end{equation}
and a reality condition 
\begin{equation}
\begin{array}{cc}
BMB^{-1}=-M^{\dagger}\,,~~~~&~~~~
B=\left(\begin{array}{ccc}
                 0&1&0\\ 
                 1&0&0\\
                 0&0&-1
              \end{array}\right)\,.
\end{array}
\label{4Mdagger}
\end{equation}
Supermatrix of the form~(\ref{4Mform}) is the general solution of 
these two equations. \newline
\indent The $4\times 4$ matrix appearing in $M$,
\begin{equation}
\left(\begin{array}{cc}
w+\textstyle{\frac{1}{2}}\lambda &-i\ta\\
-i{\rm b} & \tilde{w}-\textstyle{\frac{1}{2}}\lambda
\end{array}\right)\,,
\end{equation}
corresponds to a generator of $\mbox{O}(2,4)\cong\mbox{SU}(2,2)$ 
as demonstrated in Appendix \ref{4AppendixD}. Thus,  
the ${\cal N}\neq 4$     
superconformal group in four-dimensions may be identified  
with the supermatrix group generated by  supermatrices of the 
form $M$~(\ref{4Mform}),   which is
$\mbox{SU}(2,2|{\cal N})\equiv\mbox{G}_{s}$ having dimensions 
$(15+{\cal N}^{2}|8{\cal N})$. \newline
\newline
\indent 
When   ${\cal N}=4$, similar analysis is also possible with a subtle 
modification. 
In this case, $M$ is a $(4,4)$ supermatrix satisfying the reality 
condition~(\ref{4Mdagger}) and, instead of eq.(\ref{Mstr}),
\begin{equation}
\mbox{str}\:M=-2i\Omega\,.
\end{equation}
Such a supermatrix, $M$, is of the general form
\begin{equation}
M={\cal M}+i\textstyle{\frac{1}{2}}\Omega\,
\left(\begin{array}{cc}0&0\\0&\delta^{a}_{~b}\end{array}\right)+
i\textstyle{\frac{1}{2}}\psi\,
\left(\begin{array}{cc}1&0\\0&\delta^{a}_{~b}\end{array}\right)\,,
\label{4Mform4}
\end{equation}
where ${\cal M}$ is of the 
form~(\ref{4Mform}) with $\psi=0$, and $\psi$ in eq.(\ref{4Mform4})  
is just an arbitrary real bosonic variable. 
Essentially we add the   $\Omega$ term to eq.(\ref{4Mform}) 
to get eq.(\ref{4Mform4}). 
Same as ${\cal N}\neq 4$ case, the matrix 
commutator of $M_{1}$ and $M_{2}$  reproduces eq.(\ref{4MMcom}) 
as in eq.(\ref{4Mcommutator}), though $\psi$ is arbitrary. \newline
As the variable, $\psi$, is auxiliary, one might be tempted 
to fix its value, or more generally let it be a function of the parameters 
of superconformal transformations, $a^{\mu},\,b^{\mu},\,\Omega,\,\cdots$ and 
try to determine the function. 
However, this is not possible. The commutator of ${\cal M}_{1}$ and 
${\cal M}_{2}$ includes $\psi$ type term  
\begin{equation}
\begin{array}{c}
[{\cal M}_{1}, {\cal M}_{2}]={\cal M}_{3}+i\textstyle{\frac{1}{2}}
\psi_{3}\,1\,,\\
{}\\
\psi_{3}=2i(\varepsilon^{a}_{2}\rho_{1a}-\varepsilon^{a}_{1}\rho_{2a}
+\bar{\rho}^{a}_{2}\bar{\varepsilon}_{1a}
-\bar{\rho}^{a}_{1}\bar{\varepsilon}_{2a})\,,
\end{array}
\end{equation}
and $\psi_{3}$ can not be expressed in terms of the parameters, 
$a^{\mu}_{3},\,b^{\mu}_{3},\,\Omega_{3},\,\cdots$,  appearing 
in eq.(\ref{4MMcom}).  Hence it is not possible to put $\psi$ as 
a function of the superconformal transformation parameters.\footnote{An alternative approach may be taken as in \cite{smf185}, where a modified 
supermatrix commutator is introduced for $\mbox{SL}(m|m)$.}  
Therefore  four-dimensional 
${\cal N}=4$   superconformal algebra is  
represented by $(4,4)$ supermatrices, $M$, satisfying the 
reality condition~(\ref{4Mdagger}) 
with an equivalence relation, $\sim$, imposed
\begin{equation}
\begin{array}{cc}
M_{1}\sim M_{2}~~~~&~~~~\mbox{if\,~}M_{1}-M_{2}=i\psi\,1~~~
\mbox{for\,~some\,~}\psi\in\Real\,.
\end{array}
\label{equivre}
\end{equation}
\indent We note that an extra condition, $\mbox{str}\:M=0$,   
defines   an invariant subalgebra of the whole 
four-dimensional ${\cal N}=4$ superconformal algebra. 
This  invariant subalgebra forms a simple Lie superalgebra. In the literature 
the four-dimensional ${\cal N}=4$ superconformal algebra is often 
identified with this simple Lie superalgebra,         
the $R$-symmetry of which is $\mbox{su}(4)$ rather than 
$\mbox{u}(4)$~\cite{nahm}, as the $\Omega$ term in eq.(\ref{4Mform4}) 
is neglected. However, we emphasize here that  the whole 
${\cal N}=4$ superconformal algebra may contain a $\mbox{u}(1)$ factor 
which has non-trivial commutator 
relations with other generators as seen in eq.(\ref{4MMcom}) 
or eq.(\ref{USU}). \newline 
\indent The ${\cal N}=4$     
superconformal group in four-dimensions, $\mbox{G}_{s}$, is now 
identified  with a quotient group of the supermatrix group, 
as it is isomorphic to  the supermatrix group   
generated by  supermatrices of the form $M$~(\ref{4Mform4}) with 
an  equivalence relation imposed on the supermatrix group element, $G$, 
from eq.(\ref{equivre})
\begin{equation}
\begin{array}{cc}
G_{1}\sim G_{2}~~~~&~~~~\mbox{if\,~}G_{1}^{-1}G_{2}=e^{i\psi}\,1~~~
\mbox{for\,~some\,~}\psi\in\Real\,.
\end{array}
\label{equivre2}
\end{equation}
We also note that the 
four-dimensional ${\cal N}=4$ superconformal group has 
dimensions $(31|32)$ and is isomorphic to 
a semi-direct product  of $\mbox{U}(1)$ and a simple Lie supergroup.   
Therefore,  by breaking the $\mbox{U}(1)$ symmetry, the 
four-dimensional ${\cal N}=4$ superconformal group can be reduced to the 
simple Lie subgroup having dimensions $(30|32)$.

\section{Coset Realization of  Transformations\label{4coset}}
In this section, we first obtain an explicit formula for  
the finite nonlinear 
superconformal transformations of the supercoordinates and discuss several representations of the superconformal group. 
We then 
construct matrix (vector) valued 
functions depending on  two   (three) points 
in superspace which transform 
covariantly under superconformal transformations.  
These  are crucial variables  for  
obtaining two-point, three-point and general $n$-point 
correlation functions later. 
\subsection{Superspace as a Coset}
To obtain an explicit  formula for the finite nonlinear superconformal
transformations, we first identify the superspace, $\Real^{4|4{\cal N}}$,  
as a coset, 
$\mbox{G}_{s}/\mbox{G}_{0}$,
where $\mbox{G}_{0}\subset \mbox{G}_{s}$ is the subgroup generated by
matrices, $M_{0}$, 
of the form~(\ref{4Mform}) with $a^{\mu}=0,\,\varepsilon^{a}=0$ and
depending on 
parameters $b^{\mu},\,\rho_{a},\,\bar{\rho}{}^{a},\,
\lambda,\,\Omega,\,w_{\mu\nu},\,t^{a}{}_{b}$.    
The group of supertranslations, $\mbox{G}_{T}$,  
parameterized by coordinates, 
$z^{M}\in\Real^{4|4{\cal N}}$,   has been defined by general
elements as in eq.(\ref{4generalelement}), with the group property
given by eqs.(\ref{4groupproperty1},\,\ref{4groupproperty2}). Now we may
represent it by  
supermatrices\footnote{The subscript, $T$, denotes supertranslations.}
\begin{equation}
G_{T}(z)=\mbox{exp}\left(\begin{array}{ccc}
               0&-i\tilde{{\rm x}}&2\bar{\theta}_{b}\\
               0&0&0\\
               0&2\theta^{a}&0
                \end{array}\right)=
\left(\begin{array}{ccc}
      1&  -i\tx_{+}&2\bar{\theta}_{b}\\
      0&  1&0\\
      0&  2\theta^{a}&\delta^{a}{}_{b}
        \end{array}\right)\,.
\end{equation}
Note that  $G_{T}(z)^{-1}=G_{T}(-z)$.\newline
\indent   
In general an element of $\mbox{G}_{s}$  can be uniquely decomposed  as
$G_{T}G_{0}$. Thus for any  element $G(g)\in\mbox{G}_{s}$ we may define a
superconformal transformation, $z \stackrel{g}{\longrightarrow}
z^{\prime}$, and an associated element $G_{0}(z;g)\in\mbox{G}_{0}$ by
\begin{equation}
G(g)^{-1}G_{T}(z)G_{0}(z;g)=G_{T}(z^{\prime})\,.
\label{4GTFINITE}
\end{equation}
If $G(g)\in\mbox{G}_{T}$ then clearly $G_{0}(z;g)=1$.\newline
\indent  Infinitesimally eq.(\ref{4GTFINITE}) becomes
\begin{equation}
\delta G_{T}(z)=MG_{T}(z)-G_{T}(z)\hat{M}_{0}(z)\,, 
\label{4MGL}
\end{equation}
where  $M$ is given by eq.(\ref{4Mform}) or eq.(\ref{4Mform4}) and 
$\hat{M}_{0}(z)$, the generator of $\mbox{G}_{0}$,  has the form 
\begin{equation}
\begin{array}{l}
\hat{M}_{0}(z)\\
{}\\
=\displaystyle{\left(\begin{array}{ccc}
        \hat{w}(z) +\textstyle{\frac{1}{2}}\hat{\lambda}(z)
+i\textstyle{\frac{1}{2}}\hat{\psi}(z) & 
        0&0\\
       -i{\rm b} & 
       \tilde{\hat{w}}(z)-\frac{1}{2}\hat{\lambda}(z)+
i\textstyle{\frac{1}{2}}\hat{\psi}(z) &
         2\hat{\rho}_{b}(z)\\
       2\bar{\hat{\rho}}{}^{a}(z)&0&\hat{t}{}^{a}{}_{b}(z)+
i\textstyle{\frac{2}{{\cal N}}}(\hat{\psi}(z)+\delta^{4}{}_{{\cal N}}\,\Omega)
\delta^{a}{}_{b}
        \end{array}\right)}\,.
\end{array}
\end{equation}                           
The components depending on $z$ are given by 
\begin{equation}
\begin{array}{l}
\hat{w}(z)=w-4\bt_{a}\bar{\rho}^{a}+\tx_{-}{\rm b}
+\textstyle{\frac{1}{2}}
\mbox{tr}(4\bt_{a}\bar{\rho}^{a}-\tx_{-}{\rm b})\,1\,, \\
{}\\
\tilde{\hat{w}}(z)
=\tilde{w}+4\rho_{a}\theta^{a}-{\rm b}\tx_{+}
-\textstyle{\frac{1}{2}}
\mbox{tr}(4\rho_{a}\theta^{a}-{\rm b}\tx_{+})\,1 
=-\hat{w}(z)^{\dagger}\,,        \\
{}\\
\hat{\lambda}(z)=\lambda+2b{\cdot x}+
2(\theta^{a}\rho_{a}+\bar{\rho}^{a}\bt_{a})
=\textstyle{\frac{1}{4}}\partial_{\mu}h^{\mu}(z)\,,\\
{}\\
\hat{\psi}(z)=\psi+2\theta^{a}{\rm b}\bt_{a}
+2i(\theta^{a}\rho_{a}-\bar{\rho}^{a}\bt_{a})\,,\\
{}\\
\hat{t}{}^{a}{}_{b}(z)=t^{a}{}_{b}+4i\theta^{a}{\rm b}\bt_{b}+4
(\bar{\rho}^{a}\bt_{b}-\theta^{a}\rho_{b})-\textstyle{\frac{1}{{\cal N}}}(
4i\theta^{c}{\rm b}\bt_{c}+4
\bar{\rho}^{c}\bt_{c}-4\theta^{c}\rho_{c})\delta^{a}{}_{b}\,,\\
{}\\
\hat{\rho}_{a}(z)=\rho_{a}-i{\rm b}\bt_{a}=-i\textstyle{\frac{1}{4}}
\sigma^{\mu}\partial_{\mu}\bar{\lambda}_{a}(z)\,,\\
{}\\
\bar{\hat{\rho}}{}^{a}(z)=\bar{\rho}{}^{a}+i\theta^{a}{\rm b}
=\hat{\rho}_{a}(z)^{\dagger}\,.
\end{array}
\label{4hats}
\end{equation}
Writing $\delta G_{T}(z)={\cal L}G_{T}(z)$ we may verify that ${\cal
L}$ is identical with eq.(\ref{4calL}).\newline
\newline
\indent $\hat{w}(z),\,\tilde{\hat{w}}(z)$ can be also written as 
$\hat{w}(z)
=\textstyle{\frac{1}{4}}
\hat{w}_{\mu\nu}(z)\tilde{\sigma}^{\mu}\sigma^{\nu},\,
\tilde{\hat{w}}(z)
=\textstyle{\frac{1}{4}}
\hat{w}_{\mu\nu}(z)\sigma^{\mu}\tilde{\sigma}^{\nu}$, with 
\begin{equation}
\hat{w}_{\mu\nu}(z)=w_{\mu\nu}+4x_{[\mu}b_{\nu]}+\theta^{a}
\sigma_{[\mu}\tilde{\sigma}_{\nu]}(2\rho_{a}-i{\rm b}\bt_{a})-
(2\bar{\rho}_{a}+i\theta^{a}{\rm b})\tilde{\sigma}_{[\mu}
\sigma_{\nu]}\bt_{a}
=-\partial_{[\mu}h_{\nu]}(z)\,.
\end{equation}
The definitions\,(\ref{4hats}) can be summarized by 
\begin{equation}
D_{b\beta}\lambda^{a\alpha}=
\textstyle{\frac{1}{2}}\delta_{b}^{~a}\delta_{\beta}^{~\alpha}
(\hat{\lambda}(z)+i\hat{\Omega}(z))
-\delta_{b}^{~a}\tilde{\hat{w}}_{\beta}{}^{\alpha}(z)+
\delta_{\beta}^{~\alpha}\hat{t}{}^{a}_{~b}(z)\,,
\end{equation}
and they  give
\begin{equation}
[D_{a\alpha},{\cal L}]=
\textstyle{\frac{1}{2}}(\hat{\lambda}(z)
+i\hat{\Omega}(z))D_{a\alpha}
-\tilde{\hat{w}}_{\alpha}{}^{\beta}(z)
D_{a\beta}+\hat{t}{}^{b}{}_{a}(z)D_{b\alpha}\,,
\label{4comDL}
\end{equation}
where
\begin{equation}
\hat{\Omega}(z)=(\textstyle{\frac{4}{{\cal N}}}-1)\hat{\psi}(z)
+\delta^{4}{}_{{\cal N}}\,\Omega\,.
\label{hatOmega}
\end{equation}
For later use we note 
\begin{equation}
\begin{array}{l}
D_{a\alpha}\hat{w}_{\mu\nu}(z)=
2(\sigma_{[\mu}\tilde{\sigma}_{\nu]})_{\alpha}{}^{\beta}
\hat{\rho}_{a\beta}(z)\,,\\
{}\\
D_{a\alpha}\hat{\lambda}(z)=2\hat{\rho}_{a\alpha}(z)\,,\\
{}\\
D_{a\alpha}\hat{\psi}(z)=2i\hat{\rho}_{a\alpha}(z)\,,\\
{}\\
D_{a\alpha}\hat{t}{}^{b}{}_{c}(z)=-4\delta_{a}{}^{b}
\hat{\rho}_{c\alpha}(z)+\textstyle{\frac{4}{{\cal N}}}\delta^{b}{}_{c}
\hat{\rho}_{a\alpha}(z)\,.
\end{array}
\label{4Ddelta}
\end{equation}
\indent  The above analysis  can be simplified by reducing $G_{0}(z;g)$. To
achieve this we let 
\begin{equation}
Z_{0}=\left(\begin{array}{cc}
              0&0\\
              1&0\\
              0&1
            \end{array}\right)\,,
\end{equation}
and then
\begin{equation}
\begin{array}{cc}
M_{0}Z_{0}=Z_{0}H_{0}\,,~~~~&~~~~
H_{0}=
\left(\begin{array}{cc}
           \tilde{w}
-\textstyle{\frac{1}{2}}\lambda+i\textstyle{\frac{1}{2}}\psi & 2\rho_{b}\\
                             0&t^{a}{}_{b}+i\textstyle{\frac{2}{{\cal N}}}
                       (\psi+\delta^{4}{}_{{\cal N}}\,\Omega)\delta^{a}{}_{b} 
                          \end{array}\right)\,.
\end{array}
\label{4M0Z0}
\end{equation}
Now if we define
\begin{equation}
Z(z)\equiv G_{T}(z)Z_{0}=
\left(\begin{array}{cc}
        -i\tx_{+}&2\bt_{b}\\
        1&0\\
        2\theta^{a}&\delta^{a}_{~b}
        \end{array}\right)\,,
\label{4defZ}
\end{equation}
then $Z(z)$ transforms  
under infinitesimal  superconformal transformations as
\begin{equation}
\delta Z(z)={\cal L}Z(z)=MZ(z)-Z(z)H(z)\,, 
\label{4infZ}
\end{equation}
where $H(z)$ is given by  
\begin{equation}
\begin{array}{ll}
\hat{M}_{0}(z)Z_{0}=Z_{0}H(z)\,,&
H(z)=
\left(\begin{array}{cc}
\tilde{\hat{w}}(z)-\frac{1}{2}\hat{\lambda}(z)+i
\textstyle{\frac{1}{2}}\hat{\psi}(z)
&2\hat{\rho}_{b}(z)\\
0&\hat{t}{}^{a}_{~b}(z)
+i\textstyle{\frac{2}{{\cal N}}}
      (\hat{\psi}(z)+\delta^{4}{}_{{\cal N}}\,\Omega)\delta^{a}{}_{b}
\end{array}\right).
\end{array}
\label{4defH}
\end{equation}
\indent  From eqs.(\ref{4hcommutator},\,\ref{4Mcommutator}) considering 
\begin{equation}
[{\cal L}_{2},{\cal L}_{1}]Z(z)={\cal L}_{3}Z(z)\,,
\end{equation}
we get
\begin{equation}
H_{3}(z)={\cal L}_{2}H_{1}(z)-{\cal L}_{1}H_{2}(z)+[H_{1}(z),H_{2}(z)]\,,
\label{4Kcom}
\end{equation}
which gives separate equations for
$\tilde{\hat{w}},\,\hat{\lambda},\,\hat{\psi},\,\hat{\rho}_{a}$ 
and $\hat{t}{}^{a}{}_{b}$, thus 
$\hat{\lambda}_{3}={\cal L}_{2}\hat{\lambda}_{1}-{\cal 
L}_{1}\hat{\lambda}_{2}$, and so on.\newline
\indent    As a conjugate of  $Z(z)$ we   define  $\bar{Z}(z)$ by
\begin{equation}
\bar{Z}(z)=\left(\begin{array}{cc}
               1&0\\
                0&-1 \end{array}\right)Z(z){}^{\dagger}B=
              \left(\begin{array}{ccc}
                 1&i\tx_{-}&-2\bt_{b}\\
                 0&-2\theta^{a}&\delta^{a}_{~b}\end{array}\right)\,.
\label{4defbarZ}
\end{equation}
This satisfies
\begin{equation}
\bar{Z}(z)=\bar{Z}(0)G_{T}(z)^{-1}\,,
\end{equation}
and corresponding to eq.(\ref{4infZ}) we have
\begin{equation}
\delta \bar{Z}(z)={\cal L}\bar{Z}(z)
=\bar{H}(z)\bar{Z}(z)-\bar{Z}(z)M \,,
\label{4infbarZ}
\end{equation}
where
\begin{equation}
\bar{H}(z)=\left(\begin{array}{cc}
\hat{w}(z)+\textstyle{\frac{1}{2}}\hat{\lambda}(z)
+i\textstyle{\frac{1}{2}}\hat{\psi}(z)&0\\
2\bar{\hat{\rho}}{}^{a}(z)&\hat{t}{}^{a}{}_{b}(z)+
+i\textstyle{\frac{2}{{\cal N}}}
          (\hat{\psi}(z)+\delta^{4}{}_{{\cal N}}\,\Omega)\delta^{a}{}_{b}
\end{array}\right)\,.
\end{equation}

\subsection{Finite Transformations}
Finite superconformal transformations can be  obtained by exponentiation  
of infinitesimal transformations. To obtain a superconformal
transformation, $z\stackrel{g}{\longrightarrow}z^{\prime}$, we
therefore solve the differential equation
\begin{equation}
\begin{array}{ccc}
\displaystyle{
\frac{{\rm d}~}{{\rm d}t}z^{M}_{t}={\cal L}^{M}(z_{t})}\,,~~~~&~~~~z_{0}=z\,,
~~~&~~~z_{1}=z^{\prime}\,,
\end{array}
\label{zt01}
\end{equation}
where, with ${\cal L}$ given in eq.(\ref{4calL}), ${\cal L}^{M}(z)$ is defined by 
\begin{equation}
{\cal L}={\cal L}^{M}(z)\partial_{M}\,.
\end{equation}
From eq.(\ref{4infZ}) we get
\begin{equation}
\displaystyle{
\frac{{\rm d}~}{{\rm d}t}Z(z_{t})=MZ(z_{t})-Z(z_{t})H(z_{t}) }\,,
\end{equation}
which integrates to
\begin{equation}
Z(z_{t})=e^{tM}Z(z)K(z,t)\,,
\label{4finites}
\end{equation}
where $K(z,t)$ satisfies
\begin{equation}
\begin{array}{cc}
\displaystyle{
\frac{{\rm d}~}{{\rm d}t}K(z,t)=-K(z,t)H(z_{t})}\,,~~~~&~~~~
K(z,0)=\left(\begin{array}{cc}
               1&0\\
               0&1
             \end{array}\right)\,.
\end{array}
\end{equation}
Hence for $t=1$ with $K(z,1)\equiv K(z;g)$  the superconformal
transformation,~$z\stackrel{g}{\longrightarrow}z^{\prime}$, from 
eq.(\ref{4finites}) becomes
\begin{equation}
\begin{array}{cc}
Z(z^{\prime})=G(g)^{-1}Z(z)K(z;g)\,,~~~~&~~~~G(g)^{-1}=e^{M}\,.
\end{array}
\label{4Ztr}
\end{equation}
$G_{0}(z;g)$ in eq.(\ref{4GTFINITE}) is related to 
$K(z;g)$ from eq.(\ref{4Ztr})  by
\begin{equation}
G_{0}(z;g)Z_{0}=Z_{0}K(z;g)\,.
\end{equation}
\indent  In general  $K(z;g)$ is of the form
\begin{equation}
K(z;g)=\left(\begin{array}{cc}
                L(z_{+};g)&2\Sigma_{b}(z;g)\\
                 0&u^{a}{}_{b}(z;g)
                \end{array}\right)\,.
\label{4Kform}
\end{equation}
From
\begin{equation}
\tilde{\hat{w}}(z)-\textstyle{\frac{1}{2}}
\hat{\lambda}(z)+i\textstyle{\frac{1}{2}}
\hat{\psi}(z)=\tilde{w}-\textstyle{\frac{1}{2}}\lambda+i\textstyle{\frac{1}{2}}
\psi+4\rho_{a}\theta^{a}-{\rm
b}\tx_{+}\,,
\end{equation}
$L(z_{+};g)$ is defined  on chiral superspace, and  
since $L(z_{+};g)$ is a $2\times 2$ matrix, we have 
\begin{equation}
L(z_{+};g)\epsilon L(z_{+};g)^{t}=\det L(z_{+};g)\,\epsilon\,.
\end{equation}
Infinitesimally this is consistent with   eq.(\ref{4wt}).\newline
We decompose $L(z_{+};g)$ as
\begin{subeqnarray}
\label{hatLO}
&\hat{L}(z_{+};g)=\Omega_{+}(z_{+};g)^{-\frac{1}{2}}L(z_{+};g)\,,&\\
&{}&\nonumber\\
&\Omega_{+}(z_{+};g)=\det L(z_{+};g)\,,&
\end{subeqnarray}
where $\hat{L}\in\mbox{SL}(2,\Com)$, 
the $2\times 2$ matrices with determinant one.\newline
\indent Since $\frac{\rm{d}~}{\rm{d}t}u(z,t)=-u(z,t)(\hat{t}(z_{t})+
i\textstyle{\frac{2}{{\cal N}}}\hat{\psi}(z_{t})
 +i\textstyle{\frac{1}{2}}\delta^{4}{}_{{\cal N}}\,\Omega)$, 
$u^{\dagger}=u^{-1}$ and hence  $u\in \mbox{U}({\cal N})$. \newline
If we write
\begin{equation}
\hat{u}(z;g)
=\displaystyle{\frac{u(z;g)}{(\det u(z;g))^{\frac{1}{{\cal N}}}}}\,,
\end{equation}
then $\hat{u}\in\mbox{SU}({\cal N})$.\newline 
\indent From eq.(\ref{4Ztr}) $\bar{Z}$ transforms as
\begin{equation}
\bar{Z}(z^{\prime})=\bar{K}(z;g)\bar{Z}(z)G(g)\,,
\label{4barZtr}
\end{equation}
where 
\begin{equation}
\begin{array}{cc}
\multicolumn{2}{c}{\bar{K}(z;g)=\left(\begin{array}{cc}
                1&0\\
                0&-1 \end{array}\right)K(z;g){}^{\dagger}
\left(\begin{array}{cc}
                1&0\\
                0&-1 \end{array}\right)
=\left(\begin{array}{cc}
               \bar{L}(z_{-};g)&0\\
                -2\bar{\Sigma}(z;g)&u^{-1}(z;g) \end{array}\right)\,,}\\
{}&{}\\
~~~~~~~~~~~\bar{L}(z_{-};g)=L(z_{+};g)^{\dagger}\,,~~~&~~~\bar{\Sigma}(z;g)
=\Sigma(z;g)^{\dagger}\,.
\end{array}
\label{4Kbarform}
\end{equation}
In a similar fashion to eq.(\ref{hatLO}) we write
\begin{subeqnarray}
&\hat{\bar{L}}(z_{-};g)=\Omega_{-}(z_{-};g)^{-\frac{1}{2}}\bar{L}(z_{-};g)
=\hat{L}(z_{+};g)^{\dagger}\in\mbox{SL}(2,\Com)\,,&\\
&{}&\nonumber\\
&\Omega_{-}(z_{-};g)=\det \bar{L}(z_{-};g)=\Omega_{+}(z_{+};g)^{\ast}\,.&
\end{subeqnarray}
\indent    If we define for superinversion, 
$z\stackrel{i_{s}}{\longrightarrow}z^{\prime}$,~(\ref{4superinversion})  
\begin{equation}
\begin{array}{cc}
G(i_{s})^{-1}=\left(\begin{array}{ccc}
\bar{\epsilon}&0&0\\
0&\epsilon^{-1}&0\\
0&0&-\bar{\zeta}_{ab}
\end{array}\right)\,,~~~&~~~
K(z;i_{s})=\left(\begin{array}{cc}
i(\bar{\epsilon}\tx_{+})^{-1}&2i\tx^{-1}_{-}\bar{\theta}_{c}\zeta^{cb}\\
0&-v^{a}_{~c}(z)\zeta^{cb}
\end{array}\right)\,,
\end{array}
\end{equation}
an analogous formula  to eq.(\ref{4Ztr}) can be obtained 
\begin{equation}
G(i_{s})^{-1}Z(z)K(z;i_{s})
=\left(\begin{array}{cc}
1&0\\
i\tx^{\prime t}_{-}&-2\theta^{\prime a}{}^{t}\\
2\bar{\theta}_{b}^{\prime t}& \delta_{b}^{~a}
\end{array}\right)=
\bar{Z}(z^{\prime})^{t}\,.
\label{4tZp1}
\end{equation}
Similarly we have
\begin{equation}
\bar{K}(z;i_{s})\bar{Z}(z)G(i_{s})=Z(z^{\prime})^{t}\,,
\label{4tZp2}
\end{equation}
where
\begin{equation}
\bar{K}(z;i_{s})=\left(\begin{array}{cc}
-i(\tx_{-}\epsilon)^{-1}&0\\
2i\bar{\zeta}_{ac}\theta^{c}\tx^{-1}_{+}&-\bar{\zeta}_{ac}v^{-1c}{}_{b}(z)
\end{array}\right)\,.
\end{equation}
\indent   For later use,    we also 
define with eq.(\ref{hatOmega},\,\ref{zt01}) 
\begin{equation}
\begin{array}{cc}
\displaystyle{
\frac{{\rm d}~}{{\rm d}t}\Upsilon(z,t)=i\hat{\Omega}(z_{t})
\Upsilon(z,t)}\,,~~~~&~~~~\Upsilon(z,0)=1\,,\\
{}&{}\\
\multicolumn{2}{c}{\Upsilon(z;g)\equiv\Upsilon(z,1)\,,}
\end{array}
\end{equation}
and 
\begin{equation}
\Omega(z;g)\equiv\sqrt{\Omega_{+}(z_{+};g)\Omega_{-}(z_{-};g)}\,.
\label{Omega2}
\end{equation}
Note that
\begin{equation}
\begin{array}{cc}
\Omega(z;g)^{\ast}=\Omega(z;g)\,,
~~~~&~~~~~\Upsilon(z;g)^{\ast}=\Upsilon(z;g)^{-1}\,.
\end{array}
\end{equation}
Since 
\begin{equation}
\displaystyle{\mbox{sdet}\:G=\mbox{exp}(\mbox{str}\:\mbox{ln}\:G})\,,
\end{equation}
when ${\cal N}=4$, $\Upsilon(z;g)$ is  related to the superdeterminant of 
$G(g)\in\mbox{G}_{s}$ 
\begin{equation}
\Upsilon(z;g)=\displaystyle{e^{i\Omega}}=\sqrt{\mbox{sdet}\:G(g)}\,.
\label{Upsilon4}
\end{equation}
\indent   If
\begin{equation}
\sigma^{\nu}R^{~\mu}_{\nu}(z;g)=L(z_{+};g)\sigma^{\mu}\bar{L}(z_{-};g)\,,
\label{4green1}
\end{equation}
then $R^{~\mu}_{\nu}(z;g)$ is identical to the
definition~(\ref{4ehomogeneous}), since infinitesimally 
\begin{equation}
\hat{\lambda}(z)\sigma^{\mu}-\tilde{\hat{w}}(z)\sigma^{\mu}
+\sigma^{\mu}\hat{w}(z)=\hat{\lambda}(z)\sigma^{\mu}
-\sigma^{\nu}\hat{w}_{\nu}^{~\mu}(z)=\sigma^{\nu}\partial_{\nu}h^{\mu}(z)\,,
\end{equation}
which agrees with
eq.(\ref{4infR}).  Furthermore eq.(\ref{4green1}) shows that the
definition\,(\ref{4R2}) of $\Omega(z;g)$  
is consistent with eq.(\ref{Omega2}).   
We may  normalize $R^{~\mu}_{\nu}(z;g)$ as well
\begin{equation}
\hat{R}^{~\mu}_{\nu}(z;g)=\Omega(z;g)^{-1}R^{~\mu}_{\nu}(z;g)
\in \mbox{SO}(1,3)\cong
\mbox{SL}(2,\Com)/\Z_{2}\,.
\end{equation}

\subsection{Representations\label{subrep}}
Based on the results in the previous subsection, it is easy to show that  
the matrix, ${\cal R}_{M}{}^{N}(z;g)$, given in 
eq.(\ref{4calR})  is of the form
\begin{equation}
\begin{array}{l}
{\cal R}_{M}{}^{N}(z;g)\\
{}\\
=\left(
\begin{array}{ccc}
R_{\mu}^{~\nu}(z;g)&
i(\bar{\Sigma}^{b}(z;g)\tilde{\sigma}_{\mu}L(z_{+};g))^{\beta}&
-i(\bar{L}(z_{-};g)\tilde{\sigma}_{\mu}\Sigma_{b}(z;g))^{\db}\\
0&(\Omega(z;g)\Upsilon(z;g))^{\frac{1}{2}}
\hat{L}_{\alpha}^{~\beta}(z_{+};g)\hat{u}^{-1}{}^{b}_{~a}(z;g)&0\\
0&0&\displaystyle{\left(\frac{\Omega(z;g)}{\Upsilon(z;g)}\right)}^{\frac{1}{2}}
\hat{\bar{L}}{}^{\db}_{~\da}(z_{-};g)\hat{u}^{a}{}_{b}(z;g)
\end{array}\right)\,.
\label{4calRR}
\end{array}
\end{equation}
Since  ${\cal R}_{M}{}^{N}(z;g)$ is a representation of the four dimensional $N$-extended superconformal group,  
each of the following also forms a representation of the group, though it is  
not a faithful representation
\begin{equation}
\begin{array}{ccc}
\Omega(z;g)\in\mbox{D}\,,~~~&~~~\Upsilon(z;g)\in\mbox{U}(1)\,,
~~~&~~~\hat{R}(z;g)\in\mbox{SO}(1,3)\,,\\
{}&{}&{}\\
\hat{L}(z_{+};g)\in\mbox{SL}(2,\Com)\,,
~~~&~~~\hat{\bar{L}}(z_{-};g)\in\mbox{SL}(2,\Com)\,,~~~&~~~
\hat{u}(z;g)\in\mbox{SU}({\cal N})\,,
\end{array}
\end{equation}
where $\mbox{D}$ is the one dimensional group of dilations. \newline
Under the 
successive superconformal transformations,~$g^{\prime\prime}: 
z\stackrel{g}{\longrightarrow}
z^{\prime}\stackrel{g^{\prime}}{\longrightarrow}z^{\prime\prime}$, they satisfy
\begin{equation}
\begin{array}{cc}
\Omega(z;g)\Omega(z^{\prime};g^{\prime})=\Omega(z;g^{\prime\prime})\,,~~~~&~~~~
\mbox{and~so~on.}
\end{array}
\end{equation}
\indent  We note that when ${\cal N}\neq 4$, $\Omega_{+}(z_{+};g)$ and 
$\Omega_{-}(z_{-};g)$ can be written as
\begin{equation}
\begin{array}{c}
\Omega_{+}(z_{+};g)=\Omega(z;g)\Upsilon(z;g)^{\frac{{\cal N}}{{\cal N}-4}}\,,\\
{}\\
\Omega_{-}(z_{-};g)=\Omega(z;g)\Upsilon(z;g)^{-\frac{{\cal N}}{{\cal N}-4}}\,.
\end{array}
\label{N4OO}
\end{equation}
Hence, they also form representations of the  ${\cal N}\neq 4$  
superconformal group. On the other hand, in the case of ${\cal N}=4$,   
due to the arbitrariness of $\psi$ in eq.(\ref{4Mform4}) 
$L(z_{+};g)$ and $L(z_{-};g)$ do not form representations. 
They do so only if  the equivalence relation\,(\ref{equivre2}) 
is imposed, but   
this  will give just  $\hat{L}(z_{+};g)\,,\hat{\bar{L}}(z_{-};g)$ and 
$\Omega(z;g)$.

\subsection{Functions of Two Points\label{F2}}
In this subsection,  we construct matrix  valued   
functions depending on  two points, $z_{1}$ and $z_{2}$, 
in superspace which transform covariantly like a product of two tensors 
at $z_{1}$ and $z_{2}$  under superconformal transformations.  \newline
\indent   If  $F(z)$ is defined for $z\in\Real^{4|4{\cal N}}$ by
\begin{equation}
F(z)=\bar{Z}(0)G_{T}(z)Z(0)
=\left(\begin{array}{cc}
              -i\tx_{+}&2\bt_{b}\\
           2\theta^{a}& \delta^{a}_{~b}\end{array}\right)\,,
\label{4Fform}
\end{equation}
then  $F(z)$ satisfies
\begin{equation}
F(-z)=\left(\begin{array}{cc}
1&0\\
0&-1
\end{array}\right)F(z)^{\dagger}\left(\begin{array}{cc}
1&0\\
0&-1
\end{array}\right)
=\left(\begin{array}{cc}
              i\tx_{-}&-2\bt_{b}\\
           -2\theta^{a}& \delta^{a}_{~b}\end{array}\right)\,,
\label{4-Fform}
\end{equation}
and the superdeterminant of $F(z)$ is given by
\begin{equation}
\mbox{sdet}\:F(z)=-x_{-}^{2}\,.
\label{4sdetZZ}
\end{equation}
\indent   If we consider 
\begin{equation}
\left(\begin{array}{cc}
               1&0\\
                -2i\theta^{a}\tx^{-1}_{+}&1
\end{array}\right)F(z)
\left(\begin{array}{cc}
               1&-2i\tx^{-1}_{+}\bt_{b}\\
               0&\,1
\end{array}\right)=
\left(\begin{array}{cc}
               -i\tx_{+}&0\\
               0&v^{a}_{~b}(-z)
\end{array}\right)\,,
\label{4XV}
\end{equation}
then this defines $v^{a}_{~b}(z)$ as 
\begin{equation}
v^{a}_{~b}(z)=\delta^{a}_{~b}+4i\displaystyle{\frac{1}{x^{2}_{-}}}
\theta^{a}{\rm x}_{-}\bt_{b}\,.
\label{4Vzz}
\end{equation}
From eqs.(\ref{4sdetZZ},\,\ref{4XV}) it is evident that
\begin{equation}
\displaystyle{\det v(z)=\frac{x^{2}_{-}}{x^{2}_{+}}}\,.
\label{4vdet}
\end{equation}
It is useful to note 
\begin{equation}
v^{a}_{~b}(-z)=v^{-1}{}^{a}{}_{b}(z)
=v^{\dagger}{}^{a}_{~b}(z)=\delta^{a}_{~b}-
4i\displaystyle{\frac{1}{x^{2}_{+}}}\theta^{a}{\rm x}_{+}\bt_{b}\,.
\end{equation}
\indent   Now with the  supersymmetric interval  for $\Real^{4|4{\cal N}}$ 
superspace defined by 
\begin{equation}
\begin{array}{ccc}
\multicolumn{3}{c}{G_{T}(z_{2})^{-1}G_{T}(z_{1})=G_{T}(z_{12})\,,~~~~~~~~
z^{M}_{12}=(x^{\mu}_{12},\theta^{a}_{12},\bt_{12a})=
-z^{M}_{21}\,,}\\
{}&{}&{}\\
x_{12}^{\mu}=x_{1}^{\mu}-x_{2}^{\mu}+i\theta^{a}_{1}\sigma^{\mu}\bt_{2a}
-i\theta^{a}_{2}\sigma^{\mu}\bt_{1a}\,,
~~~~&~~~~\theta^{a}_{12}=\theta^{a}_{1}-\theta^{a}_{2}\,,~~~~&~~~~
\bt_{12a}=\bt_{1a}-\bt_{2a}\,,
\label{4susyint}
\end{array}
\end{equation}
we may write
\begin{equation}
\displaystyle{\bar{Z}(z_{2})Z(z_{1})=F(z_{12})=
              \left(\begin{array}{cc}
              i\tx_{\bar{2}1}&-2\bt_{21b}\\
                -2\theta^{a}_{21}& \delta^{a}_{~b}\end{array}\right)}\,,
\label{4ZZ}
\end{equation}
and 
\begin{equation}
\begin{array}{cc}
\mbox{sdet}\:F(z_{12})=-x^{2}_{\bar{1}2}\,,~~~~&~~~~
\det 
v(z_{21})=\displaystyle{\frac{x^{2}_{\bar{2}1}}{x^{2}_{\bar{1}2}}}\,,
\end{array}
\end{equation}
where 
\begin{equation}
\begin{array}{c}
x^{\mu}_{\bar{2}1}=x^{\mu}_{2-}-x^{\mu}_{1+}-2i
\theta^{a}_{1}\sigma^{\mu}\bar{\theta}_{2a}
=x_{21}^{\mu}+i\theta^{a}_{21}\sigma^{\mu}\bar{\theta}_{21a}
=(x_{21})_{-}^{\mu}\,,\\
{}\\
x^{\mu}_{\bar{1}2}=x^{\mu}_{1-}-x^{\mu}_{2+}-2i
\theta^{a}_{2}\sigma^{\mu}\bar{\theta}_{1a}
=x_{12}^{\mu}+i\theta^{a}_{12}\sigma^{\mu}\bar{\theta}_{12a}
=(x_{12})_{-}^{\mu}\,.
\end{array}
\end{equation}
\indent   From eqs.(\ref{4Ztr},\,\ref{4barZtr})  $F(z_{12})$ transforms as
\begin{equation}
F(z^{\prime}_{12})
=\bar{K}(z_{2};g)F(z_{12})K(z_{1};g)\,.
\label{4zz}
\end{equation}
In particular,  with eqs.(\ref{4Kform},\,\ref{4Kbarform}), this  gives 
transformation rules for $\tx^{\prime}_{\bar{1}2}$ and 
$\tx^{\prime}_{\bar{2}1}$
\begin{subeqnarray}
\label{4green2}
\tx^{\prime}_{\bar{1}2}=\bar{L}(z_{1-};g)\tx_{\bar{1}2}L(z_{2+};g)\,,\\
{}\nonumber\\
\tx^{\prime}_{\bar{2}1}=\bar{L}(z_{2-};g)\tx_{\bar{2}1}L(z_{1+};g)\,,
\end{subeqnarray}
so that
\begin{subeqnarray}
x^{\prime 2}_{\bar{1}2}=\Omega_{-}(z_{1-};g)\Omega_{+}(z_{2+};g)
x_{\bar{1}2}^{2}\,,\\
{}\nonumber\\
x^{\prime 2}_{\bar{2}1}=\Omega_{-}(z_{2-};g)\Omega_{+}(z_{1+};g)
x_{\bar{2}1}^{2}\,,
\end{subeqnarray}
and in particular
\begin{subeqnarray}
\label{4green4}
&x^{\prime 2}_{\bar{1}2}x^{\prime 2}_{\bar{2}1}
=\Omega(z_{1};g)^{2}\Omega(z_{2};g)^{2}
x_{\bar{1}2}^{2}x_{\bar{2}1}^{2}\,,&\\
{}\nonumber\\
&\displaystyle{
\left(
\frac{x^{\prime 2}_{\bar{1}2}}{x^{\prime 2}_{\bar{2}1}}
\right)^{\frac{4}{{\cal N}}-1}
=\frac{\Upsilon(z_{1};g)^{2}}{\Upsilon(z_{2};g)^{2}}\left(\frac{
x_{\bar{1}2}^{2}}{x_{\bar{2}1}^{2}}\right)^{\frac{4}{{\cal N}}-1}}\,.&
\end{subeqnarray}
From eqs.(\ref{4green1},\,\ref{4green2}) 
$\mbox{tr}(\sigma^{\mu}\tx_{\bar{1}2}\sigma^{\nu}\tx_{\bar{2}1})$ 
transforms covariantly as
\begin{equation}
\mbox{tr}(\sigma^{\mu}\tx^{\prime}_{\bar{1}2}\sigma^{\nu}
\tx^{\prime}_{\bar{2}1})=
\mbox{tr}(\sigma^{\lambda}\tx_{\bar{1}2}\sigma^{\rho}
\tx_{\bar{2}1})
R_{\lambda}{}^{\mu}(z_{1};g)
R_{\rho}{}^{\nu}(z_{2};g)\,.
\label{4green5}
\end{equation}
\indent  Since $v^{a}_{~b}(z_{21})$ transforms infinitesimally as
\begin{equation}
\delta v(z_{21})=\hat{t}(z_{2})v(z_{21})-v(z_{21})\hat{t}(z_{1})
+i\textstyle{\frac{2}{{\cal N}}}(\hat{\psi}(z_{2})-\hat{\psi}(z_{1}))
v(z_{21})\,,
\label{4infV}
\end{equation}
finitely it transforms as
\begin{equation}
v(z^{\prime}_{21})=u^{-1}(z_{2};g)v(z_{21})u(z_{1};g)\,.
\label{4green3}
\end{equation}
\indent   From eqs.(\ref{4tZp1},\,\ref{4tZp2}) $F(z_{12})$ transforms 
under superinversion as
\begin{equation}
\bar{K}(z_{2};i_{s})F(z_{12}) 
K(z_{1};i_{s})=F(-z^{\prime}_{12})^{t}\,,
\end{equation}
which gives 
\begin{equation}
\begin{array}{cc}
\tx_{2-}^{-1}\,\tx_{\bar{2}1}\,\tx_{1+}^{-1}
=-{\rm x}_{\bar{1}2}^{\prime}\,,
~~~~~&~~~~~\displaystyle{x_{\bar{1}2}^{\prime
2}=\frac{x_{\bar{2}1}^{2}}{x_{2-}^{2}\,x_{1+}^{2}}}\,, 
\end{array}
\label{4xxx}
\end{equation}
and
\begin{equation}
\bar{\zeta}v^{-1}(z_{2})v(z_{21})v(z_{1})\zeta=
v(z^{\prime}_{21})^{t}\,.
\end{equation}
Eq.(\ref{4xxx}) shows that  
eq.(\ref{4green5}) holds for superinversion as well
\begin{equation}
\mbox{tr}(\sigma^{\mu}\tx^{\prime}_{\bar{1}2}\sigma^{\nu}
\tx^{\prime}_{\bar{2}1})=
\mbox{tr}(\sigma^{\lambda}\tx_{\bar{1}2}\sigma^{\rho}\tx_{\bar{2}1})
R_{\lambda}{}^{\mu}(z_{1};i_{s})R_{\rho}{}^{\nu}(z_{2};i_{s})\,.
\label{4xxRR}
\end{equation}

\subsection{Functions of Three Points}
In this subsection,  for three points, $z_{1},z_{2},z_{3}$ in superspace,   
we construct     `tangent' vectors, 
$Z_{i}$, which transform homogeneously at $z_{i},\,i=1,2,3$. \newline
\indent  With $z_{21}\stackrel{i_{s}}{\longrightarrow}(z_{21})^{\prime},\,
z_{31}\stackrel{i_{s}}{\longrightarrow}(z_{31})^{\prime}$,   
we define $\Z_{1}^{M}=(X^{\mu}_{1},\Theta^{a}_{1},
\bar{\Theta}_{1a})\in\Real^{4|4{\cal N}}$ by
\begin{equation}
G_{T}((z_{31})^{\prime})^{-1}{}G_{T}((z_{21})^{\prime})=G_{T}(\Z_{1})\,.
\label{F3defnew}
\end{equation}
Explicit expressions for $\Z^{M}_{1}$ can be obtained by calculating
\begin{equation}
\bar{Z}((z_{31})^{\prime})Z((z_{21})^{\prime})=F(\Z_{1})=
\left(\begin{array}{cc}
-i\tX_{1+}&2\bar{\Theta}_{1b}\\
2\Theta^{a}_{1}& \delta^{a}_{~b}
\end{array}\right)\,.
\label{F3def}
\end{equation}
We get 
\begin{equation}
\begin{array}{cc}
\multicolumn{2}{c}{
\tX_{1+}=-{\rm x}^{-1}_{\bar{1}3}{\rm x}_{\bar{2}3}{\rm
x}^{-1}_{\bar{2}1}\,,}\\
{}&{}\\
\Theta^{a}_{1}=i(\tilde{\bt}{}^{a}_{21}\x_{\bar{2}1}^{-1}
-\tilde{\bt}{}^{a}_{31}\x_{\bar{3}1}^{-1})\,,~~~~&~~~~
\bar{\Theta}_{1a}=i(\x_{\bar{1}3}^{-1}\tilde{\theta}_{13a}
-\x_{\bar{1}2}^{-1}\tilde{\theta}_{12a})\,.
\end{array}
\label{4W1}
\end{equation}
Using
\begin{equation}
\x_{\bar{1}3}+\x_{\bar{2}1}+4i\tilde{\theta}_{13a}\tilde{\bt}{}^{a}_{21}
=\x_{\bar{2}3}\,,
\end{equation}
one can assure  
\begin{equation}
\begin{array}{c}
\tX_{1-}=\tX_{1+}-4i\bar{\Theta}_{1a}\Theta^{a}_{1}=
\x_{\bar{1}2}^{-1}\x_{\bar{3}2}\x_{\bar{3}1}^{-1}=\tX_{1+}^{\dagger}\,,\\
{}\\
\tX_{1}=\textstyle{\frac{1}{2}}(\tX_{1+}+\tX_{1-})=
X^{\mu}_{1}\tilde{\sigma}_{\mu}\,.
\end{array}
\label{4Ypro}
\end{equation}
It is evident from eq.(\ref{F3defnew}) that under 
$z_{2}\leftrightarrow z_{3}$,  $\Z_{1}\rightarrow - \Z_{1}$.\newline
\indent  Associated with $F(z)$  given in eq.(\ref{4Fform}) we define
$\tilde{F}(z)$ by 
\begin{equation}
\tilde{F}(z)=\left(\begin{array}{cc}
\epsilon&0\\
0&\zeta
\end{array}\right)F(z)^{t}\left(\begin{array}{cc}
\bar{\epsilon}&0\\
0&\bar{\zeta}
\end{array}\right)
=\left(\begin{array}{cc}
              i{\rm x}_{+}&2\tilde{\theta}_{b}\\
           2\tilde{\bar{\theta}}{}^{a}& \delta^{a}{}_{b}
\end{array}\right)\,.
\label{4tildeFform}
\end{equation}
With this definition  we may write
\begin{equation}
\tilde{F}(\Z_{1})=
\left(\begin{array}{cc}
i\X_{1+}&2\tilde{\Theta}_{1b}\\
2\tilde{\bar{\Theta}}{}^{a}_{1}& \delta^{a}_{~b}
\end{array}\right)\,,
\end{equation}
where
\begin{equation}
\begin{array}{cc}
\multicolumn{2}{c}{
\X_{1+}=-\tx^{-1}_{\bar{2}1}\tx_{\bar{2}3}\tx^{-1}_{\bar{1}3}\,,}\\
{}&{}\\
\tilde{\Theta}_{1a}=i(\tx^{-1}_{\bar{3}1}\bt_{31a}
-\tx^{-1}_{\bar{2}1}\bt_{21a})\,,~~~~&~~~~
\tilde{\bar{\Theta}}{}^{a}_{1}=i
(\theta^{a}_{12}\tx^{-1}_{\bar{1}2}-\theta^{a}_{13}\tx^{-1}_{\bar{1}3})\,.
\end{array}
\end{equation}
$\tilde{F}(\Z_{1})$ transforms
infinitesimally as
\begin{equation}
\begin{array}{ll}
\delta \tilde{F}(\Z_{1})=&\left(\begin{array}{cc}   
\tilde{\hat{w}}(z_{1})-\textstyle{\frac{1}{2}}\hat{\lambda}(z_{1})+i
\textstyle{\frac{1}{2}}\hat{\psi}(z_{1})&0\\
0&\hat{t}(z_{1})+i\textstyle{\frac{2}{{\cal N}}}(\hat{\psi}(z_{1})+\delta^{4}{}_{{\cal N}}\,\Omega)
\end{array}\right)
\tilde{F}(\Z_{1})\\
{}&{}\\
{}&~~-\tilde{F}(\Z_{1})
\left(\begin{array}{cc}   
\hat{w}(z_{1})+\textstyle{\frac{1}{2}}\hat{\lambda}(z_{1})+i
\textstyle{\frac{1}{2}}\hat{\psi}(z_{1})&0\\
0&\hat{t}(z_{1})+i\textstyle{\frac{2}{{\cal N}}}(\hat{\psi}(z_{1})+\delta^{4}{}_{{\cal N}}\,\Omega)
\end{array}\right)\,,
\end{array}
\end{equation}
and hence for finite transformations
\begin{equation}
\tilde{F}(\Z_{1}^{\prime})=\left(\begin{array}{cc}
L(z_{1+};g)^{-1}&0\\
0&u(z_{1};g)^{-1}
\end{array}\right) \tilde{F}(\Z_{1})
\left(\begin{array}{cc}    
\bar{L}(z_{1-};g)^{-1}&0\\
0&u(z_{1};g)
\end{array}\right)\,.
\label{4FFtrans}
\end{equation}
Thus $\Z_{1}$ transforms homogeneously at $z_{1}$, 
as    `tangent' vectors do. \newline
Explicitly we have from eq.(\ref{4FFtrans})  
\begin{equation}
\begin{array}{c}
X^{\prime}_{1+}=\Omega(z_{1};g)^{-1}\hat{L}(z_{1+};g)^{-1}
\X_{1+}\hat{\bar{L}}(z_{1-};g)^{-1}\,,\\
{}\\
\tilde{\Theta}^{\prime}_{1a}
=\Omega(z_{1};g)^{-\frac{1}{2}}\Upsilon(z_{1};g)^{-\frac{1}{2}}
\hat{L}(z_{1+};g)^{-1}\tilde{\Theta}_{1b}\hat{u}^{b}_{~a}(z_{1};g)\,,\\
{}\\
\tilde{\bar{\Theta}}{}^{\prime a}_{1}=
\Omega(z_{1};g)^{-\frac{1}{2}}\Upsilon(z_{1};g)^{\frac{1}{2}}
\hat{u}^{-1}{}^{a}_{~b}(z_{1};g)\tilde{\bar{\Theta}}{}^{b}_{1}
\hat{\bar{L}}(z_{1-};g)^{-1}\,.
\end{array}
\label{4expX}
\end{equation}
$\X_{1-}$ also transforms in the same way as  $\X_{1+}$ 
in eq.(\ref{4expX}) and hence
\begin{subeqnarray}
\label{XXXR}
&X^{\prime\mu}_{1}=\Omega(z_{1};g)^{-1}X^{\nu}_{1}\hat{R}_{\nu}{}^{\mu}(z_{1};g)\,,&
\label{XXR}\\
{}\nonumber\\
&\Theta_{1}^{\prime a}\sigma^{\mu}\bar{\Theta}^{\prime}_{1a}=
\Omega(z_{1};g)^{-1}\Theta_{1}^{a}\sigma^{\nu}\bar{\Theta}_{1a}
\hat{R}_{\nu}{}^{\mu}(z_{1};g)\,.&
\end{subeqnarray}
\indent  From eq.(\ref{4sdetZZ}) we get 
\begin{equation}
\mbox{sdet}\:\tilde{F}(\Z_{1})=\mbox{sdet}\:F(\Z_{1})=-X^{2}_{1-}=
-\displaystyle{
\frac{x^{2}_{\bar{3}2}}{x^{2}_{\bar{1}2}\,x^{2}_{\bar{3}1}}}\,.
\end{equation}
If we define a function $\tilde{v}(z)\in\mbox{U}({\cal N})$ by
\begin{equation}
\tilde{v}^{a}_{~b}(z)=(\zeta v(z)^{t}\bar{\zeta})^{a}_{~b}=
\delta^{a}_{~b}
-4i\displaystyle{\frac{1}{x^{2}_{-}}}
\tilde{\bt}{}^{a}\tx_{-}\tilde{\theta}_{b}\,,
\end{equation}
then a direct calculation leads  
\begin{equation}
\tilde{v}(\Z_{1})=v(z_{13})v(z_{32})v(z_{21})\,.
\label{4tVdecom}
\end{equation} 
Similarly for $R_{\mu}{}^{\nu}(z;i_{s})$ given in eq.(\ref{4Rzi}) we have
\begin{equation}
R(\Z_{1};i_{s})=x^{2}_{\bar{1}2}x^{2}_{\bar{2}1}
x^{2}_{\bar{3}1}x^{2}_{\bar{1}3}\,R(z_{12};i_{s})R(z_{23};i_{s})R(z_{31};i_{s})\,.
\label{41311223}
\end{equation}
From  eqs.(\ref{4green5},\,\ref{4green3})  
$\tilde{v}(\Z_{1}),R(\Z_{1};i_{s})$ transform homogeneously at $z_{1}$  
under superconformal transformation, 
$z\stackrel{g}{\longrightarrow}z^{\prime}$, 
\begin{subeqnarray}
&\tilde{v}(\Z_{1}^{\prime})=
u^{-1}(z_{1};g)\tilde{v}(\Z_{1})u(z_{1};g)\,,&\\
{}\nonumber\\
&R(\Z^{\prime}_{1};i_{s})=\Omega(z_{1};g)^{2}
R^{-1}(z_{1};g)R(\Z_{1};i_{s})R(z_{1};g)\,.&
\end{subeqnarray}
\indent  Under superinversion,   
$z_{j}\stackrel{i_{s}}{\longrightarrow}z^{\prime}_{j},\,j=1,2,3$,  \,
$\Z_{1}$ transforms to $\Z_{1}^{\prime}$,  
from eq.(\ref{4xxx}), as  
\begin{equation}  
\begin{array}{c}   
\tX^{\prime}_{1+}=\tx_{1+}\X_{1-}\tx_{1-}\,,\\   
{}\\   
\Theta^{\prime a}_{1}
=iv^{-1}{}^{a}{}_{b}(z_{1})\tilde{\bar{\Theta}}{}^{b}_{1}\tx_{1-}\,,
\end{array}
\label{4superinvZ1}
\end{equation}
and hence
\begin{subeqnarray}
\label{XXXIR}
&X^{\prime\mu}_{1}
=\Omega(z_{1};i_{s})^{-2}X^{\nu}_{1}R_{\nu}{}^{\mu}(z_{1};i_{s})\,,&\\
{}\nonumber\\
&\Theta^{\prime a}_{1}\sigma^{\mu}\bar{\Theta}{}^{\prime}_{1a}=-
\Omega(z_{1};i_{s})^{-2}\Theta^{a}_{1}\sigma^{\nu}\bar{\Theta}_{1a}
R_{\nu}{}^{\mu}(z_{1};i_{s})\,.&
\label{4-sign}
\end{subeqnarray}
Note the minus sign in eq.(\ref{4-sign}).\newline
\indent  By taking cyclic permutations of   
$z_{1},z_{2},z_{3}$ in eq.(\ref{4W1})  
we may define $\Z_{2},\Z_{3}$.   
We find $\Z_{2},\Z_{3}$ are related to $\Z_{1}$ in a simple form
\begin{equation}
\begin{array}{l}
F((\Z_{2})^{\prime})=\left(\begin{array}{cc}
                 -i\tx_{\bar{2}1}&0\\
                 0&v(z_{21})
            \end{array}\right)\tilde{F}(\Z_{1})
\left(\begin{array}{cc}
                 -i\tx_{\bar{1}2}&0\\
                 0&v(z_{12})
            \end{array}\right)\,,\\
{}\\
F(\Z_{3})=\left(\begin{array}{cc}
                 -i\x_{\bar{3}1}^{-1}&0\\
                 0&\tilde{v}(z_{13})
            \end{array}\right) \tilde{F}((\Z_{1})^{\prime})
\left(\begin{array}{cc}
                 -i\x_{\bar{1}3}^{-1}&0\\
                 0&\tilde{v}(z_{31})
            \end{array}\right)\,,
\end{array}
\end{equation}
where
$(\Z)^{\prime}=((X)^{\prime},(\Theta)^{\prime}{}^{a},
(\bar{\Theta})_{a}^{\prime})$ is defined by superinversion, 
$\Z\stackrel{i_{s}}{\longrightarrow}(\Z)^{\prime}$.\newline
Explicitly we have
\begin{subeqnarray}
\label{4Z2Z3}
&(X_{2})^{\prime}_{+}{\cdot\tilde{\sigma}}
=\tx_{\bar{2}1}\X_{1+}\tx_{\bar{1}2}\,,~~~~~&~~~~
(\Theta_{2})^{\prime}{}^{a}
=-iv^{a}{}_{b}(z_{21})\tilde{\bar{\Theta}}{}_{1}^{b}\tx_{\bar{1}2}\,,\\
{}&{}\nonumber\\
&\tX_{3+}=\x_{\bar{3}1}^{-1}(X_{1+})^{\prime}{\cdot\sigma}\,
\x_{\bar{1}3}^{-1}\,,~~~&~~~~
\Theta^{a}_{3}=-i\tilde{v}^{a}{}_{b}(z_{13})(\tilde{\bar{\Theta}}_{1})
{}^{\prime}{}^{b}\x_{\bar{1}3}^{-1}\,.
\end{subeqnarray}
\indent  From eqs.(\ref{4W1},\,\ref{4Ypro}) we get
\begin{subeqnarray}
&\displaystyle{
\frac{X_{1+}^{2}}{X_{1-}^{2}}=\frac{x_{\bar{1}2}^{2}x_{\bar{2}3}^{2}
x_{\bar{3}1}^{2}}{x_{\bar{2}1}^{2}x_{\bar{1}3}^{2}
x_{\bar{3}2}^{2}}}\,,&\\
{}\nonumber\\
&\displaystyle{
\frac{X_{1+}{\cdot X_{1-}}}{\sqrt{X_{1+}^{2}X_{1-}^{2}}}}
=-\displaystyle{
\frac{\,\mbox{tr}(\x_{\bar{2}1}\tx_{\bar{2}3}
\x_{\bar{1}3}\tx_{\bar{1}2}\x_{\bar{3}2}\tx_{\bar{3}1})}{
\,2\sqrt{x_{\bar{1}2}^{2}x_{\bar{2}1}^{2}x_{\bar{2}3}^{2}x_{\bar{3}2}^{2}
x_{\bar{3}1}^{2}x_{\bar{1}3}^{2}}}}\,.&
\end{subeqnarray}
These expressions are invariant under cyclic permutations of 
$z_{1},z_{2},z_{3}$ and hence
\begin{subeqnarray}
\label{Park}
&\displaystyle{
\frac{X_{1+}^{2}}{X_{1-}^{2}}=
\frac{X_{2+}^{2}}{X_{2-}^{2}}=\frac{X_{3+}^{2}}{X_{3-}^{2}}}\,,&\\
{}\nonumber\\
&\displaystyle{
\frac{X_{1+}{\cdot X_{1-}}}{\sqrt{X_{1+}^{2}X_{1-}^{2}}}=
\frac{X_{2+}{\cdot X_{2-}}}{\sqrt{X_{2+}^{2}X_{2-}^{2}}}=
\frac{X_{3+}{\cdot X_{3-}}}{\sqrt{X_{3+}^{2}X_{3-}^{2}}}}\,.&
\end{subeqnarray}
From eq.(\ref{XXXR}) these are invariants  for  
any continuous superconformal transformation and furthermore  
from eq.(\ref{XXXIR}) the latter is invariant under superinversion along with
\begin{equation}
\displaystyle{
\frac{X_{1+}^{2}}{X_{1-}^{2}}+
\frac{X_{1-}^{2}}{X_{1+}^{2}}}\,.
\end{equation}
Note that such invariants, depending on three points, 
do not exist in ordinary conformal theories and 
that in the case of 
${\cal N}=1$ due to the identity (\ref{N=1sigma}) 
those two variables are not independent~\cite{paper1,9808041}.

\section{Superconformal Invariance of Correlation Functions
\label{4correlation}}
In this section  we discuss the    
superconformal invariance of correlation 
functions for  quasi-primary superfields and exhibit general forms of   
two-point, three-point and $n$-point functions.

\subsection{Quasi-primary Superfields}
We first   assume\footnote{In \cite{paper1} it was explicitly 
shown that 
the chiral/anti-chiral superfields and 
supercurrents in some ${\cal N}=1$ theories are quasi-primary.}  
that there exist quasi-primary 
 superfields, 
$\Psi^{I}(z)$,  which   under the superconformal transformation, 
$z\stackrel{g}{\longrightarrow}z^{\prime}$,  transform as
\begin{equation}
\begin{array}{cc}
\Psi^{I}\longrightarrow\Psi^{\prime}{}^{I}\,,~~~&~~~ 
\Psi^{\prime}{}^{I}(z^{\prime})
=\Psi^{J}(z)D^{~I}_{J}(z;g)\,.
\end{array}
\label{4primary}
\end{equation}
$D(z;g)$ obeys the group property   
so that under the 
successive superconformal transformations,~$g^{\prime\prime}: 
z\stackrel{g}{\longrightarrow}
z^{\prime}\stackrel{g^{\prime}}{\longrightarrow}z^{\prime\prime}$, it 
satisfies 
\begin{equation}
D(z;g)D(z^{\prime};g^{\prime})=D(z;g^{\prime\prime})\,,
\label{4Drep}
\end{equation}
and hence also 
\begin{equation}
D(z;g)^{-1}=D(z^{\prime};g^{-1})\,.
\end{equation}
We choose here $D(z;g)$ to be a representation of 
$\mbox{SL}(2,\Com)\times\mbox{SU}({\cal N})\times\mbox{U}(1)\times
\mbox{D}$, which is a subgroup of the stability group at $z=0$,   
and so  we decompose the spin index, $I$, of superfields into 
$\mbox{SL}(2,\Com)$ index, $\rho$, and $\mbox{SU}({\cal N})$ index, $r$, as 
$\Psi^{I}\equiv \Psi^{\rho}{}_{r}$. Now  $D_{J}^{~I}(z;g)$  
is factorized as 
\begin{equation}
D_{J}^{~I}(z;g)=D_{\rho}^{~\sigma}(\hat{L}(z_{+};g))D^{r}_{~s}(\hat{u}(z;g))
\Omega(z;g)^{-\eta}\Upsilon(z;g)^{-\kappa}\,,
\label{4DLU}
\end{equation}
where 
$D_{\rho}^{~\sigma}(\hat{L}),\,D^{r}_{~s}(\hat{u})$ are  representations of 
$\mbox{SL}(2,\Com),\,\mbox{SU}({\cal N})$ respectively,  while    
$\eta$ and $\kappa$ are the scale dimension and $R$-symmetry charge of 
$\Psi^{\rho}{}_{r}$ respectively. \newline
Infinitesimally 
\begin{equation}
\delta\Psi^{\rho}{}_{r}(z)=
-({\cal L}+\eta\hat{\lambda}(z)+i\kappa\hat{\Omega}(z))\Psi^{\rho}{}_{r}(z)
-\Psi^{\sigma}{}_{r}(z)\textstyle{\frac{1}{2}}(s_{\mu\nu})_{\sigma}{}^{\rho}
\hat{w}^{\mu\nu}(z)
-\Psi^{\rho}{}_{s}(z)\textstyle{\frac{1}{2}}(s^{a}_{~b})^{s}_{~r}
\hat{t}{}^{b}_{~a}(z)\,,
\label{4deltaPsi}
\end{equation}
where $s_{\mu\nu},\,s^{a}_{~b}$ are matrix generators of
$\mbox{SO}(1,3),\,\mbox{SU}({\cal N})$ satisfying
\begin{equation}
\begin{array}{c}
[s_{\mu\nu},s_{\lambda\rho}]=-\eta_{\mu\lambda}s_{\nu\rho}+\eta_{\mu\rho}
s_{\nu\lambda}+\eta_{\nu\lambda}s_{\mu\rho}-\eta_{\nu\rho}s_{\mu\lambda}\,,\\
{}\\
{[}{s^{a}_{~b}},{}{s^{c}_{~d}}{]}=
2(\delta^{a}_{~d}s^{c}_{~b}-\delta^{c}_{~b}s^{a}_{~d})\,,
\end{array}
\label{4stcom}
\end{equation}
and hence
\begin{equation}
[\textstyle{\frac{1}{2}}s^{a}_{~b}t_{1}^{b}{}_{a},
\textstyle{\frac{1}{2}}s^{c}_{~d}t_{2}^{d}{}_{c}]=
\textstyle{\frac{1}{2}}s^{a}_{~b}[t_{1},t_{2}]^{b}_{~a}\,.
\end{equation}
From eqs.(\ref{4defH},\,\ref{4Kcom}) using eq.(\ref{4stcom}) we have
\begin{equation}
\delta_{3}\Psi^{\rho}{}_{r}=[\delta_{2},\delta_{1}]\Psi^{\rho}{}_{r}\,.
\end{equation}
\indent  
It is useful to consider the complex conjugate superfield 
of $\Psi^{\rho}{}_{r}$ 
\begin{equation}
\bar{\Psi}^{\rho r}(z)=\Psi^{\rho}{}_{r}(z)^{\dagger}\,.
\end{equation}
$\bar{\Psi}^{\rho r}(z)$ transforms as
\begin{equation}
\bar{\Psi}^{\prime\rho r}(z^{\prime})=\Omega(z;g)^{-\eta}\Upsilon(z;g)^{\kappa}
\bar{D}^{\rho}_{~\sigma}(\hat{\bar{L}}(z_{-};g))
D^{r}_{~s}(\hat{u}(z;g)^{-1})\bar{\Psi}^{\sigma  s}(z)\,.
\end{equation} 
\indent  Superconformal invariance for a general $n$-point  function requires
\begin{equation}
\langle
\Psi_{1}^{\prime I_{1}}(z_{1})\Psi_{2}^{\prime I_{2}}(z_{2}) 
\cdots\Psi_{n}^{\prime I_{n}}(z_{n})\rangle
=\langle \Psi_{1}^{I_{1}}(z_{1})\Psi_{2}^{I_{2}} (z_{2})
\cdots \Psi_{n}^{I_{n}}(z_{n})\rangle\,.
\label{4Green}
\end{equation}
\indent  In superconformal field theories on chiral superspace, 
the representation of $\mbox{U}(1)\times\mbox{D}$ is given by 
\begin{equation}
\Omega_{+}(z_{+};g)^{-\eta}\,,
\end{equation}
so that for ${\cal N}\neq 4$,  
$\eta$ and $\kappa$ are related by,  from eq.(\ref{N4OO}),
\begin{equation}
\eta+\textstyle{(\frac{4}{{\cal N}}-1)}\kappa=0\,.
\label{Nn4ek}
\end{equation}
On the other hand when ${\cal N}=4$, as shown in subsection \ref{subrep},  
$\Omega_{+}(z_{+};g)$ does not form a representation of the ${\cal N}=4$ 
superconformal group,  and hence  there is no 
conventional way of defining quasi-primary chiral/anti-chiral superfields  
in ${\cal N}=4$ superconformal theories.  
We speculate that this fact makes it difficult to 
construct four-dimensional 
${\cal N}=4$ superconformal theories on chiral superspace.

\subsection{Two-point Correlation Functions}
The  solution for 
the two-point  function of the quasi-primary superfields,~$
\Psi^{\rho}{}_{r},\bar{\Psi}^{\rho r}$,  has the general  form\footnote{See 
subsection \ref{subn} for a proof.}   
\begin{equation}
\displaystyle{
\langle \bar{\Psi}^{\rho r}(z_{1})\Psi^{\sigma}{}_{s}(z_{2})\rangle = 
C_{\Psi}\frac{I^{\rho\sigma}(\hat{\tx}_{\bar{1}2})I^{r}{}_{s}(
\hat{v}(z_{12}))}{
(x^{2}_{\bar{1}2})^{\frac{1}{2}(\eta-(\frac{4}{{\cal N}}-1)\kappa)}
(x^{2}_{\bar{2}1})^{\frac{1}{2}(\eta+(\frac{4}{{\cal N}}-1)\kappa)}}}\,,
\label{42gen}
\end{equation}
where we define 
\begin{equation}
\begin{array}{c}
\displaystyle{\hat{\tx}_{\bar{1}2}=
\frac{\tx_{\bar{1}2}}{(x^{2}_{\bar{1}2})^{\frac{1}{2}}}\in\mbox{SL}(2,\Com)
}\,,\\
{}\\
\displaystyle{\hat{v}^{a}{}_{b}(z_{12})=
\left(\frac{x^{2}_{\bar{2}1}}{x^{2}_{\bar{1}2}}\right)^{\frac{1}{{\cal N}}}
\,(\delta^{a}{}_{b}+4i\theta_{12}^{a}\tx^{-1}_{\bar{1}2}\bt_{12b})
\in\mbox{SU}({\cal N})}\,,
\end{array}
\end{equation}
and   $I^{\rho\sigma}(\hat{\tx}_{\bar{1}2}),\,I^{r}{}_{s}(\hat{v}(z_{12}))$ 
are   tensors transforming covariantly according to the    
appropriate representations of $\mbox{SL}(2,\Com),\,\mbox{SU}({\cal N})$ 
which  are formed by decomposition of tensor products of   
$\hat{\tx}_{\bar{1}2},\,
\hat{v}(z_{12})$.\newline 
Under superconformal transformations,  
$I^{\rho\sigma}(\hat{\tx}_{\bar{1}2})$ and  
$I^{r}_{~s}(v(z_{12}))$ satisfy from 
eqs.(\ref{4green2},\,\ref{4green3})
\begin{subeqnarray}
\label{4IIpro}
&\bar{D}(\hat{\bar{L}}(z_{1-};g))I(\hat{\tx}_{\bar{1}2})D(\hat{L}(z_{2+};g))
=I(\hat{\tx}{}^{\prime}_{\bar{1}2})\,,&\\
{}\nonumber\\
&D(\hat{u}(z_{1};g)^{-1})I(\hat{v}(z_{12}))D(\hat{u}(z_{2};g))
=I(\hat{v}(z^{\prime}_{12}))\,.&
\end{subeqnarray}
\indent  As  examples,  we first consider  
the chiral/anti-chiral  scalar and 
spinorial fields, \newline
$S(z_{+}),\,\bar{S}(z_{-}),\,\phi^{\alpha}(z_{+}),\,\bar{\phi}^{\da}(z_{-})$ 
in ${\cal N}\neq 4$ theories  which transform as 
\begin{subeqnarray}
&S^{\prime}(z_{+}^{\prime})=\Omega_{+}(z_{+};g)^{-\eta}S(z_{+})\,,&\\
{}\nonumber\\
&\bar{S}^{\prime}(z_{-}^{\prime})
=\Omega_{-}(z_{-};g)^{-\eta}\bar{S}(z_{-})\,,&\\
{}\nonumber\\
&\phi^{\prime\alpha}(z_{+}^{\prime})=\Omega_{+}(z_{+};g)^{-\eta}
\phi^{\beta}(z_{+})
\hat{L}_{\beta}^{~\alpha}(z_{+};g)\,,&\\
{}\nonumber\\
&\bar{\phi}^{\prime\da}(z_{-}^{\prime})=\Omega_{-}(z_{-};g)^{-\eta}
\hat{\bar{L}}{}^{\da}_{~\db}(z_{-};g)\bar{\phi}^{\db}(z_{-})\,,&
\end{subeqnarray}
so that from eq.(\ref{Nn4ek}) 
$\eta+(\textstyle{\frac{4}{{\cal N}}}-1)\kappa=0$ and  $s_{\mu\nu}
\rightarrow\textstyle{\frac{1}{2}}\sigma_{[\mu}\tilde{\sigma}_{\nu]}
$ for the spinorial fields.\newline
The two-point functions of  them are 
\begin{eqnarray}
&\displaystyle{
\langle \bar{S}(z_{1-})S(z_{2+})\rangle = 
C_{S}\frac{1}{(x^{2}_{\bar{1}2})^{\eta}}}&\,,\\
{}\nonumber\\
&\displaystyle{
\langle \bar{\phi}^{\da}(z_{1-})\phi^{\alpha}(z_{2+})\rangle =
C_{\phi}i\frac{(\hat{\tx}_{\bar{1}2})^{\da\alpha}}
{(x^{2}_{\bar{1}2})^{\eta}}}&\,.
\end{eqnarray}
\indent  For a  real vector field, $V^{\mu}(z)$,  where the representation
of $\mbox{SL}(2,\Com)$  is given by
$\hat{R}_{\mu}^{~\nu}(z;g)$ and the $R$-symmetry charge  is zero,
$\kappa=0$, we have
\begin{equation}
\begin{array}{cc}
\displaystyle{
\langle V^{\mu}(z_{1})V^{\nu}(z_{2})\rangle =
C_{V}\frac{I^{\mu\nu}(z_{12})}{
(x^{2}_{\bar{1}2}x^{2}_{\bar{2}1})^{\frac{1}{2}\eta}}}\,,~~~~&~~~~
I^{\mu\nu}(z_{12})=\textstyle{\frac{1}{2}}\mbox{tr}
(\sigma^{\mu}\hat{\tx}_{\bar{1}2}\sigma^{\nu}\hat{\tx}_{\bar{2}1})\,.
\end{array}
\label{4VV}
\end{equation}
From eq.(\ref{4ese}) one can show
\begin{equation}
I_{\mu\nu}(z_{12})
=\textstyle{\frac{1}{2}}\mbox{tr}
(\sigma_{\mu}\hat{\tx}_{\bar{1}2}\sigma_{\nu}\hat{\tx}_{\bar{2}1})=
\textstyle{\frac{1}{2}}\mbox{tr}
(\tilde{\sigma}_{\mu}\hat{\x}_{\bar{2}1}
\tilde{\sigma}_{\nu}\hat{\x}_{\bar{1}2})\,,
\label{4IIAB}
\end{equation}
where 
\begin{equation}
\displaystyle{
\hat{\x}=\frac{\x}{(x^{2})^{\frac{1}{2}}}=\hat{\tx}^{-1}}\,.
\end{equation}
Hence $I^{\mu\nu}(z_{12})$ satisfies
\begin{equation}
I^{\mu\nu}(z_{12})I_{\lambda\nu}(z_{12})
=\delta^{\mu}_{~\lambda}\,.
\end{equation}
Note that $I(z_{12})\propto R(z_{12};i_{s})$, 
where $R(z;i_{s})$ is given by eq.(\ref{4Rzi}).  \newline
\indent  For   gauge fields,  $\psi_{a}(z),\,\bar{\psi}^{a}(z)$,  
which transform as
\begin{subeqnarray}
&\psi^{\prime}_{a}(z^{\prime})=\Omega(z;g)^{-\eta}\Upsilon(z;g)^{-\kappa}
\psi_{b}(z)\hat{u}{}^{b}_{~a}(z;g)\,,\\
{}\nonumber\\
&\bar{\psi}^{\prime a}(z^{\prime})=\Omega(z;g)^{-\eta}\Upsilon(z;g)^{\kappa}
\hat{u}{}^{-1}{}^{a}_{~b}(z;g)\bar{\psi}^{b}(z)\,,
\end{subeqnarray}
the two-point function of them is 
\begin{equation}
\displaystyle{
\langle \bar{\psi}^{a}(z_{1})\psi_{b}(z_{2})\rangle = 
C_{\psi}
\frac{\hat{v}{}^{a}{}_{b}(z_{12})}
{(x^{2}_{\bar{1}2})^{\frac{1}{2}(\eta-(\frac{4}{{\cal N}}-1)\kappa)}
(x^{2}_{\bar{2}1})^{\frac{1}{2}(\eta+(\frac{4}{{\cal N}}-1)\kappa)}}}\,.
\end{equation}
Note that  to have
non-vanishing two-point correlation functions,  the 
 scale dimensions,$\,\eta$, 
of the two fields must be equal and the $R$-symmetry charges 
must have the same absolute value with  opposite 
signs, $\kappa,\,-\kappa$,  as shown in subsection \ref{subn} later.  \newline
\indent For a real vector superfield, $V^{\mu}(z)$, if we define
\begin{equation}
{\rm V}_{\alpha\da}(z)=\sigma_{\mu\alpha\da}V^{\mu}(z)\,,
\end{equation}
from eq.(\ref{4VV}) we get
\begin{equation}
\displaystyle{
\langle {\rm V}_{\alpha\da}(z_{1}){\rm V}_{\beta\db}(z_{2})\rangle =
2C_{V}\frac{(\hat{\x}_{\bar{2}1})_{\alpha\db}(\hat{\x}_{\bar{1}{2}}
)_{\beta\da}}{
(x^{2}_{\bar{1}2}x^{2}_{\bar{2}1})^{\frac{1}{2}\eta}}}\,.
\end{equation}
From 
\begin{equation}
\begin{array}{cc}
D_{a\alpha}(z_{1})x^{\mu}_{\bar{2}1}=2i(\sigma^{\mu}\bt_{12a})_{\alpha}\,,
~~~~&~~~~~D_{a\alpha}(z_{1})x^{\mu}_{\bar{1}2}=0\,,
\end{array}
\end{equation}
with  $\tilde{D}^{a\alpha}=\zeta^{ab}
\epsilon^{-1}{}^{\alpha\beta}D_{b\beta}$ we get 
\begin{equation}
\tilde{D}^{a\alpha}(z_{1})\langle {\rm V}_{\alpha\da}(z_{1}){\rm
V}_{\beta\db}(z_{2})\rangle=4iC_{V}(\eta-3)\displaystyle{
\frac{\tilde{\bar{\theta}}{}^{a}_{12\db}(\x)_{\beta\da}}{
(x^{2}_{\bar{1}2}x^{2}_{\bar{2}1})^{\frac{1}{2}(\eta+1)}}}\,,
\end{equation}
and hence  
$\langle {\rm V}_{\alpha\da}(z_{1}){\rm V}_{\beta\db}(z_{2})\rangle$
is conserved if\,  $\eta=3$
\begin{equation}
\begin{array}{cc}
\tilde{D}^{a\alpha}(z_{1})\langle {\rm V}_{\alpha\da}(z_{1}){\rm
V}_{\beta\db}(z_{2})\rangle=0~~~~~&~~~~~~\mbox{if~~}\eta=3\,.
\end{array}
\label{4cons1}
\end{equation}
The anti-commutator relation for  
$D_{a\alpha},\bar{D}{}^{b}_{\da}$~(\ref{4anticomD}) implies also 
\begin{equation}
\begin{array}{cc}
\displaystyle{
\frac{\partial~}{\partial x^{\mu}_{1}}}
\langle V^{\mu}(z_{1})V^{\nu}(z_{2})\rangle=0~~~~~~~
&~~~~~~~~~~~\mbox{if~~}\eta=3\,.
\end{array}
\label{4cons2}
\end{equation}
\subsection{Three-point Correlation Functions}
The  solution for 
the three-point correlation  function of the quasi-primary
superfields,~$\Psi^{\rho}{}_{r}$, 
has the general  form\footnote{See 
subsection \ref{subn} for a proof.}   
\begin{equation}
\begin{array}{l}
\displaystyle{
\langle \Psi_{1}^{\rho}{}_{r}(z_{1})\Psi_{2}^{\sigma}{}_{s}(z_{2})
\Psi_{3}^{\tau}{}_{t}(z_{3})\rangle }\\
{}\\
=\displaystyle{ 
\frac{H^{\rho}{}_{r\sigma^{\prime}s^{\prime}\tau^{\prime}t^{\prime}}
(\Z_{1})I^{\sigma^{\prime}\sigma}(\hat{\tx}_{\bar{1}2})
I^{\tau^{\prime}\tau}(\hat{\tx}_{\bar{1}3})I^{s^{\prime}}{}_{s}(
\hat{v}(z_{12}))
I^{t^{\prime}}{}_{t}(\hat{v}(z_{13}))}
{(x^{2}_{\bar{1}2})^{\frac{1}{2}(\eta_{2}-(\frac{4}{{\cal N}}-1)\kappa_{2})}
(x^{2}_{\bar{2}1})^{\frac{1}{2}(\eta_{2}+(\frac{4}{{\cal N}}-1)\kappa_{2})}
(x^{2}_{\bar{1}3})^{\frac{1}{2}(\eta_{3}-(\frac{4}{{\cal N}}-1)\kappa_{3})}
(x^{2}_{\bar{3}1})^{\frac{1}{2}(\eta_{3}+(\frac{4}{{\cal N}}-1)\kappa_{3})}}
}\,,
\end{array}
\label{43gen}
\end{equation}
where  $\Z_{1}{}^{M}
=(X_{1}^{\mu},\Theta_{1}^{a},\bar{\Theta}_{1a})
\in\Real^{4|4{\cal N}}$ is given by eq.(\ref{F3defnew}). \newline
Superconformal invariance~(\ref{4Green}) is  now 
equivalent to   
\begin{subeqnarray}
\label{4equiv}
\begin{array}{c}
H^{\rho^{\prime}}{}_{r\sigma s\tau  t}(\Z)
D_{\rho^{\prime}}{}^{\rho}(\hat{L})
=H^{\rho}{}_{r\sigma^{\prime} s\tau^{\prime}  t}(\Z^{\prime})
\bar{D}^{\sigma^{\prime}}{}_{\sigma}(\hat{\bar{L}})
\bar{D}^{\tau^{\prime}}{}_{\tau}(\hat{\bar{L}})\,,\\
{}\\
\Z^{\prime M}=( X^{\nu}\hat{R}_{\nu}{}^{\mu}(\hat{L}),\,
\Theta^{a}\hat{L},\,\hat{\bar{L}}\bar{\Theta}_{a})\,,
\end{array}\\
{~}\nonumber\\
{~}\nonumber\\
\begin{array}{c}
H^{\rho}{}_{r^{\prime}\sigma s^{\prime}\tau t^{\prime}}(\Z)
D^{r^{\prime}}{}_{r}(\hat{u})D^{s^{\prime}}{}_{s}(\hat{u})
D^{t^{\prime}}{}_{t}(\hat{u})
=H^{\rho}{}_{r\sigma s\tau  t}(\Z^{\prime\prime})\,,\\
{}\\
\Z^{\prime\prime M}=( X^{\mu},\,
\Theta^{b}(\bar{\zeta}\hat{u}\zeta)_{b}{}^{a},\,
(\bar{\zeta}\hat{u}^{-1}\zeta)_{a}{}^{b}
\bar{\Theta}_{b})\,,
\end{array}\label{4Rs1}\\
{~}\nonumber\\
{~}\nonumber\\
\begin{array}{c}
H^{\rho}{}_{r\sigma s\tau  t}(\Z)
=\lambda^{\eta_{2}+\eta_{3}-\eta_{1}}{}
H^{\rho}{}_{r\sigma s\tau
t}(\Z^{\prime\prime\prime})\,,\\
{}\\
\Z^{\prime\prime\prime M}=( \lambda
X^{\mu},\,
\lambda^{\frac{1}{2}}\Theta^{a},\,\lambda^{\frac{1}{2}}\bar{\Theta}_{a})\,,
\end{array}~~~~~~~~~~\\
{~}\nonumber\\
{~}\nonumber\\
\begin{array}{c}
H^{\rho}{}_{r\sigma s\tau  t}(\Z)
=e^{-i(\kappa_{1}+\kappa_{2}+\kappa_{3})\Omega}
H^{\rho}{}_{r\sigma s\tau
t}(\Z^{\prime\prime\prime\prime})\,,\\
{}\\
\Z^{\prime\prime\prime\prime M}=(X^{\mu},\,
e^{i\frac{1}{2}\Omega}
\Theta^{a},\,e^{-i\frac{1}{2}\Omega}\bar{\Theta}_{a})\,.
\end{array}~~~~~~~\label{4Rs2}
\end{subeqnarray}
Note that 
$\hat{L}\in\mbox{SL}(2,\Com),
\,\hat{u}\in\mbox{SU}({\cal N}),\,\lambda\in\Real,\,\Omega\in S^{1}$
and $\hat{R}_{\nu}{}^{\mu}(\hat{L})$ is given from eq.(\ref{4green1}) by
\begin{equation}
\hat{R}_{\nu}{}^{\mu}(\hat{L})=\textstyle{\frac{1}{2}}
\mbox{tr}(\tilde{\sigma}_{\nu}\hat{L}\sigma^{\mu}\hat{\bar{L}})\,.
\end{equation}
\indent In general there are  a finite number of linearly independent  
solutions of eq.(\ref{4equiv}), and this number may  be    
reduced by imposing  extra restrictions on the correlation function.  
As an example, we consider the three-point correlation function of a 
real vector superfield, $V^{\mu}(z)$,  where $\kappa=0$. 
From eq.(\ref{43gen}) we may write
\begin{equation}
\displaystyle{
\langle V^{\mu}(z_{1})V^{\nu}(z_{2})V^{\lambda}(z_{3})\rangle = 
\frac{H^{\mu\nu^{\prime}\lambda^{\prime}}
(\Z_{1})I_{\nu^{\prime}}{}^{\nu}(z_{12})
I_{\lambda^{\prime}}{}^{\lambda}(z_{13})}
{(x^{2}_{\bar{1}2}x^{2}_{\bar{2}1}
x^{2}_{\bar{3}1}x^{2}_{\bar{1}3})^{\frac{1}{2}\eta}}}\,.
\label{43V}
\end{equation}
Since eq.(\ref{4equiv}) is obtained by 
considering invariance  under continuous superconformal transformations, 
invariance   under superinversion  which is a discrete map   
may give an  extra restriction.  
Besides the superconformal invariance, the three-point function has
additional symmetry under permutations of the superfields.   
Furthermore, for supercurrents 
we may require the correlation function to satisfy the conservation equations 
like eqs.(\ref{4cons1}\,,\ref{4cons2}).  More explicitly, 
under superinversion we may require    $V^{\mu}(z)$ transforms to
\begin{equation}
V^{\prime\mu}(z^{\prime})=-V^{\nu}(z)\hat{R}_{\nu}{}^{\mu}(z;i_{s})
\Omega(z;i_{s})^{-\eta}\,.
\end{equation}
The occurrence of the minus sign in ${\cal N}=1$ Wess-Zumino model and 
vector superfield theory  was verified in \cite{paper1}.   
Invariance under superinversion, 
$z_{j}\stackrel{i_{s}}{\longrightarrow}z^{\prime}_{j},\,j=1,2,3$,\,    
implies using   eqs.(\ref{4Oi},\,\ref{4xxx},\,\ref{4xxRR})
\begin{equation}
H^{\mu^{\prime}\nu^{\prime}\lambda^{\prime}}(\Z_{1})
\hat{R}_{\mu^{\prime}}{}^{\mu}(z_{1};i_{s})
\hat{R}_{\nu^{\prime}}{}^{\nu}(z_{1};i_{s})
\hat{R}_{\lambda^{\prime}}{}^{\lambda}(z_{1};i_{s})
=-\Omega^{-\eta}(z_{1};i_{s})H^{\mu\nu\lambda}(\Z_{1}^{\prime})\,,
\label{4Invi}
\end{equation}
which also implies using eqs.(\ref{4superinvZ1},\,\ref{4Z2Z3}) 
\begin{equation}
\Omega(z_{12};i_{s})^{\eta}H^{\mu^{\prime}\nu^{\prime}\lambda^{\prime}}(\Z_{1})
\hat{R}_{\mu^{\prime}}{}^{\mu}(z_{12};i_{s})
\hat{R}_{\nu^{\prime}}{}^{\nu}(z_{12};i_{s})
\hat{R}_{\lambda^{\prime}}{}^{\lambda}(z_{12};i_{s})
=-H^{\mu\nu\lambda}(-(\Z_{2})^{\prime})\,,
\label{4Invi2}
\end{equation}
where $(\Z_{2})^{\prime}$  is given by superinversion, 
$\Z_{2}\stackrel{i_{s}}{\longrightarrow}(\Z_{2})^{\prime}$.\newline
From $\langle V^{\mu}(z_{1})V^{\nu}(z_{2})V^{\lambda}(z_{3})\rangle
=\langle V^{\mu}(z_{1})V^{\lambda}(z_{3})V^{\nu}(z_{2})\rangle$ we
have
\begin{equation}
H^{\mu\nu\lambda}(\Z)=H^{\mu\lambda\nu}(-\Z)\,,
\label{423}
\end{equation}
and from $\langle V^{\mu}(z_{1})V^{\nu}(z_{2})V^{\lambda}(z_{3})\rangle
=\langle V^{\nu}(z_{2})V^{\lambda}(z_{3})V^{\mu}(z_{1})\rangle$ we
have using eqs.(\ref{41311223},\,\ref{4Invi2}) with
$\Z\stackrel{i_{s}}{\longrightarrow}(\Z)^{\prime}$
\begin{equation}
H^{\mu\nu\lambda}(-(\Z)^{\prime})
=-\Omega(\Z;i_{s})^{-\eta}
H^{\nu\lambda^{\prime}\mu}(\Z)
\hat{R}_{\lambda^{\prime}}{}^{\lambda}(\Z;i_{s})\,.
\label{4123}
\end{equation}
Imposing these extra conditions it was shown that the 
three-point correlation 
functions  of supercurrents in ${\cal N}=1$ theories 
have two linearly independent forms \cite{9808041}. 
Similarly, the three-point functions  of real scalar superfields  
in ${\cal N}=1$ theories have also  
two linearly independent solutions \cite{paper1}.\newline
\indent Invariance under $R$-symmetry transformations~(\ref{4Rs1},d) 
implies that $H^{\mu\nu\lambda}(\Z)$ is a function of $X^{\mu},
\Theta^{a\alpha}\bar{\Theta}{}^{\da}_{a}$ or equivalently       
$X^{\mu},\,\Theta^{a}\sigma^{\mu}\bar{\Theta}_{a}$,  as demonstrated in 
subsection \ref{subn2},  and hence we may put
\begin{equation}
H^{\mu\nu\lambda}(\Z)=H^{\mu\nu\lambda}(
X^{\lambda},\Theta^{a}\sigma^{\lambda}\bar{\Theta}_{a})\,.
\end{equation}
\subsection{$n$-point Correlation Functions - in general\label{subn}}
In this subsection we show that the solution for  
$n$-point  correlation functions  of the quasi-primary superfields,~$
\Psi^{\rho}{}_{r}$,  has the general  form  
\begin{equation}
\begin{array}{l}
\langle \Psi^{\rho_{1}}_{1}{}_{r_{1}}(z_{1})\cdots\Psi^{\rho_{n}}_{n}{}_{r_{n}}(z_{n})\rangle\\
{}\nonumber\\
= H^{\rho_{1}}{}_{r_{1}\rho_{2}^{\prime}r_{2}^{\prime}\cdots
\rho_{n}^{\prime}r_{n}^{\prime}}(\Z_{1(1)},\cdots,\Z_{1(n-2)})
\displaystyle{\prod_{k=2}^{n}
\frac{I^{\rho^{\prime}_{k}\rho_{k}}(\hat{\tx}_{\bar{1}k})
I^{r^{\prime}_{k}}{}_{r_{k}}(
\hat{v}(z_{1k}))}{
(x^{2}_{\bar{1}k})^{\frac{1}{2}(\eta_{k}-(\frac{4}{{\cal N}}-1)\kappa_{k})}
(x^{2}_{\bar{k}1})^{\frac{1}{2}(\eta_{k}+(\frac{4}{{\cal N}}-1)\kappa_{k})}}
}\,,
\end{array}
\label{4ngen}
\end{equation}
where in a similar fashion to eq.(\ref{F3defnew}) 
$\Z_{1(1)},\cdots,\Z_{1(n-2)}$ are  given,  with   
$z_{k1}\stackrel{i_{s}}{\longrightarrow}\widetilde{z_{k1}},\,k\geq 2$,   by 
\begin{equation}
\begin{array}{cc}
G_{T}(\widetilde{z_{n1}})^{-1}{}G_{T}(\widetilde{z_{j1}})=G_{T}(\Z_{1(j-1)})\,,
~~~~~&~~~~
j=2,3,\cdots,n-1\,.
\end{array}
\end{equation}
We note that all of them are `tangent' vectors at $z_{1}$.\newline
Superconformal invariance~(\ref{4Green}) is  
equivalent to
\begin{subeqnarray}
\label{4equivgen}
&\begin{array}{c}
H^{\rho^{\prime}_{1}}{}_{r_{1}\rho_{2}r_{2}\cdots\rho_{n}r_{n}}(\Z_{(1)},\cdots,\Z_{(n-2)})D_{\rho^{\prime}_{1}}{}^{\rho_{1}}(\hat{L})
=H^{\rho_{1}}{}_{r_{1}\rho^{\prime}_{2}r_{2}\cdots\rho^{\prime}_{n}r_{n}}
(\Z^{\prime}_{(1)},\cdots,\Z^{\prime}_{(n-2)})\displaystyle{\prod_{k=2}^{n}
\bar{D}^{\rho^{\prime}_{k}}{}_{\rho_{k}}(\hat{\bar{L}})}\,,\\
{}\\
\Z^{\prime M}_{(j)}=(X_{(j)}^{\nu}\hat{R}_{\nu}{}^{\mu}(\hat{L}),\,
\Theta_{(j)}^{a}\hat{L},\,\hat{\bar{L}}\bar{\Theta}_{(j)a})\,,
\end{array}&\label{4Rs1ge}\\
{}\nonumber\\
{}\nonumber\\
&\begin{array}{c}
H^{\rho_{1}}{}_{r^{\prime}_{1}\rho_{2}r^{\prime}_{2}\cdots\rho_{n}
r^{\prime}_{n}}(\Z_{(1)},\cdots,\Z_{(n-2)})\displaystyle{\prod_{k=1}^{n}
D^{r^{\prime}_{k}}{}_{r_{k}}(\hat{u})}=
H^{\rho_{1}}{}_{r_{1}\rho_{2}r_{2}\cdots\rho_{n}r_{n}}(\Z^{\prime\prime}_{(1)},\cdots,\Z^{\prime\prime}_{(n-2)})\,,\\
{}\\
\Z_{(j)}^{\prime\prime M}=( X^{\mu}_{(j)},\,
\Theta_{(j)}^{b}(\bar{\zeta}\hat{u}\zeta)_{b}{}^{a},\,
(\bar{\zeta}\hat{u}^{-1}\zeta)_{a}{}^{b}\bar{\Theta}_{(j)b})\,,
\end{array}&\label{4Rs1gen}\\
{}\nonumber\\
{}\nonumber\\
&\begin{array}{c}
H^{\rho_{1}}{}_{r_{1}\rho_{2}r_{2}\cdots\rho_{n}r_{n}}
(\Z_{(1)},\cdots,\Z_{(n-2)})
=\lambda^{-\eta_{1}+\eta_{2}+\cdots+\eta_{n}}{}
H^{\rho_{1}}{}_{r_{1}\rho_{2}r_{2}\cdots\rho_{n}r_{n}}(
\Z^{\prime\prime\prime}_{(1)},\cdots,\Z^{\prime\prime\prime}_{(n-2)})\,,\\
{}\\
\Z_{(j)}^{\prime\prime\prime M}=( \lambda
X^{\mu}_{(j)},\,
\lambda^{\frac{1}{2}}\Theta^{a}_{(j)},\,\lambda^{\frac{1}{2}}
\bar{\Theta}_{(j)a})\,,
\end{array}&\label{4Rs2dil}\\
{}\nonumber\\
{}\nonumber\\
&\begin{array}{c}
H^{\rho_{1}}{}_{r_{1}\rho_{2}r_{2}\cdots\rho_{n}r_{n}}
(\Z_{(1)},\cdots,\Z_{(n-2)})
=e^{-i(\kappa_{1}+\cdots+\kappa_{n})\Omega}
H^{\rho_{1}}{}_{r_{1}\rho_{2}r_{2}\cdots\rho_{n}r_{n}}(
\Z^{\prime\prime\prime\prime}_{(1)},\cdots,
\Z^{\prime\prime\prime\prime}_{(n-2)})\,,\\
{}\\
\Z_{(j)}^{\prime\prime\prime\prime M}=(X_{(j)}^{\mu},\,
e^{i\frac{1}{2}\Omega}
\Theta^{a}_{(j)},\,
e^{-i\frac{1}{2}\Omega}\bar{\Theta}_{(j)a})\,.
\end{array}&\label{4Rs2gen}
\end{subeqnarray}
Thus, in general $n$-point 
functions  reduce to one  unspecified $(n-2)$-point  function 
which must transform  homogeneously  under the 
rigid transformations,  \newline
$\mbox{SL}(2,\Com)\times\mbox{SU}({\cal N})
\times\mbox{U}(1)\times\mbox{D}$.\newline
\newline
\indent\textit{Proof}\footnote{The key idea in this proof  
first appeared in \cite{paper1}.}\newline
Without loss of generality, using the supertranslational invariance   
we can put the $n$-point  function  with 
$z_{k1}\stackrel{i_{s}}{\longrightarrow}\widetilde{z_{k1}},\,k\geq 2$ as
\begin{equation}
\begin{array}{l}
\langle \Psi^{\rho_{1}}_{1}{}_{r_{1}}(z_{1})\cdots\Psi^{\rho_{n}}_{n}{}_{r_{n}}(z_{n})\rangle\\
{}\\
={\cal H}^{\rho_{1}}{}_{r_{1}\rho_{2}^{\prime}r_{2}^{\prime}\cdots
\rho_{n}^{\prime}r_{n}^{\prime}}(\widetilde{z_{21}},\widetilde{z_{31}},\cdots,\widetilde{z_{n1}})
\displaystyle{\prod_{k=2}^{n}
\frac{I^{\rho^{\prime}_{k}\rho_{k}}(\hat{\tx}_{\bar{1}k})
I^{r^{\prime}_{k}}{}_{r_{k}}(
\hat{v}(z_{1k}))}{
(x^{2}_{\bar{1}k})^{\frac{1}{2}(\eta_{k}-(\frac{4}{{\cal N}}-1)\kappa_{k})}
(x^{2}_{\bar{k}1})^{\frac{1}{2}(\eta_{k}+(\frac{4}{{\cal N}}-1)
\kappa_{k})}}}\,.
\end{array}
\end{equation}
The superconformal invariance of the 
correlation function~(\ref{4Green}), using
eqs.(\ref{4green4},\,\ref{4expX},\,\ref{4IIpro}), implies
\begin{equation}
\begin{array}{l}
{\cal H}^{\rho_{1}}{}_{r_{1}\rho_{2}^{\prime}r_{2}\cdots
\rho_{n}^{\prime}r_{n}}(\widetilde{z^{\prime}_{21}},
\widetilde{z^{\prime}_{31}},\cdots,\widetilde{z^{\prime}_{n1}})
\displaystyle{\prod_{k=2}^{n}
\bar{D}^{\rho^{\prime}_{k}}{}_{\rho_{k}}(\hat{\bar{L}}(z_{1-};g))}\\
{}\\
=\Omega(z_{1};g)^{-\eta_{1}+\eta_{2}+\cdots+\eta_{n}}
\Upsilon(z_{1};g)^{-(\kappa_{1}+\kappa_{2}+\cdots+\kappa_{n})}
{\cal H}^{\rho^{\prime}_{1}}{}_{r^{\prime}_{1}\rho_{2}r^{\prime}_{2}\cdots
\rho_{n}r^{\prime}_{n}}(\widetilde{z_{21}},
\widetilde{z_{31}},\cdots,\widetilde{z_{n1}})\\
{}\\
~~~~~\times D_{\rho_{1}^{\prime}}{}^{\rho_{1}}(\hat{L}(z_{1+};g))
\displaystyle{\prod_{k=1}^{n}
D^{r^{\prime}_{k}}{}_{r_{k}}(\hat{u}(z_{1};g))}\,.
\end{array}
\label{dGGgen}
\end{equation}
Now  we consider a superconformal transformation, 
$z\stackrel{g_{u}}{\longrightarrow}z^{\prime\prime}$, defined by
\begin{equation}
\begin{array}{c}
G_{T}(z^{\prime\prime})=G_{T}(z^{\prime\prime}_{1})G_{T}(z_{s}(z^{\prime};u))\,,\\
{}\\
G_{T}(z^{\prime})=G_{T}(z_{1})^{-1}G_{T}(z)\,,
\end{array}
\label{specialspecial}
\end{equation}
where $z_{s}(z^{\prime};u)$ is a special superconformal 
transformation given in eq.(\ref{specialz}) and 
$z^{\prime\prime}_{1}$ can be arbitrary.   Since
\begin{equation}
\begin{array}{ll}
\multicolumn{2}{c}{G_{T}(\widetilde{z^{\prime\prime}_{k1}})
=G_{T}(u)G_{T}(\widetilde{z_{k1}})~~~~~~
\mbox{~for}~k\geq 2\,,}\\
{}&{}\\
\hat{L}(z_{1+};g_{u})=\hat{\bar{L}}(z_{1-};g_{u})=1\,,~~~~&~~~~
\Omega(z_{1};g_{u})=\Upsilon(z_{1};g_{u})=1\,,
\end{array}
\end{equation}
 ${\cal H}^{\rho_{1}}{}_{r_{1}\rho_{2}r_{2}\cdots
\rho_{n}r_{n}}(\widetilde{z_{21}},\widetilde{z_{31}},\cdots,\widetilde{z_{n1}})$    possesses a supertranslational invariance 
\begin{equation}
{\cal H}^{\rho_{1}}{}_{r_{1}\rho_{2}r_{2}\cdots
\rho_{n}r_{n}}(\widetilde{z_{21}},\widetilde{z_{31}},\cdots,\widetilde{z_{n1}})={\cal H}^{\rho_{1}}{}_{r_{1}\rho_{2}r_{2}\cdots
\rho_{n}r_{n}}(\widetilde{z^{\prime\prime}_{21}},\widetilde{z^{\prime\prime}_{31}},\cdots,\widetilde{z^{\prime\prime}_{n1}})\,.
\end{equation}
Thus we can write
\begin{equation}
\begin{array}{c}
{\cal H}^{\rho_{1}}{}_{r_{1}\rho_{2}r_{2}\cdots
\rho_{n}r_{n}}(\widetilde{z_{21}},\widetilde{z_{31}},\cdots,\widetilde{z_{n1}})=H^{\rho_{1}}{}_{r_{1}\rho_{2}r_{2}\cdots
\rho_{n}r_{n}}(\Z_{1(1)},\cdots,\Z_{1(n-2)})\,,\\
{}\\
G_{T}(\Z_{1(j-1)})=G_{T}(\widetilde{z_{n1}})^{-1}G_{T}(\widetilde{z_{j1}})\,.
\end{array}
\label{dGtrinv}
\end{equation}
With the transformation rule for $\Z_{1}$~(\ref{4expX}), eq.(\ref{dGtrinv}) completes our proof.~~~~~~~~\textit{Q.E.D.}\newline
\newline
\indent  We note that, in the case of $n=2$, 
$H^{\rho_{1}}{}_{r_{1}\rho_{2}r_{2}}$ is independent of $z_{1},z_{2}$ 
and eqs.(\ref{4Rs2dil},d) show that two-point functions vanish 
if $\eta_{1}\neq\eta_{2}$  or  $\kappa_{1}\neq -\kappa_{2}$. Furthermore, 
if the representation is irreducible then $H=1$ by Schur's Lemma.

\subsection{Selection Rule \& Superconformal Invariants\label{subn2}}
We begin with   fields, $\psi^{I}(x)$, depending on $x\in\Real^{4}$     
which are obtained by letting  the Grassmann coordinates inside  
quasi-primary superfields, 
$\Psi^{I}(x,\theta^{a},\bar{\theta}_{a})$, be zero
\begin{equation}
\psi^{I}(x)\equiv \Psi^{I}(x,0,0)\,.
\end{equation}
They are the lowest order term appearing 
in the power series expansions of superfields 
in Grassmann coordinates.  
The superconformal invariance under $\mbox{U}(1)$ 
transformations~(\ref{4Rs2gen}) implies for arbitrary $\Omega\in S^{1}$
\begin{equation}
\langle \psi^{I_{1}}_{1}(x_{1})\cdots\psi^{I_{n}}_{n}(x_{n})\rangle
=e^{i(\kappa_{1}+\cdots+\kappa_{n})\Omega}
\langle \psi^{I_{1}}_{1}(x_{1})\cdots\psi^{I_{n}}_{n}(x_{n})\rangle\,,
\end{equation}
hence,  
if the the sum of the $R$-symmetry charge, $\kappa_{i}$,  is  not zero then 
the correlation function   must vanish 
as exhibited in eq.(\ref{selection})
\begin{equation}
\begin{array}{cc}
\langle \psi^{I_{1}}_{1}(x_{1})\cdots\psi^{I_{n}}_{n}(x_{n})\rangle=0
~~~~&~~~~\mbox{if~~~}
\displaystyle{\sum_{i=1}^{n}\kappa_{i}\neq 0}\,.
\end{array}
\label{selection2}
\end{equation}
This selection rule can be generalized further to all the  
other  component fields in the power series expansions of superfields 
\begin{equation}
\Psi^{I}(x,\theta^{a},\bar{\theta}_{a})
=\psi^{I}(x)+\psi^{I}_{a\alpha}(x)\theta^{a\alpha}+
\bar{\theta}^{\da}_{a}\bar{\psi}^{Ia}_{\da}(x)+\,\cdots\,.
\end{equation}
If we define the $R$-symmetry charge of the component fields, 
$\psi^{I}(x)$, $\psi^{I}_{a\alpha}(x)$, $\bar{\psi}^{Ia}_{\da}(x)$, etc. as 
$\kappa$, $\kappa+\frac{1}{2}$, $\kappa-\frac{1}{2}$, etc.  respectively, 
then  the invariance under $\mbox{U}(1)$ transformation, 
\begin{equation}
\langle\Psi^{I_{1}}_{1}(x_{1},\theta^{a}_{1},\bar{\theta}_{1a})\,\cdots\,
\rangle=e^{i(\kappa_{1}+\cdots+\kappa_{n})\Omega}
\langle\Psi^{I_{1}}_{1}(x_{1},e^{i\frac{1}{2}\Omega}\theta^{a}_{1},
e^{-i\frac{1}{2}\Omega}\bar{\theta}_{1a})\,\cdots\,\rangle\,,
\end{equation}
implies that the selection rule~(\ref{selection2}) holds for 
all the component fields. \newline 
\indent The existence of this kind of selection rule  
in ${\cal N}=4$ super Yang-Mills   
theory was previously predicted  by  Intriligator 
within  the context of $AdS$/CFT correspondence, as the dual IIB supergravity contains a corresponding $\mbox{U}(1)$ symmetry~\cite{9811047}. 
Therefore our results provide a supporting evidence for the Maldacena 
conjecture, as the selection rule here is derived 
by purely considering the symmetry on CFT side 
without referring to the string side.  \newline
\indent Essentially, for ${\cal N}\neq 4$ case, the selection rule exists     
since the  four-dimensional ${\cal N}\neq 4$ superconformal group 
includes the $\mbox{U}(1)$ factor inevitably.  However in ${\cal N}=4$ case, 
as verified in subsection \ref{SCA},   the   corresponding 
superconformal group is isomorphic to a 
semi-direct product of $\mbox{U}(1)$ and a simple Lie supergroup so that  
it can be reduced to the simple Lie subgroup by breaking the  
$\mbox{U}(1)$ symmetry.   
In this case, the selection rule will not be 
applicable to the corresponding ${\cal N}=4$ superconformal theory, since the 
$\mbox{U}(1)$ representation becomes trivial, $\Upsilon(z;g)=1$,  and 
the $R$-symmetry charge is not defined.\newline
\newline
\indent Now,  we consider 
correlation functions of quasi-primary scalar superfields, $\Psi(z)$.   
From eqs.(\ref{4Rs1ge},b)    
$H(\Z_{1(1)},\cdots,\Z_{1(n-2)})$ must be 
$\mbox{SL}(2,\Com)\times\mbox{SU}({\cal N})$      
invariant  and hence it is a function of  
$\mbox{SL}(2,\Com)\times\mbox{SU}({\cal N})$   invariants. 
According to \cite{weyl}, these   invariants 
can be obtained by  
contracting in all possible ways the spinorial indices of 
\begin{equation}
\begin{array}{cccc}
\epsilon_{\alpha\beta}\,,~~~~\epsilon^{-1\alpha\beta}\,,~~~&~~
\bar{\epsilon}_{\da\db}\,,~~~~\bar{\epsilon}^{-1\da\db}\,,~~~&~~
({\rm X}_{1(i)})_{\alpha\da}\,,~~~&~~
\Theta^{a\alpha}_{1(i)}\bar{\Theta}^{\da}_{1(j)a}\,,\\
{}&{}&{}&{}\\
\multicolumn{4}{c}{\epsilon_{a_{1}\cdots a_{{\cal N}}}
\Theta_{1(i_{1})}^{a_{1}\alpha_{1}}\cdots
\Theta_{1(i_{{\cal N}})}^{a_{{\cal N}}\alpha_{{\cal N}}}\,,~~~~~~~~~~~~~
\epsilon^{a_{1}\cdots a_{{\cal N}}}
\bar{\Theta}_{1(i_{1})a_{1}}^{\da_{1}}\cdots
\bar{\Theta}_{1(i_{{\cal N}})a_{{\cal N}}}^{\da_{{\cal N}}}\,.}
\end{array}
\end{equation}
On the other hand if we write
\begin{equation}
\begin{array}{cc}
{\cal X}^{\mu}_{1(I)}=(\hat{X}^{\mu}_{1(i)},\,
\hat{\Theta}{}^{a}_{1(j)}\sigma^{\mu}\hat{\bar{\Theta}}{}_{1(k)a})\,,~~~~&~~~~
1\leq I\leq (n-1)(n-2)\,,
\end{array}
\end{equation}
where $\hat{\Z}_{1(j)}^{M}=(\hat{X}^{\mu}_{1(j)},
\hat{\Theta}{}^{a}_{1(j)},\hat{\bar{\Theta}}_{1(j)a})$ are normalized 
$\Z^{M}_{1(j)}$
\begin{equation}
\begin{array}{ccc}
\hat{X}^{\mu}_{1(j)}=\displaystyle{\frac{X^{\mu}_{1(j)}}{(X_{1(1)+}^{2}X_{1(1)-}^{2})^{\frac{1}{4}}}}\,,~~&~~~
\hat{\Theta}{}^{a}_{1(j)}=
\displaystyle{\frac{\Theta^{a}_{1(j)}}{(X_{1(1)+}^{2}X_{1(1)-}^{2})^{\frac{1}{8}}}}\,,~~&~~~
\hat{\bar{\Theta}}_{1(j)a}=\displaystyle{\frac{\bar{\Theta}_{1(j)a}}{(X_{1(1)+}^{2}X_{1(1)-}^{2})^{\frac{1}{8}}}}\,,
\end{array}
\end{equation}
then using eq.(\ref{N=1sigma2}) one can show that   
${{\cal X}}_{1(I)}{\cdot{{\cal X}}_{1(J)}}$    
are all the invariants for 
$\mbox{SL}(2,\Com)\times\mbox{SU}({\cal N})
\times\mbox{U}(1)\times\mbox{D}$ and hence invariants for 
the whole ${\cal N}$-extended 
superconformal group. Note that    
from eq.(\ref{4-sign}),  some of them are pseudo invariants 
under superinversion. \newline
Explicitly, we may reproduce  
the invariants depending on three points~(\ref{Park}) as
\begin{equation}
\begin{array}{cc}
\displaystyle{
(\hat{X}_{1(1)+}^{2})^{2}=\frac{X_{1(1)+}^{2}}{X_{1(1)-}^{2}}}
\,,~~~~&~~~~
\displaystyle{\hat{X}_{1(1)+}{\cdot\hat{X}_{1(1)-}}=
\frac{X_{1(1)+}{\cdot X_{1(1)-}}}{\sqrt{X_{1(1)+}^{2}
X_{1(1)-}^{2}}}}\,,
\end{array}
\end{equation}
and using
\begin{equation}
\displaystyle{
X_{1(l-1)+}^{2}=\frac{x_{\bar{l}n}^{2}}{x_{\bar{l}1}^{2}
x_{\bar{1}n}^{2}}}\,,
\end{equation}
\newline
\begin{equation}
\begin{array}{c}
\displaystyle{X_{(l,m)+}^{2}=\frac{x_{\bar{l}m}^{2}}{
x_{\bar{l}1}^{2}x_{\bar{1}m}^{2}}}\,,\\
{}\\
X_{(l,m)+}^{\mu}=X^{\mu}_{1(l-1)+}-X^{\mu}_{1(m-1)-}+
2i\Theta^{a}_{1(l-1)}\sigma^{\mu}\bar{\Theta}_{1(m-1)a}\,,
\end{array}
\label{Xml+}
\end{equation}
we may  also obtain cross ratio type invariants 
depending on four points,  $z_{r},z_{s},z_{t},z_{u}$ 
\begin{equation}
\displaystyle{\frac{x^{2}_{\bar{r}s}x^{2}_{\bar{t}u}}{
x^{2}_{\bar{r}u}x^{2}_{\bar{t}s}}}\,.
\label{crossratio}
\end{equation}
\indent 
From $\hat{X}_{1(1)+}^{2}\hat{X}_{1(1)-}^{2}=1$, 
the number of different ${\cal X}_{1(I)}{\cdot{\cal X}_{1(J)}},\,\,
1\leq I,J\leq (n-1)(n-2)$ is  
$\frac{1}{2}(n^{2}-3n+4)(n^{2}-3n+1)$. However,  
a vector, $\alpha$, in $d$-dimensions may be specified by $d$ 
equations, $\alpha{\cdot\beta_{i}}=c_{i};1\leq i\leq d$. 
Thus the number of independent superconformal invariants  or 
${\cal X}_{1(I)}{\cdot{\cal X}_{1(J)}}$ in four-dimensions, $\#_{n}$, is  
\begin{equation}
\#_{n}=\left\{\begin{array}{ll}
2~~~~&\mbox{for}~n=3\,,\\
{}&{}\\
4n^{2}-12n+1~~~~&\mbox{for}~n\geq 4\,.
\end{array}\right.
\end{equation}
This result holds for ${\cal N}\geq 2$, and for ${\cal N}=1$ 
due to the identity~(\ref{N=1sigma})  the number of independent 
${\cal X}_{1(I)}{\cdot{\cal X}_{1(J)}}$  reduces further to\footnote{Similar analysis for ${\cal N}=1$ was  done by Osborn~\cite{9808041}. } 
\begin{equation}
\#_{n}=\left\{\begin{array}{ll}
1~~~~&\mbox{for}~n=3\,,\\
{}&{}\\
4n^{2}-12n~~~~~~~&\mbox{for}~n\geq 4\,.
\end{array}\right.
\end{equation}
\indent  In the case of $\sum_{i}\kappa_{i}=0$,   
$H(\hat{\Z}_{1(1)},\cdots,\hat{\Z}_{1(n-2)})$ must be    
$\mbox{SL}(2,\Com)\times\mbox{SU}({\cal N})\times
\mbox{U}(1)\times\mbox{D}$ invariant. Hence it is a function of 
${{\cal X}}_{1(I)}{\cdot{{\cal X}}_{1(J)}}$ and  
the $n$-point correlation function of scalar superfields 
reduces to an arbitrary $\#_{n}$ variable function.   
The  $n$-point function in this case  
may have the following general form 
\begin{equation}
\langle \Psi_{1}(z_{1})\cdots\Psi_{n}(z_{n})\rangle=
\displaystyle{
\frac{F({\cal X}_{1(I)}{\cdot{\cal X}_{1(J)}})}{
\displaystyle{
\prod_{l\neq m}
(x_{\bar{l}m}^{2})^{\Delta_{lm}}}}}\,,
\label{nscalar}
\end{equation}
where
\begin{equation}
\Delta_{lm}=-\textstyle{\frac{1}{2(n-1)(n-2)}}\displaystyle{
\sum_{i=1}^{n}\eta_{i}
+\textstyle{\frac{1}{2(n-2)}}(\eta_{l}+\eta_{m})+
\textstyle{\frac{1}{2n}}(\textstyle{\frac{4}{{\cal N}}}-1)
(\kappa_{l}-\kappa_{m})}\,.
\end{equation}
$F({\cal X}_{1(I)}{\cdot{\cal X}_{1(J)}})$ is related to  
$H(\Z_{1(1)},\cdots,\Z_{1(n-2)})$ by 
\begin{equation}
F({\cal X}_{1(I)}{\cdot{\cal X}_{1(J)}})
=H(\hat{\Z}_{1(1)},\cdots,\hat{\Z}_{1(n-2)})
\displaystyle{\prod_{2\leq l\neq m}
(\hat{X}_{1(l,m)+}^{2})^{\Delta_{lm}}}\,.
\end{equation}
This relation can be derived  using  eq.(\ref{4Rs2dil})  and 
the following identity which  
holds when  $(\frac{4}{{\cal N}}-1)\,\sum_{i=1}^{n}{}\kappa_{i}=0$ 
\begin{equation}
\begin{array}{c}
\displaystyle{\left(\prod_{l\neq m}
(x_{\bar{l}m}^{2})^{\Delta_{lm}}\right)
\left(\prod_{k=2}^{n}
(x^{2}_{\bar{1}k})^{-\frac{1}{2}(\eta_{k}-(\frac{4}{{\cal N}}-1)\kappa_{k})}
(x^{2}_{\bar{k}1})^{-\frac{1}{2}(\eta_{k}+(\frac{4}{{\cal N}}-1)\kappa_{k})}
\right)}\\
{}\\
=\lambda^{-\frac{1}{2}(-\eta_{1}+\eta_{2}+\cdots+\eta_{n})}
\displaystyle{\prod_{2\leq l\neq m}
\left(\frac{\lambda x_{\bar{l}m}^{2}}{x_{\bar{l}1}^{2}x_{\bar{1}m}^{2}}
\right)^{\Delta_{lm}}}\,,
\end{array}
\label{identityPPP}
\end{equation}
where $\lambda\in\Real$.

\subsection{Non-supersymmetric Case\label{nonsusy}} 
In particular, here, we consider the 
non-supersymmetric case, i.e. ${\cal N}=0$. 
Quasi-primary fields, 
$\Psi^{\rho}(x)$, transform under the conformal transformations, 
$x\stackrel{g}{\longrightarrow} x^{\prime}$, as
\begin{equation}
\Psi^{\rho}(x)~~~\longrightarrow~~~
\Psi^{\prime \rho}(x^{\prime})=\Psi^{\sigma}(x)
D_{\sigma}{}^{\rho}(\hat{L}(x;g))\Omega(x;g)^{-\eta}\,.
\label{dfieldtr}
\end{equation}
$n$-point function has the general form
\begin{equation}
\langle \Psi^{\rho_{1}}_{1}(x_{1})\cdots\Psi^{\rho_{n}}_{n}(x_{n})\rangle
=\displaystyle{\frac{
H^{\rho_{1}}{}_{\rho_{2}^{\prime}\cdots\rho_{n}^{\prime}}
(X_{1(1)},\cdots,X_{1(n-2)})
I^{\rho^{\prime}_{2}\rho_{2}}(\hat{\tx}_{12})\cdots 
I^{\rho^{\prime}_{n}\rho_{n}}(\hat{\tx}_{1n})}{
(x^{2}_{12})^{\eta_{2}}\cdots (x^{2}_{1n})^{\eta_{n}}}}\,,
\end{equation}
where  $H^{\rho_{1}}{}_{\rho_{2}\cdots \rho_{n}}(X_{1(1)},\cdots,X_{1(n-2)})$ 
is a function depending on  $n-2$ points, \newline
$X_{1(j)},\,\,1\leq j\leq n-2$,
\begin{equation}
\begin{array}{cc}
(X_{1(j)})^{\mu}=\displaystyle{\frac{x_{n1}^{\mu}}{x_{n1}^{2}}-
\frac{x_{(j+1)1}^{\mu}}{x_{(j+1)1}^{2}}}\,,~~~~&~~~~
X_{1(j)}^{2}=\displaystyle{\frac{x_{(j+1)n}^{2}}{x_{n1}^{2}x_{(j+1)1}^{2}}}\,.
\end{array}
\end{equation}
Under conformal transformations, 
$X_{1(j)}$   transforms homogeneously at $x_{1}$ 
as in eq.(\ref{XXR}).\newline 
Conformal invariance is  equivalent to   
\begin{subeqnarray}
\label{non-susy}
\begin{array}{c}
H^{\rho^{\prime}_{1}}{}_{\rho_{2}\cdots\rho_{n}}(X_{(1)},\cdots,X_{(n-2)})
D_{\rho^{\prime}_{1}}{}^{\rho_{1}}(\hat{L})
=H^{\rho_{1}}{}_{\rho^{\prime}_{2}\cdots\rho^{\prime}_{n}}
(X^{\prime}_{(1)},\cdots,X^{\prime}_{(n-2)})
\bar{D}^{\rho^{\prime}_{2}}{}_{\rho_{2}}(\hat{\bar{L}})\cdots
\bar{D}^{\rho^{\prime}_{n}}{}_{\rho_{n}}(\hat{\bar{L}})\,,\\
{}\\
X^{\prime \mu}_{(j)}=X_{(j)}^{\nu}\hat{R}_{\nu}{}^{\mu}(\hat{L})\,,
\end{array}\\
{~}\nonumber\\
{~}\nonumber\\
\begin{array}{c}
H^{\rho_{1}}{}_{\rho_{2}\cdots\rho_{n}}
(X_{(1)},\cdots,X_{(n-2)})
=\lambda^{-\eta_{1}+\eta_{2}+\cdots+\eta_{n}}{}
H^{\rho_{1}}{}_{\rho_{2}\cdots\rho_{n}}(X^{\prime\prime}_{(1)},
\cdots,X^{\prime\prime}_{(n-2)})\,,~~~~~~~~~~~~~\\
{}\\
X_{(j)}^{\prime\prime \mu}=\lambda X^{\mu}_{(j)}\,.~~~~~~~~~~
\end{array}\label{non-susyb}
\end{subeqnarray}
We note that this result holds in  arbitrary dimension, $d$.
\newline
\indent  
As an example, we consider scalar fields, $\Psi(x)$. The  
$n$-point function of them is from eq.(\ref{nscalar})
\begin{equation}
\langle \Psi_{1}(x_{1})\cdots\Psi_{n}(x_{n})\rangle=
\displaystyle{\frac{F(\hat{X}_{1(i)}{\cdot\hat{X}_{1(j)}})}{
\displaystyle{\prod_{l<m}(x_{lm}^{2})^{-\frac{1}{(n-1)(n-2)}
(\eta_{1}+\cdots+\eta_{n})+\frac{1}{n-2}(\eta_{l}+\eta_{m})}}}}\,,
\end{equation}
where
\begin{equation}
\hat{X}_{1(j)}=\displaystyle{\frac{X_{1(j)}}{\sqrt{X_{1(1)}^{2}}}}\,.
\end{equation}
Alternatively, from eq.(\ref{crossratio}), we may regard 
$F(\hat{X}_{1(i)}{\cdot\hat{X}_{1(j)}})$ as an arbitrary 
function of cross ratios $(x_{ij}^{2}x_{kl}^{2})/(x_{ik}^{2}x_{jl}^{2})$. \newline
The number of independent 
$\hat{X}_{1(i)}{\cdot\hat{X}_{1(j)}},\,\,1\leq i,j\leq n-2$ or 
cross ratios 
in $d$-dimensions, $\#_{n|d}$, is  
\begin{equation}
\#_{n|d}=\left\{\begin{array}{ll}
\textstyle{\frac{1}{2}}n(n-3)~~~~&\mbox{for}~n\leq d+2\,,\\
{}&{}\\
dn-\textstyle{\frac{1}{2}}(d+1)(d+2)~~~~&\mbox{for}~n> d+2\,.
\end{array}\right.
\end{equation}
and the $n$-point function of scalar fields reduces  
to an arbitrary $\#_{n|d}$ variable function.

\newpage
\section{Superconformally Covariant Operators\label{4operator}}
In general acting on a quasi-primary superfield, $\Psi^{\rho}{}_{r}(z)$, 
with the spinor derivative, $D_{a\alpha}$, does not lead to a
quasi-primary field.  
For a superfield, $\Psi^{\rho}{}_{r}$, from
eqs.(\ref{4comDL},\,\ref{4Ddelta}\,,\ref{4deltaPsi})  we have
\begin{equation}
\begin{array}{ll}
D_{a\alpha}\delta\Psi^{\rho}{}_{r}=&
-({\cal L}+(\eta+\textstyle{\frac{1}{2}})\hat{\lambda}+i
(\kappa+\textstyle{\frac{1}{2}})\hat{\Omega})
D_{a\alpha}\Psi^{\rho}{}_{r}\\
{}&{}\\
{}&{}+\tilde{\hat{w}}_{\alpha}{}^{\beta}D_{a\beta}\Psi^{\rho}{}_{r}
-D_{a\alpha}\Psi^{\sigma}{}_{r}
\textstyle{\frac{1}{2}}(s_{\mu\nu}\hat{w}^{\mu\nu})_{\sigma}^{~\rho}\\
{}&{}\\
{}&{}-\hat{t}{}^{b}{}_{a}D_{b\alpha}\Psi^{\rho}{}_{r}
-D_{a\alpha}\Psi^{\rho}{}_{s}
\textstyle{\frac{1}{2}}(s^{b}{}_{c}\hat{t}{}^{c}{}_{b})^{s}{}_{r}\\
{}&{}\\
{}&{}+
2\hat{\rho}_{b\beta}(\Psi Y^{b\beta}{}_{a\alpha})^{\rho}{}_{r}\,.
\end{array}
\label{4infDPSI}
\end{equation}
We may connect the generator of $\mbox{SL}(2,\Com)$ to $\mbox{SO}(1,3)$ by
\begin{equation}
s_{\alpha}{}^{\beta}\equiv -\textstyle{\frac{1}{2}}s_{\mu\nu}
(\sigma^{[\mu}\tilde{\sigma}{}^{\nu]})_{\alpha}{}^{\beta}\,,
\end{equation}
where
\begin{equation}
\begin{array}{c}
[s_{\alpha\beta},s_{\gamma\delta}]=\epsilon_{\alpha\delta}s_{\beta\gamma}
+\epsilon_{\beta\delta}s_{\alpha\gamma}
+\epsilon_{\alpha\gamma}s_{\beta\delta}
+\epsilon_{\beta\gamma}s_{\alpha\delta}\,,\\
{}\\
s_{\alpha\beta}=s_{\alpha}{}^{\gamma}\epsilon_{\gamma\beta}=s_{\beta\alpha}\,.
\end{array}
\end{equation}
$Y^{b\beta}{}_{a\alpha}$ is given by
\begin{equation}
Y^{b\beta}{}_{a\alpha}=((\textstyle{\frac{4}{{\cal N}}}-1)\kappa-\eta)
\delta^{b}{}_{a}\delta^{\beta}{}_{\alpha}+
\delta^{b}{}_{a}s_{\alpha}{}^{\beta}+2s^{b}{}_{a}\delta^{\beta}{}_{\alpha}\,.
\end{equation}
To ensure that $D_{a\alpha}\Psi^{\rho}{}_{r}$ is quasi-primary it is
necessary that the terms proportional to $\hat{\rho}$ vanish and this can be
achieved by restricting $D_{a\alpha}\Psi^{\rho}{}_{r}$ to an irreducible
representation of $\mbox{SL}(2,\Com),\mbox{SU}({\cal N})$ and choosing a
particular value of $\eta$ and $\kappa$ 
so that $\Psi Y=0$. 
The changes of the scale dimension and the $R$-symmetry charge, 
$\eta\rightarrow\eta+\frac{1}{2},\,\kappa\rightarrow\kappa
+\frac{1}{2}$,  in eq.(\ref{4infDPSI}) are   also
apparent  from eq.(\ref{4ReD}) 
\begin{equation}
D_{a\alpha}=\Omega(z;g)^{1/2}\Upsilon(z;g)^{\frac{1}{2}}
\hat{L}_{\alpha}{}^{\beta}(z;g)\hat{u}^{-1}{}^{b}{}_{a}(z;g)
D^{\prime}_{b\beta}\,.
\end{equation}
\indent  As an illustration we consider tensorial
fields, $\Psi^{b_{1}\cdots b_{n}}_{a_{1}\cdots a_{m}\alpha_{1}
\cdots\alpha_{k}\da_{1}\cdots\da_{l}}$, which transform as
\begin{equation}
\begin{array}{ll}
\delta\Psi^{b_{1}\cdots b_{n}}_{a_{1}\cdots a_{m}\alpha_{1}
\cdots\alpha_{k}\da_{1}\cdots\da_{l}}
=&\displaystyle{
-({\cal L}+\eta\hat{\lambda}+i\kappa \hat{\Omega})
\Psi^{b_{1}\cdots b_{n}}_{a_{1}\cdots a_{m}\alpha_{1}
\cdots\alpha_{k}\da_{1}\cdots\da_{l}}}\\
{}&{}\\
{}&{}+\displaystyle{\sum^{k}_{p=1}}\tilde{\hat{w}}_{\alpha_{p}}{}^{\beta}
\Psi^{b_{1}\cdots b_{n}}_{a_{1}\cdots a_{m}\alpha_{1}\cdots\beta
\cdots\alpha_{k}\da_{1}\cdots\da_{l}}
-\displaystyle{\sum^{l}_{q=1}}
\Psi^{b_{1}\cdots b_{n}}_{a_{1}\cdots a_{m}\alpha_{1}
\cdots\alpha_{k}\da_{1}\cdots\db\cdots\da_{l}}
\hat{w}^{\db}{}_{\da_{q}}\\
{}&{}\\
{}&{}-\displaystyle{\sum^{m}_{i=1}}
\Psi^{b_{1}\cdots b_{n}}_{a_{1}\cdots c\cdots a_{m}\alpha_{1}
\cdots\alpha_{k}\da_{1}\cdots\da_{l}}\hat{t}{}^{c}{}_{a_{i}}+
\displaystyle{\sum^{n}_{j=1}}\hat{t}{}^{b_{j}}{}_{d}
\Psi^{b_{1}\cdots d\cdots b_{n}}_{a_{1}\cdots a_{m}\alpha_{1}
\cdots\alpha_{k}\da_{1}\cdots\da_{l}}\,.
\end{array}
\end{equation}
Note that spinorial indices,~$\alpha,\da$ may be   raised 
by $\epsilon^{-1},\,\bar{\epsilon}^{-1}$. \newline
For $\Psi^{b_{1}\cdots b_{n}}_{a_{1}\cdots a_{m}\alpha_{1}
\cdots\alpha_{k}\da_{1}\cdots\da_{l}}$  we have
\begin{equation}
\begin{array}{ll}
(\Psi Y^{b\beta}{}_{a\alpha})^{b_{1}\cdots b_{n}}_{a_{1}\cdots a_{m}\alpha_{1}
\cdots\alpha_{k}\da_{1}\cdots\da_{l}}&=
-2\displaystyle{\sum_{p=1}^{k}}\delta^{b}_{~a}\delta^{\beta}_{~\alpha_{p}}
\Psi^{b_{1}\cdots b_{n}}_{a_{1}\cdots a_{m}\alpha_{1}
\cdots\alpha\cdots\alpha_{k}\da_{1}\cdots\da_{l}}\\
{}&{}\\
{}&{}\,\,
+2\displaystyle{\sum_{i=1}^{m}}\delta^{b}_{~a_{i}}\delta^{\beta}_{~\alpha}
\Psi^{b_{1}\cdots b_{n}}_{a_{1}\cdots a\cdots a_{m}\alpha_{1}
\cdots\alpha_{k}\da_{1}\cdots\da_{l}}\\
{}&{}\\
{}&{}\,\,
-2\displaystyle{\sum_{j=1}^{n}}\delta_{a}^{~b_{j}}\delta^{\beta}_{~\alpha}
\Psi^{b_{1}\cdots b\cdots b_{n}}_{a_{1}\cdots a_{m}\alpha_{1}
\cdots\alpha_{k}\da_{1}\cdots\da_{l}}\\
{}&{}\\
{}&{}\,\,
+\left((\frac{4}{{\cal N}}-1)\kappa-\eta+k+2\textstyle{\frac{1}{{\cal N}}}(n-m)\right)\delta^{b}_{~a}
\delta^{\beta}_{~\alpha}
\Psi^{b_{1}\cdots b_{n}}_{a_{1}\cdots a_{m}\alpha_{1}
\cdots\alpha_{k}\da_{1}\cdots\da_{l}}\,.
\end{array}
\label{4rhoY}
\end{equation}
In particular, 
eq.(\ref{4rhoY}) shows that the following are quasi-primary
\begin{subeqnarray}
\label{4DPsi}
D_{(b(\beta}\Psi_{a_{1}\cdots a_{m})
\alpha_{1}\cdots\alpha_{k})\da_{1}\cdots\da_{l}}
~~~~~~~~~~~~~~~~~\mbox{if}~~~
\eta-(\textstyle{\frac{4}{{\cal N}}}-1)
\kappa=-k+2(1-\textstyle{\frac{1}{{\cal N}}})m\,,\label{4DPsi1}~~~~~\\
{}\nonumber\\
\bar{\zeta}_{b(a}
\tilde{D}^{b\beta}\Psi_{a_{1}\cdots a_{m})\beta\da_{1}\cdots\da_{l}}
~~~~~~~~~~~~~~~~~~~~\mbox{if}~~~
\eta-(\textstyle{\frac{4}{{\cal N}}}-1)
\kappa=3+2(1-\textstyle{\frac{1}{{\cal N}}})m\,,\label{4DPsi2}~\,~~~~~~\\
{}\nonumber\\
D_{[b(\beta}\Psi_{a_{1}\cdots a_{m}]
\alpha_{1}\cdots\alpha_{k})\da_{1}\cdots\da_{l}}
~~~~~~~~~~~~~~~~~\mbox{if}~~~
\eta-(\textstyle{\frac{4}{{\cal N}}}-1)\kappa
=-k-2(1+\textstyle{\frac{1}{{\cal N}}})m\,,\label{4DPsi3}~~~~~\\
{}\nonumber\\
\bar{\zeta}_{b[a}
\tilde{D}^{b\beta}\Psi_{
a_{1}\cdots a_{m}]\beta\da_{1}\cdots\da_{l}}
~~~~~~~~~~~~~~~~~~~~\mbox{if}~~~
\eta-(\textstyle{\frac{4}{{\cal N}}}-1)\kappa
=3-2(1+\textstyle{\frac{1}{{\cal N}}})m\,,~~~~~~\,~\label{4DPsi4}
\end{subeqnarray}
where  $(\,),\,[\,]$ denote the usual
symmetrization,\,anti-symmetrization respectively and obviously 
eqs.(\ref{4DPsi3},d) are nontrivial if $m+1\leq {\cal N}$. \newline
\indent   
Now we consider the case where  more than one spinor derivative,
$D_{a\alpha}$, act on a quasi-primary superfield. In this case, 
it is useful to note 
\begin{equation}
\begin{array}{cc}
D_{(a(\alpha}D_{b)\beta)}=0\,,~~~~~&~~~~~D_{[a[\alpha}D_{b]\beta]}=0\,,
\end{array}
\end{equation}
and
\begin{equation}
D_{a\alpha}\hat{\rho}_{b\beta}=0\,.
\end{equation}
From eq.(\ref{4rhoY}) one can show that the following are
quasi-primary
\begin{subeqnarray}
\label{4DDPsi}
D_{[b_{1}(\beta_{1}}\cdots D_{b_{n}\beta_{n}}
\Psi_{\alpha_{1}\cdots\alpha_{k})a_{1}\cdots a_{m}]
\da_{1}\cdots\da_{l}}~~~\mbox{if}~~~
\eta-(\textstyle{\frac{4}{{\cal N}}}-1)\kappa=2-k-2n-2(1+\textstyle{\frac{1}{{\cal N}}})m\,,\label{4DDPsi2}~~~~~\\
{}\nonumber\\
D_{(bc}\Psi_{a_{1}\cdots a_{m})\da_{1}\cdots\da_{l}}~~~~~~~~~~~~~~~~~~~~~~~~~
\mbox{if}~~~
\eta-(\textstyle{\frac{4}{{\cal N}}}-1)\kappa
=2+2(1-\textstyle{\frac{1}{{\cal N}}})m\,,\label{4DDPsi1}~~\,~~~~~~~~~~~~~~~
\end{subeqnarray}
where
\begin{equation}
D_{bc}=\epsilon^{-1\gamma\beta}D_{b\beta}D_{c\gamma}\,.
\end{equation}
Eq.(\ref{4DDPsi2}) is nontrivial if $m+n\leq {\cal N}$.\newline
Similar analysis for $\bar{D}^{a}_{\da}$ is ready to be done 
by taking complex conjugates of the
results~(\ref{4DPsi},\,\ref{4DDPsi}). We also note that 
in ${\cal N}\neq 4$ chiral superfield theories, 
$\eta$ and
$\kappa$ are related by $\eta+(\textstyle{\frac{4}{{\cal N}}}-1)\kappa=0$ as demonstrated in eq.(\ref{Nn4ek}).\newline
\newline
\newline

\begin{center}
\large{\textbf{Acknowledgments}}
\end{center}
I am deeply indebted to  Hugh Osborn for many aspects of writing-up 
in this paper. I also wish to thank Youngjai Kiem for his helpful comments. 
Part of this work was done with the hospitality of 
Asia Pacific Center for Theoretical Physics.

\newpage
\appendix
\begin{center}
\Large{\textbf{Appendix}}
\end{center}
\section{Notations \& Useful Equations\label{AppendixA}}
With the four-dimensional  Minkowskian metric, $\eta^{\mu\nu} =
\mbox{diag}(+1,-1,-1,-1)$, the  $4\times 4$ gamma matrices,
$\gamma^{\mu},~\mu=0,1,\cdots 3$,   
satisfy the Clifford algebra
\begin{equation}
\gamma^{\mu}\gamma^{\nu}+\gamma^{\nu}\gamma^{\mu}=2\eta^{\mu\nu}\,.
\label{4gc}
\end{equation}
The gamma matrices for even dimensions can be chosen in general
  to have the form
\begin{equation}
\gamma^{\mu}=\left(\begin{array}{cc}
                 0 & \sigma^{\mu}\\
                 \tilde{\sigma}^{\mu} & 0 \end{array}\right)\,,
\end{equation} 
and to satisfy
\begin{equation}
\gamma_{5}=-i\gamma_{0}\gamma_{1}\gamma_{2}\gamma_{3}=\left(\begin{array}{cc}
                 1 & 0\\
                  0&-1 \end{array}\right)\,.
\label{4g5}
\end{equation}
We also assume the hermiticity condition
\begin{equation}
\gamma^{0}\gamma^{\mu\dagger}\gamma^{0}=\gamma^{\mu}\,.
\label{4Gdagger}
\end{equation}
The $2\times 2$ matrices, $\sigma^{\mu},\tilde{\sigma}^{\mu}$
satisfy from eqs.(\ref{4gc},\,\ref{4g5})
\begin{equation}
\sigma^{\mu}\tilde{\sigma}^{\nu}+\sigma^{\nu}\tilde{\sigma}^{\mu}
=2\eta^{\mu\nu}\,,
\label{sigti}
\end{equation}
and
\begin{subeqnarray}
&\textstyle{\frac{1}{2}}\mbox{tr}(\sigma^{\mu}\tilde{\sigma}^{\nu})
=\eta^{\mu\nu}\,,&\label{4Tr2}\\
{}\nonumber\\
&\textstyle{\frac{1}{2}}\mbox{tr}(\sigma^{\mu}\tilde{\sigma}^{\nu}
\sigma^{\lambda}\tilde{\sigma}^{\rho})=\eta^{\mu\nu}\eta^{\lambda\rho}
+\eta^{\nu\lambda}\eta^{\rho\mu}-\eta^{\mu\lambda}\eta^{\nu\rho}-
i\epsilon^{\mu\nu\lambda\rho}\,,&\label{4Tr4}
\end{subeqnarray}
where we put $\epsilon_{0123}=-\epsilon^{0123}=1$.\newline
$\sigma^{\mu}$ and $\tilde{\sigma}^{\mu}$  separately form bases 
of $2 \times 2$  matrices with the completeness relation
\begin{equation}
\sigma^{\mu}_{\alpha\dot{\alpha}}\tilde{\sigma}_{\mu}^{\dot{\beta}\beta}=
2\delta_{\alpha}^{~\beta}\delta_{\dot{\alpha}}^{~\dot{\beta}}\,.
\label{4contract}
\end{equation}
The coefficient on the right hand side 
may be determined by eq.(\ref{4Tr2}). \newline  
Charge conjugation matrix $C$ satisfies
\begin{equation}
C\gamma^{\mu}C^{-1}=-\gamma^{\mu}\,,
\label{4charge}
\end{equation}
and has the form
\begin{equation}
C=\left(\begin{array}{cc}
\epsilon&0\\
0&\bar{\epsilon}^{-1}
\end{array}\right)\,,
\end{equation}
where
$\epsilon_{\alpha\beta},\,\bar{\epsilon}_{\dot{\alpha}\dot{\beta}}$ are
the $2\times 2$ anti-symmetric matrices, 
$\epsilon_{12}=\bar{\epsilon}_{12}=1$ with inverses,\newline 
$(\epsilon^{-1})^{\alpha\beta},\,
(\bar{\epsilon}^{-1})^{\dot{\alpha}\dot{\beta}}$.\newline
Eq.(\ref{4charge}) implies 
\begin{equation}
\epsilon\tilde{\sigma}^{\mu t}\bar{\epsilon}=-\sigma^{\mu}\,.
\label{4ese}
\end{equation}
From eqs.(\ref{4contract},\,\ref{4ese}) we get
\begin{equation}
\begin{array}{c}
\sigma^{\mu}_{\alpha\da}\sigma_{\mu\beta\db}=2\epsilon_{\alpha\beta}
\bar{\epsilon}_{\da\db}\,,\\
{}\\
(\sigma^{[\mu}\tilde{\sigma}^{\nu ]})_{\alpha}{}^{\beta}
(\sigma_{[\mu}\tilde{\sigma}_{\nu ]})_{\gamma}{}^{\delta}=
4(\delta_{\alpha}^{~\beta}\delta_{\gamma}^{~\delta}
-2\delta_{\alpha}^{~\delta}\delta^{~\beta}_{\gamma})\,,\\
{}\\
(\sigma^{[\mu}\tilde{\sigma}^{\nu ]})_{\alpha}{}^{\beta}
(\tilde{\sigma}_{[\mu}\sigma_{\nu ]})^{\da}{}_{\db}=0\,.
\label{4sig4}
\end{array}
\end{equation}
It is useful to note
\begin{equation}
\delta_{\alpha}^{~\delta}\delta^{~\beta}_{\gamma}-
\delta_{\alpha}^{~\beta}\delta^{~\delta}_{\gamma}
=\epsilon_{\alpha\gamma}\epsilon^{-1}{}^{\beta\delta}\,.
\label{4dde}
\end{equation}
We may choose 
\begin{equation}
\sigma^{0}=\tilde{\sigma}^{0}=1\,,
\end{equation}
then from eqs.(\ref{4Gdagger},\,\ref{sigti})
\begin{equation}
\begin{array}{ccc}
\sigma^{i}=-\tilde{\sigma}^{i}\,,~~~~&~~~~
\sigma^{\mu}{}^{\dagger}=\sigma^{\mu}\,,~~~~&~~~~i=1,2,3\,.
\end{array}
\end{equation}
\indent  In four-dimensions there is a unique correspondence between a general
four vector, $v^{\mu}$, and a $2\times 2$ matrix, ${\rm
v}_{\alpha\da}$ or $\tilde{{\rm v}}^{\da\alpha}$,  through
\begin{equation}
\begin{array}{cc}
{\rm v}_{\alpha\da}=v^{\mu}\sigma_{\mu\alpha\da}\,,~~~~~&~~~~~
v^{\mu}=\textstyle{\frac{1}{2}}\mbox{tr}(\tilde{\sigma}^{\mu}{\rm
v})\,,\\
{}&{}\\
\tilde{{\rm v}}^{\da\alpha}=v^{\mu}\tilde{\sigma}_{\mu}^{\da\alpha}
\,,~~~~~&~~~~~
v^{\mu}=\textstyle{\frac{1}{2}}\mbox{tr}(\sigma^{\mu}\tilde{{\rm v}})\,.
\end{array}
\label{4vvt}
\end{equation}
With this notation it is convenient to  introduce the variables
\begin{equation}
\tx_{\pm}=\tx\pm2i\bar{\theta}_{a}\theta^{a}\,.
\label{4tildeX}
\end{equation}
Note that
\begin{equation}
\begin{array}{cc}
\tx_{\pm}=x^{\mu}_{\pm}
\tilde{\sigma}_{\mu}\,,~~~~~~&~~~~~~
x^{\mu}_{\pm}=x^{\mu}\mp i\theta^{a}\sigma^{\mu}\bar{\theta}_{a}\,,\\
{}&{}\\
\multicolumn{2}{c}{\tx^{\dagger}_{\pm}=\tx_{\mp}\,,}
\end{array}
\label{4Xprop}
\end{equation}
and
\begin{equation}
{\rm x}_{\pm}=x^{\mu}_{\pm}\sigma_{\mu}={\rm x}\mp
2i\tilde{\theta}_{a}\tilde{\bt}{}^{a}\,.
\end{equation}
\indent  Some useful identities relevant to the present paper  are
\begin{subeqnarray}
\label{N=1sigma}
&\theta^{\alpha}\theta^{\beta}=-\textstyle{\frac{1}{2}}
\epsilon^{-1\alpha\beta}\theta\tilde{\theta}\,,~~~~~~~~~~
\bar{\theta}^{\dot{\alpha}}\bar{\theta}^{\dot{\beta}}=
\textstyle{\frac{1}{2}}
\bar{\epsilon}^{-1\dot{\alpha}\dot{\beta}}
\tilde{\bar{\theta}}\bar{\theta}\,,&\\
&{}&\nonumber\\
&\theta\sigma^{\mu}\bar{\theta}\theta\sigma^{\nu}\bar{\theta}
=\textstyle{\frac{1}{2}}\eta^{\mu\nu}\theta\tilde{\theta}
\tilde{\bar{\theta}}\bar{\theta}\,,&\\
&{}&\nonumber\\
&\theta\sigma^{\mu}\bar{\theta}\theta\sigma_{\mu}\bar{\theta}=
\displaystyle{\frac{4}{x^{2}}}(\theta{\rm x}\bar{\theta})^{2}\,,&
\end{subeqnarray}
\begin{subeqnarray}
\label{N=1sigma2}
&\theta^{\alpha}_{1}\bar{\theta}_{2}^{\da}=\textstyle{\frac{1}{2}}
\theta_{1}\sigma^{\mu}\bar{\theta}_{2}\,
\tilde{\sigma}{}^{\da\alpha}_{\mu}\,,&\\
&{}&\nonumber\\
&\epsilon_{\mu_{1}\cdots\mu_{d}}\epsilon_{\nu_{1}\cdots\nu_{d}}=\displaystyle{\sum_{p=1}^{d!}{}\mbox{sign}(p)\,
\delta_{\mu_{1}\nu_{p_{1}}}\cdots\delta_{\mu_{d}\nu_{p_{d}}}}~~~~~p:\,\mbox{permutations}\,,&\\
&{}&\nonumber\\
&\epsilon_{\mu_{1}\cdots\mu_{d}}x_{(1)}^{\mu_{1}}\cdots x_{(d)}^{\mu_{d}}=\pm\sqrt{\epsilon^{i_{1}\cdots i_{d}}x_{(1)}{\cdot x_{(i_{1})}}\cdots x_{(d)}{\cdot x_{(i_{d})}}}\,,&
\end{subeqnarray}
\begin{equation}
\det(\theta^{a}\tilde{\theta}_{b})
=(\textstyle{\frac{1}{2}})^{{\cal N}}({\cal N}+1)!\,\displaystyle{
\prod_{a=1}^{{\cal N}}\theta^{a}\tilde{\theta}_{a}}\,.
\end{equation}

\section{Solution of Superconformal Killing Equation\label{4AppendixB}}
To solve the superconformal Killing equation~(\ref{4masternew})   
we first note that from eq.(\ref{4master}) 
\begin{equation}
D_{b\beta}D_{a\alpha}\tilde{{\rm h}}^{\dot{\gamma}\gamma}
=\textstyle{\frac{1}{4}}\delta_{\alpha}^{~\gamma}D_{b\delta}D_{a\beta}
\tilde{{\rm h}}^{\dot{\gamma}\delta}\,.
\end{equation}
Contracting $\beta$ and $\gamma$ indices gives
\begin{equation}
D_{b\delta}D_{a\alpha}\tilde{{\rm h}}^{\dot{\gamma}\delta}=0\,,
\end{equation}
and hence
\begin{equation}
D_{b\beta}D_{a\alpha}\tilde{{\rm h}}^{\dot{\gamma}\gamma}=0\,.
\label{4DDh}
\end{equation}
Now we  write a general solution of
eq.(\ref{4ordiKi}) as
\begin{equation}
h^{\mu}(z)=a^{\mu}(\theta,\bar{\theta})+\lambda(\theta,\bar{\theta})
x^{\mu}+w^{\mu}_{~\nu}(\theta,\bar{\theta})x^{\nu}
+2x{\cdot b}(\theta,\bar{\theta})x^{\mu}-x^{2}b^{\mu}(\theta,\bar{\theta})\,,
\label{4hA}
\end{equation}
where
$w_{\mu\nu}(\theta,\bar{\theta})+w_{\nu\mu}(\theta,\bar{\theta})=0.$
\newline
Then eq.(\ref{4hA}) can be written in terms of  
$\tilde{{\rm h}}=h^{\mu}\tilde{\sigma}_{\mu}$   as
\begin{equation}
\tilde{{\rm h}}(z)=\tx_{-}{\rm b}(\theta,\bar{\theta})\tx_{+}+
\tx_{-}W(\theta,\bar{\theta})
+\bar{W}(\theta,\bar{\theta})\tx_{+}
+\tilde{{\rm A}}(\theta,\bar{\theta})\,,
\label{4tildeh}
\end{equation}
where
\begin{equation}
\begin{array}{ll}
W(\theta,\bar{\theta})
=&-{\textstyle\frac{1}{4}}w_{\mu\nu}(\theta,\bar{\theta})
\sigma^{\mu}\tilde{\sigma}^{\nu}
+{\textstyle\frac{1}{2}}\lambda(\theta,\bt) -2i{\rm
b}(\theta,\bt)\bt_{a}\theta^{a}
=\bar{W}(\theta,\bt)^{\dagger}\,,\\
{}&{}\\
\tilde{{\rm A}}(\theta,\bt)=&\tilde{{\rm a}}(\theta,\bt)
-i{\textstyle\frac{1}{2}}w_{\mu\nu}(\theta,\bt)
(\bt_{a}\theta^{a}\sigma^{[\mu}\tilde{\sigma}^{\nu]}+
\tilde{\sigma}^{[\mu}\sigma^{\nu]}\bt_{a}\theta^{a})+4\theta^{a} 
{\rm b}(\theta,\bt)\bt_{b}\bt_{a}\theta^{b}=\tilde{{\rm A}}(\theta,\bt)^{\dagger}\,.
\end{array}
\end{equation}
Essentially we may regard   $W(\theta,\bt)$ as  an 
arbitrary $2\times 2$  matrix and $\tilde{{\rm A}}(\theta,\bt)$ as an arbitrary
$2\times 2$ hermitian  matrix.\newline
\indent  The variables,~$\tx_{\pm}$ defined in eq.(\ref{4tildeX}), satisfy
\begin{equation}
\begin{array}{ll}
D_{a\alpha}\tx^{\dot{\beta}\beta}_{+}=
-4i\delta_{\alpha}^{~\beta}\bar{\theta}_{a}^{\dot{\beta}}\,,~~~~&~~~~
D_{a\alpha}\tx^{\dot{\beta}\beta}_{-}=0\,,\\
{}&{}\\
\bar{D}^{a}_{\da}\tx^{\dot{\beta}\beta}_{+}=0\,,~~~~&~~~~
\bar{D}^{a}_{\da}\tx^{\dot{\beta}\beta}_{-}=4i\delta_{\da}^{~\db}
\theta^{a\beta}\,,
\end{array}
\label{4DX}
\end{equation}
which ensure that substituting  eq.(\ref{4tildeh})  into
eq.(\ref{4master}) leads  independent   equations 
for ${\rm
b}(\theta,\bt),~W(\theta,\bt),~\tilde{{\rm A}}(\theta,\bt)$. \newline
\newline
\indent  After substituting eq.(\ref{4tildeh}) into
eq.(\ref{4master}), we get from the $x^{2}$-terms
\begin{equation}
D_{a\alpha}({\rm b}(\theta,\bt)\tilde{\sigma}^{\mu}){}_{\delta}^{~\beta}
=\textstyle{\frac{1}{2}}\delta_{\alpha}^{~\beta}
D_{a\gamma}({\rm b}(\theta,\bt)\tilde{\sigma}^{\mu}){}_{\delta}^{~\gamma}\,.
\end{equation}
From eq.(\ref{4contract})
\begin{equation}
\delta_{\gamma}^{~\beta}D_{a\alpha}{\rm b}_{\delta\dot{\delta}}(\theta,\bt)
=\textstyle{\frac{1}{2}}
\delta_{\alpha}^{~\beta}D_{a\gamma}{\rm b}_{\delta\dot{\delta}}(\theta,\bt)\,,
\end{equation}
and hence 
\begin{equation}
D_{a\alpha}{\rm b}(\theta,\bt)=0\,.
\end{equation}
In a similar fashion, or taking complex conjugate,  
one can show $\bar{D}^{a}_{\dot{\alpha}}
{\rm b}(\theta,\bt)=0$. Therefore $b^{\mu}(\theta,\bt)$ is independent
of $\theta,\bt$ and eq.(\ref{4DX}) shows that 
$\tx_{-}{\rm b}\tx_{+}$ is a solution of
eq.(\ref{4master}).\newline
\newline
\indent  The remaining terms lead
\begin{equation}
\begin{array}{l}
\,(D_{a\alpha}\bar{W}(\theta,\bt)\tilde{\rm x}_{-})^{\db\beta}+
(\tx_{-} D_{a\alpha}W(\theta,\bt))^{\db\beta}+
4i(D_{a\alpha}\bar{W}(\theta,\bt)\bt_{b})^{\db}\theta^{b\beta}+D_{a\alpha}
\tilde{{\rm A}}{}^{\db\beta}(\theta,\bt)\\
{}\\
=\textstyle{\frac{1}{2}}\delta_{\alpha}^{~\beta}\{
(D_{a\gamma}\bar{W}(\theta,\bt)\tilde{\rm x}_{-})^{\db\gamma}+
(\tx_{-}D_{a\gamma}W(\theta,\bt))^{\db\gamma}+
4i(D_{a\gamma}\bar{W}(\theta,\bt)\bt_{b})^{\db}\theta^{b\gamma}+D_{a\gamma}
\tilde{{\rm A}}{}^{\db\gamma}(\theta,\bt)\}\,.
\end{array}
\end{equation}
This gives two separate equations
\begin{eqnarray}
&D_{a\alpha}(\bar{W}(\theta,\bt)\tilde{\sigma}^{\mu}+\tilde{\sigma}^{\mu}
W(\theta,\bt))^{\db\beta}=\textstyle{\frac{1}{2}}
\delta_{\alpha}^{~\beta}D_{a\gamma}
(\bar{W}(\theta,\bt)\tilde{\sigma}^{\mu}+\tilde{\sigma}^{\mu}
W(\theta,\bt))^{\db\gamma}\,,\label{4linearinx}\\
{}\nonumber\\
&D_{a\alpha}(\tilde{{\rm A}}(\theta,\bt)
+4i\bar{W}(\theta,\bt)\bt_{b}\theta^{b})^{\db\beta}=\textstyle{\frac{1}{2}}
\delta_{\alpha}^{~\beta}
D_{a\gamma}(\tilde{{\rm A}}(\theta,\bt)
+4i\bar{W}(\theta,\bt)\bt_{b}\theta^{b})^{\db\gamma}\,.\label{4zerothinx}
\end{eqnarray}
Eq.(\ref{4linearinx}) is equivalent, from eq.(\ref{4contract}),  to
\begin{equation}
\delta_{\gamma}^{~\beta}D_{a\alpha}
\bar{W}^{\db}_{~\dot{\gamma}}(\theta,\bt)+
\delta_{~\dot{\gamma}}^{\db}D_{a\alpha}
W^{~\beta}_{\gamma}(\theta,\bt)=\textstyle{\frac{1}{2}}
\delta_{\alpha}^{~\beta}(D_{a\gamma}
\bar{W}^{\db}_{~\dot{\gamma}}(\theta,\bt)
+\delta^{\db}_{~\dot{\gamma}}D_{a\delta}
W^{~\delta}_{\gamma}(\theta,\bt))\,.
\label{4linearinx1.5}
\end{equation}
By contracting $\beta$ and $\gamma$ indices one can solve 
 $D_{a\alpha}\bar{W}^{\da}_{~\db}(\theta,\bt)$ in terms of 
$D_{a\alpha}W_{\beta}^{~\gamma}(\theta,\bt)$,  and hence 
 eq.(\ref{4linearinx1.5}) is   equivalent to 
\begin{subeqnarray}
\label{4linearinx2}
&D_{a\alpha}\bar{W}^{\da}_{~\db}(\theta,\bt)=\textstyle{\frac{1}{3}}
\delta^{\da}_{~\db}(D_{a\beta}W_{\alpha}^{~\beta}(\theta,\bt)-2
D_{a\alpha}W_{\beta}^{~\beta}(\theta,\bt))
=\textstyle{\frac{1}{2}}\delta^{\da}_{~\db}D_{a\alpha}
\bar{W}^{\dg}_{~\dg}(\theta,\bt)\,,
\label{4linearinx2a}\\
{}\nonumber\\
&D_{a\alpha}W_{\beta}^{~\gamma}(\theta,\bt)
=\textstyle{\frac{1}{3}}\delta_{\alpha}^{~\gamma}(
2D_{a\delta}W_{\beta}^{~\delta}(\theta,\bt)-
D_{a\beta}W_{\delta}^{~\delta}(\theta,\bt))-
\textstyle{\frac{1}{2}}\delta_{\beta}^{~\gamma}D_{a\alpha}
\bar{W}^{\db}_{~\db}(\theta,\bt)\,.\label{4linearinx2b}
\end{subeqnarray}
Eq.(\ref{4zerothinx}) gives from eq.(\ref{4DDh}) 
\begin{equation}
0=D_{b\beta}D_{a\alpha}\tilde{{\rm A}}{}^{\dot{\gamma}\gamma}(\theta,\bt)+
4i(D_{b\beta}D_{a\alpha}\bar{W}(\theta,\bt)
\bt_{c})^{\dot{\gamma}}\theta^{c\gamma}-4i
(D_{a\alpha}\bar{W}(\theta,\bt)\bt_{b})^{\dot{\gamma}}
\delta^{~\gamma}_{\beta}\,,
\label{4DDzero}
\end{equation}
and $\{D_{a\alpha},D_{b\beta}\}=0$ implies
\begin{equation}
0=\delta_{\beta}^{~\gamma}(D_{a\alpha}
\bar{W}(\theta,\bt)\bt_{b})^{\dot{\gamma}}
+\delta_{\alpha}^{~\gamma}(D_{b\beta}
\bar{W}(\theta,\bt)\bt_{a})^{\dot{\gamma}}\,,
\end{equation}
and hence from eq.(\ref{4linearinx2a})
\begin{equation}
0=D_{a\alpha}\bar{W}^{\da}_{~\da}(\theta,\bt)\bt^{\db}_{b}\,.
\end{equation}
Therefore $\bar{W}^{\da}_{~\da}(\theta,\bt)$ is of the  form 
\begin{equation}
\begin{array}{cc}
\bar{W}^{\da}_{~\da}(\theta,\bt)
=\bar{w}(\bt)+w^{\prime}(\theta)\bt^{2{\cal N}}\,,~~~~&~~~~
\bt^{2{\cal N}}=\bt_{1}^{2}\bt_{2}^{2}\cdots\bt_{{\cal N}}^{2}\,.
\end{array}
\end{equation}
We may require $w^{\prime}(0)=0$.\newline
Substituting this expression into eq.(\ref{4linearinx2a}) gives
\begin{equation}
D_{a\alpha}\bar{W}^{\da}_{~\db}(\theta,\bt)=\textstyle{\frac{1}{2}}
\delta^{\da}_{~\db}D_{a\alpha}w^{\prime}(\theta)\bt^{2{\cal N}}\,,
\end{equation}
and hence we can put
\begin{equation}
\begin{array}{l}
\bar{W}^{\da}_{~\db}(\theta,\bt)=\textstyle{\frac{1}{2}}
\delta^{\da}_{~\db}w^{\prime}(\theta)\bt^{2{\cal N}}
+\bar{w}^{\da}_{~\db}(\bt)\,,\\
{}\\
W^{~\alpha}_{\beta}(\theta,\bt)=\textstyle{\frac{1}{2}}
\delta^{~\alpha}_{\beta}\bar{w}^{\prime}(\bt)\theta^{2{\cal N}}
+w^{~\alpha}_{\beta}(\theta)\,.
\end{array}
\end{equation}
Substituting this expression again into eq.(\ref{4linearinx2a}) gives
\begin{subeqnarray}
&D_{a\alpha}w^{\prime}(\theta)\bt^{2{\cal N}}=-\bar{w}^{\prime}(\bt)D_{a\alpha}
\theta^{2{\cal N}}\,,\\
{}\nonumber\\
&D_{a\beta}w_{\alpha}^{~\beta}(\theta)
=2D_{a\alpha}w_{\beta}^{~\beta}(\theta)\,.
\end{subeqnarray}
Thus we get 
\begin{equation}
\begin{array}{cc}
w^{\prime}(\theta)=2ic\,\theta^{2{\cal N}}\,,~~~~&~~~~
\theta^{2{\cal N}}=\theta^{2}_{1}\theta^{2}_{2}\cdots\theta^{2}_{{\cal N}}
=\bar{\theta}^{2{\cal N}}{}^{\dagger}\,,
\end{array}
\end{equation}
and from eq.(\ref{4linearinx2b})
\begin{equation}
D_{a\alpha}w_{\beta}^{~\gamma}(\theta)=\textstyle{\frac{1}{2}}
\delta_{\alpha}^{~\gamma}
D_{a\delta}w_{\beta}^{~\delta}(\theta)\,,
\end{equation}
where $c\in\Real$.\newline
From eq.(\ref{4DDh}) 
\begin{equation}
0=D_{a\alpha}D_{b\beta}w_{\gamma}^{~\delta}(\theta)\,.
\end{equation}
Hence we can put
\begin{equation}
D_{a\alpha}w_{\beta}^{~\beta}(\theta)=4\rho_{a\alpha}\,.
\end{equation}
Therefore we get the solution for $W_{\alpha}^{~\beta}(\theta,\bt)$
\begin{equation}
W(\theta,\bt)=
-ic\,\theta^{2{\cal N}}\bt^{2{\cal N}}
-4\rho_{a}\theta^{a}
-\textstyle{\frac{1}{4}}w_{\mu\nu}\sigma^{[\mu}\tilde{\sigma}^{\nu]}
+\textstyle{\frac{1}{2}}\lambda \,.
\end{equation}
However, $\tilde{{\rm h}}(z)$ is independent of $c$ and hence we can
put $c=0$. \newline
\indent 
Now, eq.(\ref{4zerothinx}) reads
\begin{equation}
\begin{array}{cc}
D_{a\alpha}\tilde{{\rm A}}{}^{\db\beta}(\theta,\bt)=
\textstyle{\frac{1}{2}}\delta_{\alpha}^{~\beta}
D_{a\gamma}\tilde{{\rm A}}{}^{\db\gamma}(\theta,\bt)\,,~~~~&~~~~~
\bar{D}^{a}_{\da}\tilde{{\rm A}}{}^{\db\beta}(\theta,\bt)=
\textstyle{\frac{1}{2}}\delta_{\da}^{~\db}
\bar{D}^{a}_{\dg}\tilde{{\rm A}}{}^{\dg\beta}(\theta,\bt)\,,
\end{array}
\end{equation}
and hence
\begin{equation}
\bar{D}^{b}_{\da}D_{a\alpha}\tilde{{\rm A}}{}^{\db\beta}(\theta,\bt)
=\textstyle{\frac{1}{4}}
\delta_{\da}^{~\db}\delta_{\alpha}^{~\beta}
\bar{D}^{b}_{\dg}D_{a\gamma}\tilde{{\rm A}}{}^{\dg\gamma}(\theta,\bt)\,.
\end{equation}
Since $D_{a\alpha}D_{b\beta}\tilde{{\rm A}}(\theta,\bt)=0$ we can put
\begin{subeqnarray}
\label{4Aprop}
&\bar{D}^{a}_{\dg}D_{b\gamma}\tilde{{\rm A}}{}^{\dg\gamma}(\theta,\bt)=
-16it^{a}_{~b}+8\Omega\delta^{a}_{~b}\,,&\\
{}\nonumber\\
&D_{a\alpha}\tilde{{\rm A}}{}^{\da\alpha}(\theta,\bt)
=-8i\bar{\varepsilon}_{a}^{\da}\,,&
\end{subeqnarray}
where $\Omega\in\Real$ and $
t\in\mbox{su}({\cal N})$ i.e. $t^{\dagger}=-t,\,\mbox{tr}\,t=0$.\newline
Eq.(\ref{4Aprop}) gives the general solution for
$\tilde{{\rm A}}(\theta,\bt)$
\begin{equation}
\tilde{{\rm A}}(\theta,\bt)=-4i\bt_{a}t^{a}_{~b}\theta^{b}
+2\Omega\bt_{a}\theta^{a}+4i(\bar{\varepsilon}_{a}\theta^{a}
-\bt_{a}\varepsilon^{a})+\tilde{{\rm a}}\,.
\label{4solutionforA}
\end{equation}
All together, we get  the general solution of 
the superconformal Killing equation~(\ref{4solutionforh}).

\section{Basis for  Superconformal Algebra\label{4AppendixC}}
We write the superconformal generators in general as 
\begin{equation}
{\cal K}{\cdot{\cal P}}=
a^{\mu}P_{\mu}+\varepsilon^{a}Q_{a}+\bar{Q}^{a}\bar{\varepsilon}_{a}+
\lambda D+i\Omega R
+\textstyle{\frac{1}{2}}w^{\mu\nu}M_{\mu\nu}+b^{\mu}K_{\mu}+
S^{a}\rho_{a}+\bar{\rho}^{a}\bar{S}_{a}+t^{a}_{~b}A^{b}_{~a}\,,
\end{equation}
for
\begin{subeqnarray}
&{\cal
K}=(a^{\mu},b^{\mu},\varepsilon^{a},\bar{\varepsilon}_{a},\rho_{a},
\bar{\rho}^{a},
\lambda,\Omega,w^{\mu}{}_{\nu},t^{a}{}_{b})\,,&\\
{}\nonumber\\
&{\cal P}=(P_{\mu},K_{\mu},Q_{a},\bar{Q}^{a},S_{a},\bar{S}^{a},D,R,
M^{\mu}{}_{\nu},A^{a}{}_{b})\,,&
\end{subeqnarray}
where the $\mbox{SU}({\cal N})$ generators, $A^{a}_{~b}$, satisfy
$A^{\dagger}=-A,~\mbox{tr}A=0$. 
The  superconformal algebra can now be obtained by imposing
\begin{equation}
[{\cal K}_{1}{\cdot{\cal P}},{\cal K}_{2}{\cdot{\cal P}}]
=-i{\cal K}_{3}{\cdot{\cal P}}\,,
\end{equation}
where ${\cal K}_{3}$ is given by  eq.(\ref{4MMcom}).  
From this expression, we can read off 
the following  superconformal algebra. 
\begin{itemize}
\item Poincar\'{e} algebra
\begin{equation}
\begin{array}{cc}
[P_{\mu},P_{\nu}]=0\,, &
[M_{\mu\nu},P_{\lambda}]=i(\eta_{\mu\lambda}P_{\nu}
-\eta_{\nu\lambda}P_{\mu})\,,\\
{}&{}\\
\multicolumn{2}{l}{[M_{\mu\nu},M_{\lambda\rho}]=
i(\eta_{\mu\lambda}M_{\nu\rho}-\eta_{\mu\rho}M_{\nu\lambda}
-\eta_{\nu\lambda}M_{\mu\rho}+\eta_{\nu\rho}M_{\mu\lambda})\,.~~~~~~~}
\end{array}
\label{4poincare}
\end{equation}
\item Supersymmetry algebra
\begin{equation}
\begin{array}{c}
\{Q_{a\alpha}, \bar{Q}^{b}_{\dot{\alpha}}\}
=2\delta_{a}^{~b}\sigma^{\mu}_{\alpha\dot{\alpha}}P_{\mu}\,,~~~~\\
{}\\
{}[M_{\mu\nu},Q_{a}]
=i\textstyle{\frac{1}{2}}\sigma_{[\mu}\tilde{\sigma}_{\nu]}Q_{a}\,,~~\\
{}\\
{}[M_{\mu\nu},\bar{Q}^{a}]
=-i\textstyle{\frac{1}{2}}\bar{Q}^{a}
\tilde{\sigma}_{[\mu}\sigma_{\nu]}\,,\\
{}\\
{}[{P}_{\mu},Q_{a\alpha}]=
[P_{\mu},\bar{Q}^{a}_{\dot{\alpha}}]=\{Q_{a\alpha},Q_{b\beta}\}=
\{\bar{Q}^{a}_{\dot{\alpha}},\bar{Q}^{b}_{\dot{\beta}}\}=0\,.
\end{array}
\end{equation}
\item Special superconformal algebra
\begin{equation}
\begin{array}{c}
[K_{\mu},K_{\nu}]=0\,,~~~~~~~
[M_{\mu\nu},K_{\lambda}]=i(\eta_{\mu\lambda}K_{\nu}
-\eta_{\nu\lambda}K_{\mu})\,,\\
{}\\
{}\{\bar{S}_{a}^{\dot{\alpha}},S^{b\alpha}\}
=2\delta_{a}^{~b}\tilde{\sigma}^{\mu\dot{\alpha}\alpha}K_{\mu}\,,~~\\
{}\\
{}[M_{\mu\nu},S^{a}]
=-i\textstyle{\frac{1}{2}}S^{a}\sigma_{[\mu}\tilde{\sigma}_{\nu]}\,,\\
{}\\
{}[M_{\mu\nu},\bar{S}_{a}]
=i\textstyle{\frac{1}{2}}\tilde{\sigma}_{[\mu}\sigma_{\nu]}\bar{S}_{a}\,,~~\\
{}\\
{}[{K}_{\mu},S^{a\alpha}]=
[K_{\mu},\bar{S}_{a}^{\dot{\alpha}}]=\{S^{a\alpha},S^{b\beta}\}=
\{\bar{S}_{a}^{\dot{\alpha}},\bar{S}_{b}^{\dot{\beta}}\}=0\,.
\end{array}
\end{equation}
\item Cross terms between $(P,Q,\bar{Q})$ and $(K,S,\bar{S})$
\begin{equation}
\begin{array}{cc}
\multicolumn{2}{c}{[P_{\mu},K_{\nu}]=2i(M_{\mu\nu}+\eta_{\mu\nu}D)\,,}\\
{}&{}\\
~~~~~~~~
\displaystyle{[P_{\mu},S^{a}]=-\bar{Q}^{a}\tilde{\sigma}_{\mu}}\,,~~~&~~~
\displaystyle{[P_{\mu},\bar{S}_{a}]=\tilde{\sigma}_{\mu}Q_{a}}\,,~~~~~~\\
{}&{}\\
{}~~~~~~[K_{\mu},Q_{a}]=\sigma_{\mu}\bar{S}_{a}\,,~~~&~~~
{}[K_{\mu},\bar{Q}^{a}]=-S^{a}\sigma_{\mu}\,,~~~\\
{}&{}\\
\multicolumn{2}{c}{\{Q_{a\alpha},S^{b\beta}\}=i\delta_{a}^{~b}
(2\delta_{\alpha}^{~\beta}D+
(\sigma^{[\mu}\tilde{\sigma}^{\nu]})_{\alpha}^{~\beta}M_{\mu\nu})
-2i\delta_{\alpha}^{~\beta}(A^{b}_{~a}
+(\textstyle{\frac{4}{{\cal N}}}-1)\delta_{a}^{~b}R)\,,}\\
{}&{}\\
\multicolumn{2}{c}{\{\bar{Q}^{a}_{\da},\bar{S}^{\db}_{b}\}
=-i\delta_{a}^{~b}
(2\delta_{~\da}^{\db}D-
(\tilde{\sigma}^{[\mu}\sigma^{\nu]})_{~\da}^{\db}M_{\mu\nu})
-2i\delta_{\da}^{~\db}(A^{a}_{~b}
+(\textstyle{\frac{4}{{\cal N}}}-1)\delta_{~b}^{a}R)\,,}\\
{}&{}\\
\multicolumn{2}{c}{\{Q_{a\alpha},\bar{S}^{\db}_{b}\}
=\{\bar{Q}^{a}_{\da},S^{b\beta}\}=0\,.}
\end{array}
\end{equation}
\item Dilations
\begin{equation}
\begin{array}{ll}
{}~~~[D,P_{\mu}]=-iP_{\mu}\,,~~~~&~~~~[D,K_{\mu}]=iK_{\mu}\,,\\
{}&{}\\
{}~~~[D,Q_{a}]=-i\textstyle{\frac{1}{2}}Q_{a}\,,~~~~&~~~~
{[D,\bar{Q}^{a}]=-i\textstyle{\frac{1}{2}}\bar{Q}^{a}}\,,\\
{}&{}\\
{}~~~[D,S^{a}]=i\textstyle{\frac{1}{2}}S^{a}\,,~~~~&~~~~
[D,\bar{S}_{a}]=i\textstyle{\frac{1}{2}}\bar{S}_{a}\,,\\
{}&{}\\
\multicolumn{2}{c}{[D,D]=[D,R]=[D,M_{\mu\nu}]=[D,A^{a}_{~b}]=0\,.}
\end{array}
\end{equation}
\item R-symmetry, $\mbox{U}(1)\times\mbox{SU}({\cal N})$
\begin{equation}
\begin{array}{ll}
{}~~~[R,Q_{a}]=-i\textstyle{\frac{1}{2}}Q_{a}\,,~~~~&~~~~
[R,\bar{Q}^{a}]=i\textstyle{\frac{1}{2}}\bar{Q}^{a}\,,\\
{}&{}\\
{}~~~[R,S^{a}]=i\textstyle{\frac{1}{2}}S^{a}\,,~~~~&~~~~
[R,\bar{S}_{a}]=-i\textstyle{\frac{1}{2}}\bar{S}_{a}\,,\\
{}&{}\\
\multicolumn{2}{c}{[R,R]=[R,P_{\mu}]=[R,K_{\mu}]
=[R,M_{\mu\nu}]=[R,A^{a}_{~b}]=0\,,}~~~\\
{}&{}\\
\multicolumn{2}{c}{
[A^{a}_{~b},A^{c}_{~d}]
=2i(\delta^{c}_{~b}A^{a}_{~d}-\delta^{a}_{~d}A^{c}_{~b})\,,}\\
{}&{}\\
{}~~~[A^{a}_{~b},Q_{c}]=-2i\delta^{a}_{~c}Q_{b}\,,~~~~&~~~~
[A^{a}_{~b},\bar{Q}^{c}]=2i\delta_{b}^{~c}\bar{Q}^{a}\,,\\
{}&{}\\
{}~~~[A^{a}_{~b},S^{c}]=2i\delta_{b}^{~c}S^{a}\,,~~~~&~~~~
[A^{a}_{~b},\bar{S}_{c}]=-2i\delta^{a}_{~c}\bar{S}_{b}\,,\\
{}&{}\\
\multicolumn{2}{c}{
[A^{a}_{~b},P_{\mu}]=[A^{a}_{~b},K_{\mu}]=[A^{a}_{~b},M_{\mu\nu}]=0\,.}
\end{array}
\label{USU}
\end{equation}
\end{itemize}

\section{Realization of 
$\mbox{O}(2,4)\stackrel{{}_{{}_{\displaystyle{\sim}}}}{\displaystyle{=}}
\mbox{SU}(2,2)$ structure in $M$  \label{4AppendixD}}
We exhibit explicitly the relation of the four-dimensional conformal 
group to\newline 
$\mbox{O}(2,4)\cong\mbox{SU}(2,2)$ 
by introducing six-dimensional gamma
matrices with $A=0,1,\cdots,5$
\begin{equation}
\left(\begin{array}{cc}
      0&\Sigma^{A}\\
   \tilde{\Sigma}^{A}&0
\end{array}\right)\,.
\label{48gamma}
\end{equation}
$\Sigma^{A},\tilde{\Sigma}^{A}$ satisfy
\begin{equation}
\Sigma^{A}\tilde{\Sigma}^{B}+
\Sigma^{B}\tilde{\Sigma}^{A}=
2G^{AB}\,,
\end{equation}
where $G^{AB}
=\mbox{diag}(+1,-1,-1,-1,-1,+1)$. In particular, here we
choose $\Sigma^{A},\tilde{\Sigma}^{A}$ as
\begin{equation}
\begin{array}{cll}
\Sigma^{\mu}=\left(\begin{array}{cc}
                 \tilde{\sigma}^{\mu} &0\\
                     0&\sigma^{\mu}
                  \end{array}\right)\,,~~~~&~~~
\Sigma^{4}=\left(\begin{array}{cc}
                     0&i\\
                     -i&0
                  \end{array}\right)\,,~~~~&~~~
\Sigma^{5}=\left(\begin{array}{cc}
                     0&i\\
                      i&0
                  \end{array}\right)\,,\\
{}&{}&{}\\
\tilde{\Sigma}^{\mu}=\left(\begin{array}{cc}
                      \sigma^{\mu} &0\\
                     0&\tilde{\sigma}^{\mu}
                  \end{array}\right)\,,~~~~&~~~
\tilde{\Sigma}^{4}=-\Sigma^{4}\,,~~~~&~~~
\tilde{\Sigma}^{5}=-\Sigma^{5}\,.
\end{array}
\end{equation}
$\Sigma_{A},\tilde{\Sigma}_{A}$ satisfy
\begin{equation}
\left(\begin{array}{cc}
        0&1\\
         1&0
        \end{array}\right)\Sigma_{A}\left(\begin{array}{cc}
        0&1\\
         1&0
        \end{array}\right)=\tilde{\Sigma}_{A}{}^{\dagger}=\Sigma^{A}\,.
\label{4bc}
\end{equation}
For the supermatrix,~$M$, given in eq.(\ref{4Mform}), we may now express 
the $4\times 4$ part in terms of $\Sigma^{AB}\equiv
\textstyle{\frac{1}{2}}\Sigma^{[A}\tilde{\Sigma}^{B]}$ as
\begin{equation}
m\equiv
\left(\begin{array}{cc}
w+\textstyle{\frac{1}{2}}\lambda &-i\ta\\
-i{\rm b} & \tilde{w}-\textstyle{\frac{1}{2}}\lambda
\end{array}\right)
=\textstyle{\frac{1}{2}}w_{AB}
\Sigma^{AB}\,,
\end{equation}
where $w_{45},\,w_{\mu 4},\,w_{\mu 5}$ are given by
\begin{equation}
\begin{array}{ccc}
\omega_{45}=\lambda\,,~~~~&~~~~\omega_{\mu 4}=a_{\mu}-b_{\mu}\,,~~~~&~~~~
\omega_{\mu 5}=a_{\mu}+b_{\mu}\,.
\end{array}
\end{equation}
$\Sigma^{AB}$ 
generates the Lie algebra of $\mbox{O}(2,4)$
\begin{equation}
[\Sigma^{AB},\Sigma^{CD}]=
-G^{AC}\Sigma^{BD}+G^{AD}\Sigma^{BC}
+G^{BC}\Sigma^{AD}-G^{BD}\Sigma^{AC}\,.
\end{equation}
Eq.(\ref{4Mdagger}), the condition on $M$, is  
satisfied partially by eq.(\ref{4bc}).\newline
\indent  In general, $m$ can be defined as a  
$4\times 4$ matrix subject to $\mbox{tr}\:m=0$ and a reality condition 
\begin{equation}
\begin{array}{cc}
bm+m^{\dagger}b=0\,,~~~~~&~~~~
b=\left(\begin{array}{cc}
        0&1\\
         1&0
        \end{array}\right)\,.
\end{array}
\label{4mcon}
\end{equation}
Now, if we write
\begin{equation}
\begin{array}{ll}
\tilde{m}=p^{-1}{}m{}p\,,~~~~~&~~~~b=p{}j{}p^{-1}\,,\\
{}&{}\\
p=\textstyle{\frac{1}{\sqrt{2}}}\left(\begin{array}{cr}
        1&1\\
         1&-1
        \end{array}\right)\,,~~~~~&~~~~
j=\left(\begin{array}{cr}
        1&0\\
         0&-1
        \end{array}\right)\,,
\end{array}
\end{equation}
then $\mbox{tr}\:\tilde{m}=0$ and 
eq.(\ref{4mcon}) is equivalent to 
\begin{equation}
j\tilde{m}+\tilde{m}^{\dagger}j=0\,.
\end{equation}
Hence $\tilde{m}\in\mbox{su}(2,2)$.

\bibliographystyle{unsrt}
\bibliography{reference}

\end{document}